\documentclass[a4paper,11pt,openany]{book}
\usepackage[english]{babel}
\usepackage{lettrine}         
\usepackage[Lenny]{fncychap}   

\usepackage{latexsym}
\usepackage{amssymb}
\usepackage{amsmath}
\usepackage{amsfonts}
\usepackage{eucal}
\usepackage{mathrsfs}
\usepackage{empheq}

\usepackage{graphicx}
\usepackage{psfrag}
\usepackage{pstricks}
\usepackage{pst-node}
\usepackage{pst-plot}
\usepackage{url}
\usepackage[bf, hang]{subfigure}
\usepackage[bf, hang]{caption}

\usepackage{fancyhdr}
\fancypagestyle{plain}{%
  \fancyhead{}
  
}

\usepackage[colorlinks=true,linktocpage=false,pagebackref=true,citecolor=blue]{hyperref}

\begin{document}

\begin{titlepage}

 \begin{center}
     \vspace{2em}
     \includegraphics[height=.15\textheight]{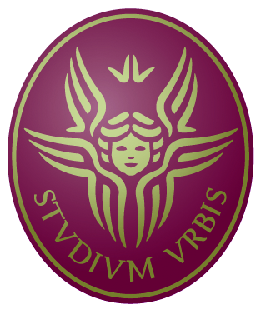}\\
     \vspace{1em}
     {\textsc{Sapienza Universit\`{a} di Roma}}\\
     {\textsc{Dottorato di Ricerca in Fisica}}\\
     {\textsc{Scuola di dottorato ``Vito Volterra''}}\\
     \vspace{4em}
     {\huge \textbf{Competition between Superconductivity and Charge Density Waves:\\ the Role of Disorder}}\\
     \vspace{4em}
     {\textsc{Thesis submitted to obtain the degree of}}\\
     {\textit{``Dottore di Ricerca'' -- Philosophi}\ae{} \textit{Doctor}}\\
     {\textsc{PhD in Physics -- XXI cycle -- October 2008}}\\
     \vspace{4em}
     {\textsc{by}}\\
     \vspace{1em}
     {\Large \textbf{Alessandro Attanasi}}\\
     \vspace{6em}
 \end{center}

  \begin{center}
    \begin{tabular}{c c c c c c c c c c}
      \textbf{Program Coordinator} & & & & ~~ & & & & & \textbf{Thesis Advisors} \\[0.2cm]
      \large{Prof. Enzo Marinari} & & & & ~~ & & & & & \large{Prof. Carlo Di Castro}\\[0.2cm]
      & & & & ~~ & & & & & \large{Dr. Jos\'e Lorenzana}\\[0.2cm]
      & & & & ~~ & & & & & \large{Dr. Andrea Cavagna} \\
    \end{tabular}
  \end{center}

\end{titlepage}

  \clearpage{\pagestyle{empty}\cleardoublepage}

  \pagenumbering{roman}
  \pagestyle{empty}

\vspace{3cm}

\begin{flushright}
\begin{minipage}[t]{3cm}
\begin{center}
\emph{To my princess}\newline
\emph{Mim\`i}
\end{center} 
\end{minipage}
\end{flushright}

  \clearpage{\pagestyle{empty}\cleardoublepage}
  \tableofcontents
  \clearpage{\pagestyle{empty}\cleardoublepage}
 
  \pagestyle{fancy}
  \newpage

  \setcounter{page}{1}
  \chapter*{List of Abbreviations}
  \addcontentsline{toc}{chapter}{List of Abbreviations}
  \markboth{List of Abbreviations}{List of Abbreviations}
  \begin{center}
\begin{tabular}{lll}
\textbf{ARPES} &  & Angle-Resolved-PhotoEmission-Spectroscopy \\ 
\textbf{BCS} &  &  Bardeen-Cooper-Schrieffer\\ 
\textbf{CDW} &  &  Charge Density Waves\\ 
\textbf{DOS} &  &  Density Of States\\ 
\textbf{GPE} &  &  Giant Proximity Effect\\ 
\textbf{HTS} &  &  High-Temperature Superconductors\\ 
\textbf{LDOS} &  &  Local Density Of States\\ 
\textbf{SC} &  &  Superconductivity\\ 
\textbf{SDW} &  &  Spin Density Waves\\ 
\textbf{STM} &  & Scanning Tunnelling Microscopy\\
{$\mathbf{Bi2212}$} & & $Bi_2Sr_2CaCu_2O_{8+\delta}$\\
 $\mathbf{Y123}$ & & $YBa_2Cu_3O_7$
\end{tabular}
\end{center}

  \clearpage{\pagestyle{empty}\cleardoublepage}

  \chapter*{Introduction}
  \addcontentsline{toc}{chapter}{Introduction}
  \markboth{Introduction}{Introduction}
  The understanding of the High-Temperature Superconductors (HTS) is one of the most challenging problem in the field of the condensed matter physics. Since their discovery in $1986$ \cite{Bednorz_1986}, many experimental and theoretical points are been clarified, but a global comprehension of their features is far away. 
\\
The main focus of this thesis deals with the competition between Superconductivity (SC) and Charge Density Waves (CDW), trying to outcrop the role of quenched disorder in superconductors with short coherence length. We will develop and analyze Ginzburg-Landau like phenomenological models of this competitions with an eye on the physics of underdoped HTS. This competition has been studied in the context of bismuthates \cite{Aharony_1993} and we will extend and deepen these studies for an application to cuprates.

Cuprates are very complex materials so we will concentrate ourselves on simplified models which capture at least qualitatively their physics. In this regard  several recent experiments show a qualitatively different behaviors for HTS in respect to traditional superconductors and we will try to find a qualitative explanation to at least some of them. In particular the following experiments will be addressed:  
\begin{itemize}
 \item {\bf Giant Proximity Effect:} It has been observed \cite{Bozovic_2004} that in a $SS'S$ junction, where $S$ and $S'$ are two different HTS with $T_c(S')<T_c(S)$, there is an anomalous proximity effect in the temperature range where $T_c(S')<T<T_c(S)$. The anomaly is that Josephson effect is observed even if the intermediate layer has a width which is several times the superconductivity coherence length, in strong contrast with BCS theory.  
\item {\bf Precursor diamagnetism above $\mathbf T_c$ :} Tunneling experiments \cite{Iguchi_2001,Iguchi_2002} as well as susceptibility experiments
  \cite{Panagopoulos_2004,Panagopoulos_2004_1,Cabo_2006,Thisted_2003,Thisted_2004,Lascialfari_2003,Bodi_2002} have revealed anomalous diamagnetic activity above $T_c$.
\item {\bf Anomalous Nernst effect:} Nernst transport experiments
\cite{Podolsky_2007,Mukerjee_2004,Ussishkin_2002,Wang_2001,Ong_2004,Xu_2000,Wang_2006} have revealed the presence of vortex like excitations for temperatures much larger than $T_c$.  
\item {\bf Modulations in the vortex core:} Conductance modulations in Scanning Tunneling Miscroscopy (STM) has been interpreted as CDW formation at the vortex core \cite{Hoffman_2002}. 
\end{itemize} 

More recently STM experimets present firm evidence of some kind of charge modulation in underdoped cuprates\cite{Kohsaka_2007}. The peculiar observations of the above experiments are located in the so called ``pseudo-gap'' region of the phase diagram, just over the ``superconducting-dome''.

The model that will be used captures in a simple way the idea that the ``pseudo-gap'' phase is formed of bound fermion pairs which are close to a CDW instability but generally do not have long range order due
to quenched  disorder. Thus the charge degrees of freedom will be modeled by an Ising order parameter in the presence of quenched disorder, so representing a charge glassy phase. This glassy phase will be in competition with a superconducting phase modeled by a complex order parameter. 

In order to derive the model the additional simplified assumption has been made, that the order parameter behaves similarly as in a large negative $U$ Hubbard model close to half-filling. This last condition can appear rather arbitrary but can be at least qualitatively justified in some microscopic ``stripe'' like  models. The model is clearly oversimplified in that it ignores some basic features of cuprates like the d-wave symmetry of the superconducting gap and the complexity of the possible CDW textures.  However it
captures in a simple way the competition between CDW and SC. The two order parameters can be embodied in a single $SO(3)$ order parameter with the order along the $z$-axis corresponding to CDW order and the order in the $xy$-plane corresponding to superconducting order. Thus the model can be written in the lattice as an anisotropic Heisenberg model. Also we are interested in the long wave length physics where quantum effects can be taken into account as renormalization of the parameters, thus the model reads:
\begin{equation}\label{eq:heis}
 H=-J\sum_{\langle i,j\rangle}\vec{S}_i\cdot\vec{S}_j-G\sum_i(S_i^z)^2+\frac{W}{2}\sum_ih_iS_i^z
\end{equation} 
where $\vec{S}_i=\{S_i^x,S_i^y,S_i^z\}$ is a classical Heisenberg pseudospin (hereafter spin)  with
$|\vec{S}|=1$, $J$ is a positive coupling constant. The first term represents the nearest neighbor interaction of the order parameter. For simplicity we are assuming that this term is isotropic in the spin space. The second term breaks the symmetry in the spin space with $G>0$ favoring a CDW and $G<0$ favoring a superconducting state; $h_i$ are statistical indipendent quenched random variables with a flat probability distribution between $-1$ and $+1$; also $W>0$. These random fields would mimic impurities always present in the real samples. 

As said above this effective Hamiltonian is inspired on the negative $U$ Hubbard model at strong coupling. Indeed if we start from a one band generalized attractive Hubbard model at half-filling: 
\begin{eqnarray}\label{eq:hm}
H=\sum_{<i,j>}\sum_{\sigma}t_{ij}c_{i\sigma}^\dag c_{j\sigma}-\frac{1}{2}|U|\sum_{i,\sigma}n_{i\sigma}n_{i,\overline{\sigma}}+\nonumber\\
+\frac{1}{2}\sum_{<i,j>}\sum_{\sigma,\sigma'}W_{ij}n_{i\sigma}n_{j\sigma'}+\sum_{i,\sigma}E_in_{i\sigma}
\end{eqnarray}
where $W_{ij}$ is the interatomic interaction and $E_i$ is a random single site energy, one can do the following transformations. First, an attraction-repulsion transformation in order to map the starting Hamiltonian into a half-filled positive Hubbard model; second, in the strong coupling limit $(U\gg t_{ij})$ it is possible to perform a canonical transformation that maps the positive Hubbard model into an antiferromagnetic quantum Heisenberg model. At long wave lengths the antiferromagnetic  order parameter behaves classically in the sense that the only effect of quantum fluctuations is to renormalize the original parameters (renormalized classical regime). Thus the spin is treated as a classical variable and at this point the antiferromagnetic model can be mapped trivially on the ferromagnetic one just by using the staggered magnetization as a variable. This gives a slightly  different version of the model with anisotropy in $J$ induced by the $W$ term in Eq.~(\ref{eq:hm}). But we prefer Eq.~(\ref{eq:heis}) which has essentially the same symmetries but is easier to analyze. 

As a first step we will study the model in one dimension without disorder to familiarize  with it and also to search for a possible explanation of the Giant Proximity Effect (GPE). In the continuum limit the model reads:
\begin{eqnarray}
 F[\theta(x),\phi(x)] & = & 
\int dx \Big\{\frac{\rho}{2}\Big[\Big(\frac{d \theta}{dx}\Big)^2+\cos^2\theta\Big(\frac{d\phi}{dx}\Big)^2\Big]-g\sin^2\theta\Big\} 
\end{eqnarray}
where $\theta(x)$ is the azimuthal angle of our order parameter (controlling the CDW-SC competition), $\phi(x)$ is its angle in the $xy$-plane, i.e. the superconducting phase, while $\rho$ and $g$ are the coupling constants in the continuum limit with $\rho$ playing the role of a $SO(3)$ stiffness.  

To address the Giant Proximity Effect the model will be studied in a Josephson junction geometry of the type S-CDW-S where S represent a superconductor and the role of the Josephson barrier is played by the
  CDW phase. The idea is that the $S'$ superconductor of the experiment done at $T_c(S')<T<T_c(S)$ is in reality a CDW which condensates in a superconductor at $T<T_c(S')$.  In practice one takes $g>0$ ($g<0$) inside (outside) the barrier region. 

After that we will investigate numerically the ground state properties of the $2$-dimensional lattice model Eq. (\ref{eq:heis}), focusing our attention especially on the stiffness and magnetization of the system in the $xy$ plane. We will show that disorder can induce superconductivity in a CDW phase. This picture is really interesting because it could show how an insulating system can produce a superconducting phase thank to the interplay with impurites.

\section*{Outline of the Thesis}
\begin{description}
\item[] In Chapter (\ref{chapter1}) we will review the basic ideas regarding superconductors in a simple pedagogical way.
\item[] In Chapter (\ref{chapter2}), as in the previous one, we will introduce the fundamental concepts of the Charge Density Wave state.
\item[] In Chapter (\ref{chapter3}) we will illustrate the model Eq. (\ref{eq:heis}) starting from the microscopic Hubbard model.
\item[] In Chapter (\ref{chapter4}) we will solve the model (without disorder) in 1-dimension and in the continuum limit, in order to discuss the problem of the Giant Proximity Effect.
\item[] In Chapter (\ref{chapter5})  we will describe the results obtained from the numerical study of the 2-dimensional model with quenched disorder.
\item[] In the appendicies (\ref{appendix3}), (\ref{appendix4}), (\ref{appendix5}) it is possible to find more details about calculations performed into the chapters (\ref{chapter3}), (\ref{chapter4}) and (\ref{chapter5}).
\end{description}

  \clearpage{\pagestyle{empty}\cleardoublepage}
  
  \newpage
  \pagenumbering{arabic}
  \setcounter{page}{1}  

  \chapter{Superconductivity: an Overview}
  \label{chapter1}
\vspace{3cm}
In this chapter we want to review some basic concepts regarding the Superconductivity (SC), introducing symbols and notations we shall use hereafter. Following in a some way a historical line, we shall describe what is the penetration depth $\lambda$, the coherence length $\xi$, the superconducting stiffness $\rho_s$, and what are the main differences between the so called Type $I$ and Type $II$ superconductors. We will also speak about of High Temperature Superconductors (HTS), particularly with an eye on some specific experiments that are our starting point in this research thesis, motivating some choices for the model that we will study later.

\section{The Discovery of Superconductivity}
 In $1911$ the Dutch physicist H. Kamerlingh-Onnes discovered a new fascinating natural phenomenon after called Superconductivity\cite{Onnes_1911}\footnote{The true discovery was due to a Onnes' student, who did not appear into the pubblication.}. He wanted to measure the electrical resistance of a substance when it was cooled and purified as much as possible; in great astonishment he observed that the electric resistance of mercury at a temperature below $4.15$ $K$ was zero. This temperature at which the jump of the resistance is observed is called the \emph{critical temperature} $T_c$. After this discovery new challenges and new puzzles broke in the world of science.

\subsection{The Meissner effect}
The progress of superconductivity studies was very slow, because one had to cool metals down to very low temperatures and this task was not so simple; these studies had to be carried out with liquid helium getting to few Kelvin degrees. But the liquefaction of helium was itself another both interesting and difficult problem. It was twenty-two years after Onnes' discovery that the second fundamental property of superconductors was revealed. W. Meissner and R. Ochsenfeld \cite{Meissner_1933} observed that a superconducting sample was able to force a constant, but not very strong, magnetic field out of it; now we refer to this effect as the \emph{Meissner effect}. To prove the existence of superconductivity it is necessary, at least, that both fall of resistence and Meissner effect be observed. At this point we could put a simple but also important question: why is a superconductor characterized by both zero resistance and Meissner effect, what is the real difference between a superconductor and a perfect conductor, that has only a zero resistance? To answer this question we describe three ideal experiments following the next pictures with their captions:
\begin{figure}[htb!]
\begin{center}
\includegraphics[clip=true,scale=0.9]{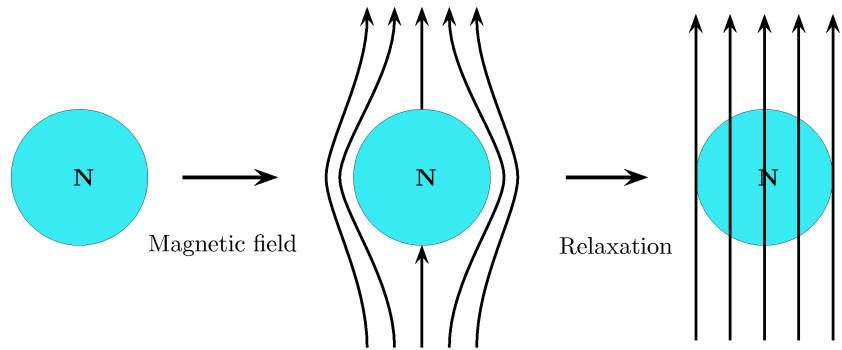}
\caption{{ A normal conductor (N) has a finite resistance at any temperature, so when it is immersed in a magnetic field, currents arise to preserve field's flux, according to the laws of electromagnetic induction. But since the resistance is nonzero, these currents decay and the field penetrates in the ball.}}
\label{fig:uno}
\end{center}
\end{figure}  

\clearpage
\newpage

\begin{figure}[htb!]
\begin{center}
 \includegraphics[clip=true,scale=0.52]{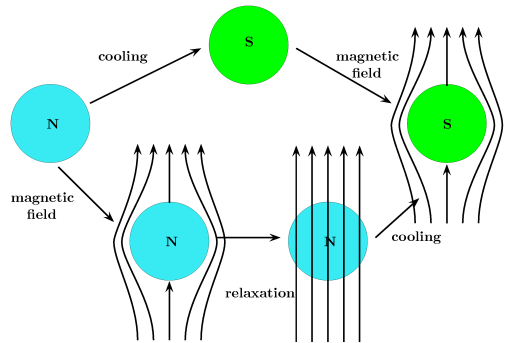}
\caption{\small{ If a normal metal is cooled below its critical temperature $T_c$ becoming a superconductor (S), and then we apply a magnetic field, it is expelled from the sample according to the laws of electromagnetic induction. But because of its zero resistance this situation does not change over the time. On the other way, if we start with a normal metal and we apply a magnetic field we see (as described in the previous picture) that this field penetrates into the sample; but now if we lower the temperature below $T_c$, the field is expelled from the sample. This is precisely the Meissner effect.}}
\label{fig:due}
\end{center}
\end{figure}  

\begin{figure}[htb!]
\begin{center}
 \includegraphics[clip=true,scale=0.52]{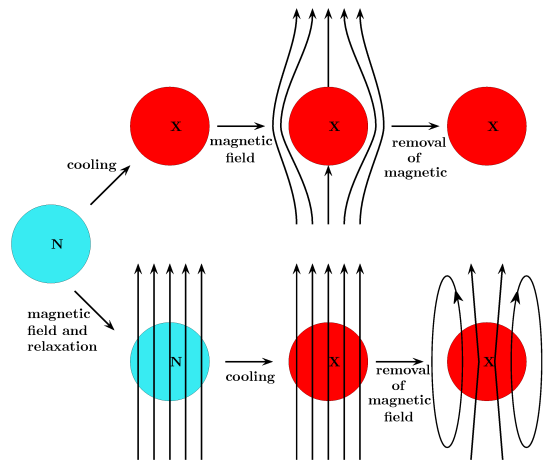}
\caption{\small{If a hypothetical ideal conductor (zero resistance) does not exhibit the Meissner effect, we would see a different behavior. If the temperature is lowered in the presence of the field, when the resistanceless state (X) is achieved the magnetic field is conserved, even if the external field is removed. However, such a state has never been observed in experiments.}}
\end{center}
\label{fig3-cap1}
\end{figure}  

\clearpage
\newpage

\begin{figure}[htb!]
\begin{center}
\includegraphics[clip=true,scale=0.8]{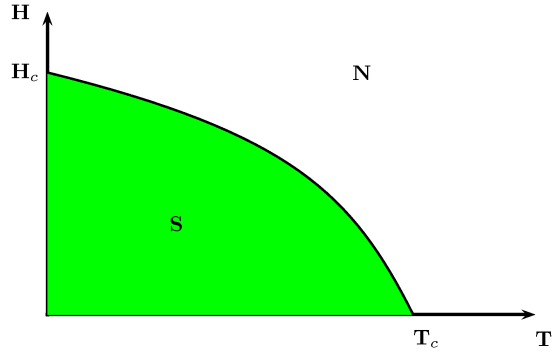}
\caption{Phase diagram of a transition from the normal (N) to the superconductor (S) state.}
\label{fig:HT_phasediagram}
\end{center}
\end{figure}

The Meissner effect is really important, and it proves that the ``superconducting state'' is a reversible equilibrium state, a stable thermodynamic one, while for a simple perfect conductor the magnetic field history is important. The reversibility of the expulsion of a magnetic field from a superconductor implies that the transition between normal and superconducting state is reversible in temperature $T$ and magnetic field $H$; thus there are two phases separated by a critical curve $H=H_c(T)$ as skecthed in Fig: (\ref{fig:HT_phasediagram}). Here we are referring to so-called Type $I$ superconductors, defined more precisely below.  
The critical field $H_c$ can be related to the free energy difference between the normal and the superconducting state.To see this we have to define the thermodynamic potential density of both normal and superconducting state in presence of a magnetic field. Imposing their equality at the transition line we can obtain the expression that defines $H_c$. The thermodynamic potential energy in a magnetic field is the Gibbs free energy, given by:
\begin{equation}
 G(H,T)=F(T)-\frac{1}{4\pi}\int_0^H B(H')dH'
\end{equation}
where $F(T)$ is the Helmholtz free energy. In a superconductor, negleticing surface effects, $B=0$, then:
\begin{equation}
 G_s(H,T)\equiv F_s(T)
\end{equation}
while neglecting the much smaller response of the normal state, we can have $B=H$, then:
\begin{equation}
 G_n(H,T)=F_n(T)-\frac{H^2}{8\pi}
\end{equation}
Along the transition line $H=H_c(T)$ the two thermodynamic potentials are equal, so we obtain:
\begin{equation}
 F_n(T)-F_s(T)=\frac{H_c^2(T)}{8\pi}
\end{equation}
This expression defines the \emph{thermodynamic critical field} $H_c(T)$ as a function of the difference of free energy density between the normal and the superconducting phases. 

It was found empirically that $H_c(T)$ is well approximated by a parabolic law:
\begin{equation}
 H_c(T)\approx H_c(0)[1-(T/T_c)^2]
\end{equation}
and while the transition in zero field at $T_c$ is of second order, the transition in the presence of a field is of first order as we can see observing the jump in the variation of the specific heat between the two phases:
\begin{equation}
 C_n-C_s=T\Big[\frac{\partial^2 G_s}{\partial T^2}-\frac{\partial^2G_n}{\partial T^2}\Big]_H = -\frac{T}{8\pi}\frac{d^2(H_c^2)}{dT^2}
\end{equation}
We have also to remember that for $T=0$ the free energy densities difference between the two states is called condensation energy, and as we show better later it is realted to a kind of condensation of the electrons near the Fermi surface:
\begin{equation}
F_n(0)-F_s(0)=\frac{H_c^2(0)}{8\pi}
\end{equation}

Finally we want to come back to Meissner effect, observing that if a sample is in a magnetic field, the transition to a superconductor requires energy expenditure to expell the magnetic filed outside the bulk of the sample. But if the magnetic field is too strong it is impossible to the sample to gain the superconductovity state at any temperature. There exists also another critical parameter which obstructs the occurrence of superconductivity; it is a \emph{critical current} $j_c$. We known that if there exists an external magnetic field, there is a screening current running along the sample surface and providing the Meissner effect. But we could ask ourselves what happend when a generic transport current runs through a superconductor. This current generates a magnetic field and if it runs into the bulk of the superconducting sample, due to the Meissner effect the magnetic field is forced out of the bulk superconductor. But it is equivalent to say that the transport currents must run on the surface. So all currents are on the surface. Obviously it is impossible for these currents to flow in a zero thick layer, but they are distributed over a certain thickness; so the magnetic field penetrates inside the superconductor too. Both currents and magnetic fields decrease with depth into the material and the tipical length scale over which they go to zero is called \emph{London penetration depth} $\lambda_{\scriptscriptstyle L}$, to be defined later. It is important to note that the above assertions are valid for the so called Type $I$ superconductors, which we shall  describe more deeply later; up to now we can say that these superconductors were the first historically  discovered and their physics is well understood.
 
\subsection{The London equation}
In $1935$ F. London and H. London \cite{London_1935} gave the first theoretical description of the behavior of a superconductor in a magnetic field. 

In an approximate way we can write down the following free energy for a superconductor into a magnetic field:
\begin{equation}
F=\int F_s(0)d\vec{r} + E_{kin} + E_{mag} 
\end{equation}
where $F_s(0)$ is the free energy density of the condensed state, $E_{kin}$ is the kinetic energy related to the current $\vec{j}_s$, and $E_{mag}$ is the magnetic energy due to the field $\vec{h}$. But we have:
\begin{eqnarray}
 \int F_s(0)d\vec{r} &\equiv& E_0 = {\mathrm{constant}}\\
 E_{kin} &=& \int \frac{1}{2}mv^2n_s\,d\vec{r}\label{eq:kin_ener}\\
 E_{mag} &=& \int \frac{h^2(\vec{r})}{8\pi}\,d\vec{r}
\end{eqnarray}
where $n_s$ is the superconducting electron density, and Eq: (\ref{eq:kin_ener}) is valid only if the velocity (and thus the current $\vec{j}_s$) is a spatially slow function. Now we have to remember the definition of the current and the Maxwell equation relating the magnetic field with the current:
\begin{eqnarray}
 \vec{j}_s &=& n_se\vec{v}(\vec{r})\\
\vec{\nabla}\cdot\vec{h} &=& \frac{4\pi}{c}\vec{j}_s
\end{eqnarray}
where $e$ is the electron charge; thus we can write the kinetic energy as:
\begin{eqnarray}
 E_{kin}&=&\int \frac{1}{2}m\Big(\frac{j^2_s}{n_s^2e^2}\Big)n_s\,d\vec{r}\nonumber\\
&=& \int \frac{1}{2}\frac{m}{n_se^2}\frac{c^2}{(4\pi)^2}|\vec{\nabla}\cdot\vec{h}|^2\,d\vec{r}
\end{eqnarray}
Now we can rewrite the free energy for the superconductor as:
\begin{equation}
 F = E_0 +\frac{1}{8\pi}\int\Big[h^2+\lambda_{\scriptscriptstyle L}^2|\vec{\nabla}\cdot\vec{h}|^2\Big]\,d\vec{r}
\end{equation}
where
\begin{equation}
 \lambda_{\scriptscriptstyle L}\equiv\sqrt{\frac{mc^2}{4\pi n_se^2}}
\end{equation}
is the so called London length. Now minimizing the free energy with respect to the variation of the field distributions $\delta\vec{h}(\vec{r})$ we can obtain the equilibrium state:
\begin{eqnarray}
 \delta F &=& \frac{1}{4\pi}\int[\vec{h}\cdot\delta\vec{h}+\lambda^2_{\scriptscriptstyle L}\vec{\nabla}\times\vec{h}\cdot\vec{\nabla}\times\delta\vec{h}]\,d\vec{r}\nonumber\\
&=&\frac{1}{4\pi}\int[\vec{h}+\lambda^2_{\scriptscriptstyle L}\vec{\nabla}\times\vec{\nabla}\times\vec{h}]\cdot\delta\vec{h}\,d\vec{r}\equiv0
\end{eqnarray}
thus the so called London equation is:
\begin{equation}
\label{eq:London}
 \vec{h}+\lambda^2_{\scriptscriptstyle L}\vec{\nabla}\times\vec{\nabla}\times\vec{h}=0
\end{equation}
and it allows, within the Maxwell equations, to obtain the magnetic field and current distributions.

\begin{figure}[htb!]
\begin{center}
\includegraphics[clip=true,scale=0.8]{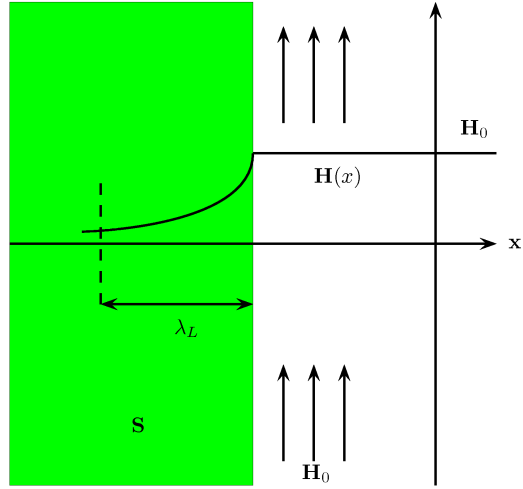}
\caption{The boundary between a superconductor and a magnetic field. The field falls exponentially with increasing depth.}
\label{fig:londondepth}
\end{center}
\end{figure}

We can see at work the London theory investigating the properties of a flat semi-infinite superconducting sample parallel to the $XY$ plane. Taking in mind the Maxwell equations:
\begin{eqnarray}
\vec{\nabla}\times\vec{h} &=& \frac{4\pi}{c}\vec{j}_s\label{eq:maxwell}\\
\vec{\nabla}\cdot\vec{h} &=& 0
\end{eqnarray}
and assuming the magnetic field $\vec{h}$ orthogonal to the $z$ axis\footnote{If the magnetic field is parallel to the $z$ axis, using the Maxwell equations and the London equation it is possible to see that the magnetic field cannot penetrate into the superconductor.}, and without loss of generality parallel to the $x$ axis, the Maxwell equation Eq: (\ref{eq:maxwell}) becomes:
\begin{equation}
 \frac{dh}{dz}=\frac{4\pi}{c}\vec{j}_s
\end{equation}
and then the London equation reads:
\begin{equation}
 h-\lambda_{\scriptscriptstyle L}^2\frac{d^2h}{dz^2}=0
\end{equation}
and its solution is:
\begin{equation}
 h(z)=h(0)e^{-z/\lambda_{\scriptscriptstyle L}}
\end{equation}
This result show that the magnetic field is able to penetrate into the superconductor over a length of the order of the London length $\lambda_{\scriptscriptstyle L}$, falling off exponentially (see Fig. (\ref{fig:londondepth})).
The penetration depth $\lambda_{\scriptscriptstyle L}$ depends on the properties of the material and its order of magnitude, for many common materials such as Aluminium, Mercury, Niobium, is roughly $10^2$ \AA. It has to underline that $\lambda_{\scriptscriptstyle L}$ is temperature-dependent, so as the temperature is increased from zero to a critical value, $\lambda_{\scriptscriptstyle L}$ increases too. And we can imagine the loss of superconductivity upon heating as the increase of $\lambda_{\scriptscriptstyle L}$ until it covers the whole of the sample. 

\subsection{The coherence legth $\xi$ and the energy gap $\Delta$}
Pippard gaves an estimate for $\xi$ using the uncertainty principle; the correlation distance of the superconducting electrons $\xi$ is related to the range of momentum $\delta p$ by:
\begin{equation}
 \xi\delta p\sim\hslash
\end{equation}
In the condensation process the electrons involved are those whitin  distance $K_BT_c$ of the Fermi surface, i.e.
\begin{equation}
 K_BT_c\sim v_{\scriptscriptstyle F}\delta p
\end{equation}
where $v_{\scriptscriptstyle F}$ is the Fermi velocity. The cooherence length is such that:
\begin{equation}
 \xi\sim\frac{\hslash v_{\scriptscriptstyle F}}{K_BT_c}
\end{equation}

Another important feature is the existence of a gap in the low energy excitations, in which the electrons are bound in so called Cooper pairs of size $\xi$. A gap $\Delta$ appears in the excitation spectra, and it is of the order of the energy to break a Cooper pair.
Another relation for $\xi$ can be deduced from the knowledge of $\Delta$, indeed to create a Cooper pair, the important momentum range is given by:
\begin{equation}
 \delta p\approx \frac{2\Delta}{v_{\scriptscriptstyle F}}
\end{equation}
thus using again the uncertainty principle we have:
\begin{equation}
 \xi=\frac{\hslash v_{\scriptscriptstyle F}}{\pi\Delta}
\end{equation}
where the factor $1/\pi$ is arbitrary and chosen for convenience. It is really interesting to observe also that the energy gap $\Delta$ and the critical temperature $T_c$ are proportional each other.

\subsection{Type $I$ and Type $II$ superconductors}
Up to now we have described general features of superconductors whitout specifying if we can apply them to all superconducting materials. The existence of the coherence length $\xi$ and of the London penetration depth $\lambda_{\scriptscriptstyle L}$ leads to a natural classification of superconductors into two categories, which result to have very different properties:
\begin{enumerate}
 \item Type $I$ superconductors for which $\lambda_{\scriptscriptstyle L}\ll\xi$
 \item Type $II$ superconductors for which $\lambda_{\scriptscriptstyle L}\gg\xi$
\end{enumerate}

\begin{figure}[htb!]
\begin{center}
\includegraphics[clip=true,scale=0.8]{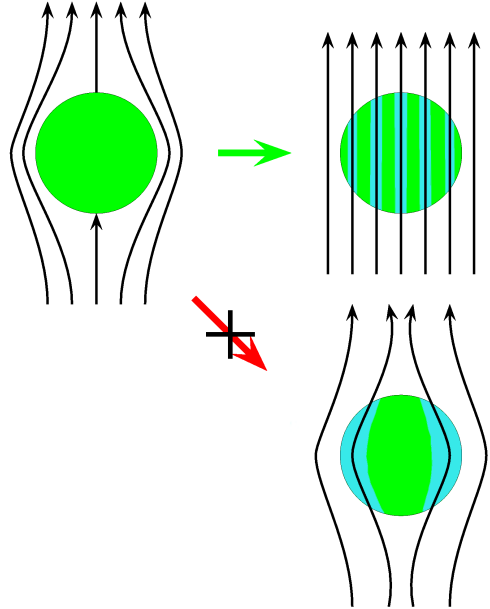}
\caption{A superconducting ball in a magnetic field. In a sufficiently strong magnetic field an intermediate state occurs; the normal regions are traversed by magnetic field lines.}
\label{fig:intermediate}
\end{center}
\end{figure}

Type $I$ superconductors, also called Pippard superconductors, behave roughly as ``ideal'' superconductors, they are described not by the London theory but by the nonlocal Pippard theory. From the microscopic point of view, the properties of Type $I$ superconductors are well explained by the BCS theory. We have to point out also that the behavior of the magnetic field could be more sophisticated in a Type $I$ superconductor than described up to here.  For example if we have a sample with a dimension $d$ less than the penetration depth $\lambda_{\scriptscriptstyle L}$, or with a peculiar geometry, we can see an unobvious penetration of the magnetic field in the sample where some regions are superconducting and others are normal. In these situations we shall speak of the \emph{intermediate state}. To clear this situation we can imagine a superconducting ball (whose radius is much larger than $\lambda_{\scriptscriptstyle L}$) immersed in a weak uniform magnetic field. We assume that initially the magnetic field is completely forced out of the ball, so now the magnetic field is nonuniform; its magnitude near the ``poles'' is smaller, while near the ``equator'' is greater. If we strengthen the magnetic field it reaches the critical value firstly near the equator, so this field destroys superconductivity penetrating inside. But when the field penetrates inside the ball its magnitude decreases, it will become lower than critical value and it therefore cannot obstruct superconductivity. So now we are in a paradoxical situation;  the solution is that the sample splits into alternating normal and superconducting zones and ``transmits'' the field through its normal zones (see Fig. (\ref{fig:intermediate})). Such a state is the intermediate state.
So in the Type $I$ superconducting phase diagram we have not only the thermodynamic field transition line $H_c(T)$, but also another transition line $H_{c1}(T)$. The critical field $H_{c1}$ is smaller than the termodynamic critical field $H_c$, and for field $H$ smaller then $H_{c1}$ we observe a perfect Meissner effect with the total exclusion of the field from the sample, while for $H_{c1}\,<\,H\,<\,H_c$ in the sample there is the establishment of the ``intermediate state'', as described above. Type $II$ superconductors follows the London equation for small fields, and they are completely different from Type $I$ superconductors. The concept of a ``thermodynamic critical field'' $H_c$ for a Type $II$ superconductor can be introduced, but it is only a convenient concept; for this kind of superconductors we have to define two other different critical field\footnote{Here we are neglecting surface superconductivity, and then the possibility to define the critical field $H_{c3}$, that is greater then $H_{c2}$}, $H_{c1}$ and $H_{c2}$. For field $H$ smaller then $H_{c1}$ we observe a perfect Meissner effect, while for $H_{c1}\,<\,H\,<\,H_{c2}$ the field penetrates the sample but this situation is completely different from the intermediate state of Type $I$ superconductors. For these field values in Type $II$ superconductors, eddy currents spontaneously appear in the sample and a vortex state is created as it was theoretically predicted by A. A. Abrikosov \cite{Abrikosov_1957}. The vortex is formed by a normal state core of diameter of the order $\xi$, and screening currents in the region of finite magnetic field of size $\lambda_L$. The diameter is perfectly determined and does not depend on the external magnetic flux, and it does not reach the ordinary dimensions of the intermediate regions in type $I$ superconductors. In a type $II$ superconductor, the vortices are oriented parallel to the magnetic field and also they interact each other with a repulsive force forming a regular triangular lattice. Vortices occur if the external magnetic field strength reaches the so-called lower critical field $H_{c1}$; at this value, vortices penetrate superconductor and if the field strengthens their number increase as their density all superconducting diamagnetic effects are destroyed. This happens when the so-called upper critical field $H_{c2}$ is reached. 

Now we can describe better the vortex state starting from the situation for which the magnetic field is small; this implies that only few field lines penetrate into the sample that has only few regions into the normal state. In this vortex state the superconducting electron density $n_s$ is zero at the centre of the vortex and, over a length of the order of $\xi$, it reaches the maximum value, while the field lines extend themselves for a distance of the odrer of $\lambda$, that in this situation is greater than the coherence length itself. We have to point out also that orthogonal to the vortex line there are eddy currents, and for a single vortex the magnetic flux enclosed by a circle of radius $r\gg\lambda$ around the vortex is quantized:
\begin{equation}
\phi_0 =\frac{ch}{2e}
\end{equation}
where $\phi_0$ is the elemental quantum flux. Now we can study the energy of a single vortex line, and the shape of the magnetic field, for the case $\lambda\gg\xi$; in this situation the vortex core is really small and we can neglect its contribution to the energy, so writing for the vortex energy the following expression:
\begin{equation}
 {\mathcal E} = \int_{r>\xi}\frac{1}{8\pi}(h^2+\lambda^2|\vec{\nabla}\cdot\vec{h}|^2)\,d\vec{r}
\end{equation}
Minimizing the above expression we find the London equation (Eq. (\ref{eq:London})), that is valid outside the vortex core. In order to include the core contribution we can, thanks to small dimension of the core itself, use a delta function, and write down:
\begin{equation}
\label{eq:vortex}
 \vec{h}+\lambda^2\vec{\nabla}\cdot\vec{\nabla}\cdot\vec{h}=\vec{\phi}_0\delta(\vec{r})
\end{equation}
where $\vec{\phi}_0$ is a vector parallel to the vortex line, with an intensity equal to the magnetic flux through the vortex itself. If we now integrate Eq. (\ref{eq:vortex}) over a disk $C$ with radius $r$ concentric to the vortex, we obtain:
\begin{equation}
\label{eq:vortex2}
 \int_C \vec{h}\cdot d\vec{\sigma}+\lambda^2\oint_C\vec{\nabla}\times\vec{h}\cdot\,d\vec{l}=\phi_0
\end{equation}
Using the Maxwell equation $\vec{\nabla}\cdot\vec{h}=0$ it is possible to find the solution of Eq.  (\ref{eq:vortex2}):
\begin{equation}
 h=\frac{\phi_0}{2\pi\lambda^2}K_0(r/\lambda)
\end{equation}
where $K_0$ is a Bessel function; and using the asymptotic expression of $K_0$ we have:
\begin{equation}
 h=\frac{\phi_0}{2\pi\lambda^2}\ln(\lambda/r) \qquad {\mathrm for} \qquad \xi<r\ll\lambda
\end{equation}
\begin{equation}
 h=\frac{\phi_0}{2\pi\lambda^2}\sqrt{\frac{\pi\lambda}{2r}}\exp(-r/\lambda) \qquad {\mathrm for} \qquad r\gg\lambda
\end{equation}

Using the explicit expression of the magnetic field it is possible to calculate the energy of the vortex line:
\begin{equation}
 {\mathcal E}=\Big(\frac{\phi_0}{4\pi\lambda}\Big)^2\ln(\lambda/\xi)
\end{equation}
This expression shows more clearly that $\phi_0$ is the elemental quantum flux, because if for example we  double the flux, it is better to have two distinct vortex lines each one with a flux $\phi_0$ than a single flux line with the flux equal to $2\phi_0$. 

The physics of the vortex lines is really rich and interesting, for example these vortex line can interact each other rearranging themselves into a triangular lattice as predicted by Abrikosov. 

\section{The Landau-Ginzburg theory}
In $1950$ L. D. Landau and V. L. Ginzburg \cite{Ginzburg_1950} suggested a more general theory of superconductivity that is used to present day too. It is based on the more general theory of the second order phase transition developed by Landau himself; in a few words we can say that if there exists an order parameter $\psi$ which goes to zero at the transition temperature $T_c$, the free energy may be expanded in powers of $\psi$, and the coefficients of the expansion are regular functions of the temperature. Thus the free energy density is written as:
\begin{equation}
\label{eq:LG1}
 F=F_n+\alpha(T)|\psi|^2+\frac{\beta(T)}{2}|\psi|^4+\cdots
\end{equation}
where $F_n$ is the free energy density of the normal state. The equation (\ref{eq:LG1}) is limited to the case where the order parameter $\psi$ is a constant throughout the specimen. If $\psi$ has a spatial variation, then the spatial derivative of $\psi$ must be added to Eq: (\ref{eq:LG1}), and at the leading order we can write:
\begin{equation}
\label{eq:LG2}
 F=F_n+\alpha(T)|\psi|^2+\frac{\beta(T)}{2}|\psi|^4+\gamma|\nabla\psi|^2+\cdots
\end{equation}

Eq: (\ref{eq:LG2}) would not have been of a great help in the understanding of the properties of superconductors if Ginzburg and Landau had not proposed an extension to describe the superconductors in the presence of a magnetic field. With a great physical insight, they considered the order parameter $\psi$ as a kind of ``wave function'', and in order to ensure the gauge invariance they wrote the free energy density as:
\begin{equation}
 \label{eq:LG}
 F=F_n+\alpha|\psi|^2+\frac{\beta}{2}|\psi|^4+\frac{1}{2m}\Big|\Big(-i\hslash\nabla-\frac{e^*\vec{A}}{c}\Big)\psi\Big|^2+\frac{h^2}{8\pi}
\end{equation}
where $\vec{A}$ is the vector potential for the magnetic field $\vec{h}$, and $e^*$ for Landau and Ginzburg ``had no reason to be different from the electron charge''. Only thanks to the microscopic theory we know that $e^*=2e$.

\subsection{The Ginzburg-Landau equations}
Starting from Eq: (\ref{eq:LG}) and minimizing the free energy $f=\int F\,d\vec{r}$ for variations of the order parameter $\psi$ and of the magnetic field $\vec{h}$, we obtain the famous Ginzburg-Landau equations:
\begin{eqnarray}
\label{eq:LGprep}
\delta f &=& \int d\vec{r}\Big\{\delta\psi^*\Big[\alpha\phi+\beta|\psi|^2\psi+\frac{1}{2m}\Big(-i\hslash\nabla-\frac{2e\vec{A}}{c}\Big)^2\psi\Big]+c.c.\Big\}+\nonumber\\
&&+\int\Big\{\delta\vec{A}\cdot\Big[\frac{\vec{\nabla}\times\vec{h}}{8\pi}-\frac{e}{mc}\psi^*\Big(-i\hslash\nabla-\frac{2e\vec{A}}{c}\Big)\psi \Big]+c.c.\Big\}
\end{eqnarray}
where $c.c.$ means the complex conjugate. For $f$ to be a minimum, i.e. $\delta f=0$, Eq: (\ref{eq:LGprep}) yields the two equations of Ginzburg and Landau:
\begin{equation}
 \label{eq:LG_1}
\frac{1}{2m}\Big(-i\hslash\nabla-\frac{2e\vec{A}}{c}\Big)^2\psi+\alpha\psi+\beta|\psi|^2\psi=0
\end{equation}
\begin{equation}
 \label{eq:LG_2}
\frac{\vec{\nabla}\times\vec{h}}{4\pi}=\frac{\vec{j}}{c}=\frac{e\hslash}{imc}(\psi^*\nabla\psi-\psi\nabla\psi^*)-\frac{4e^2}{mc^2}\psi^*\psi\vec{A}
\end{equation}

\subsection{The two characteristic lengths $\xi(T)$, $\lambda(T)$}
The two Ginzburg-Landau equations (\ref{eq:LG_1}) and (\ref{eq:LG_2}) have two special and obvious solutions:
\begin{enumerate}
 \item $\psi\equiv0$, where $\vec{A}$ is determined only by applied field $\vec{H}$ (that is different from the internal field $\vec{h}$). This solution describes the normal state.
 \item $\psi\equiv\psi_0=\sqrt{-\alpha/\beta}$ and $\vec{A}=0$. This solution describes the ordinary superconducting state with perfect Meissner effect. $\psi_0$ corresponds to the lowest free energy when $\alpha<0$.
\end{enumerate}

In the case of a very weak field, $\psi$ is expected to vary very slowly, close to the value $\psi_0$. The range of variation of $\psi$ can be deduced by the first Ginzburg-Landau equation by setting $\vec{A}=0$. Introducing the rescaled variable:
\begin{equation}
 \Psi=\psi/\psi_0
\end{equation}
Eq: (\ref{eq:LG_1}) is written as:
\begin{equation}
\label{eq:land}
 -\frac{\hslash^2}{2m}\nabla^2\Psi+\alpha\Psi-\alpha\Psi^3=0
\end{equation}
In order to have a rigth dimensional equation, it is natural to introduce the length $\xi(T)$ such that:
\begin{equation}
 \xi^2(T)=-\frac{\hslash^2}{2m\alpha}
\end{equation}
which gives the range of variation of $\psi$. This characteristic length $\xi(T)$\footnote{This length is certainly not the same length as Pippard's $\xi$, since this $\xi(T)$ diverges at the critical temperature, whereas the electrodynamic $\xi$ is essentially constant} is called the temperature dependent coherence length. Its physical meaning is clear: if we depress the superconducting order parameter at one point, $\xi$ represents the distance over which the order parameter is recovevered (indeed if we solve Eq:  (\ref{eq:land}) neglecting the non linear term, we can easily find a solution for which the order parameter decays exponentially over a distance of the order of $\xi$).

If we now eamine Eq: (\ref{eq:LG_2}) for the current, to the first order in $h$, $|\psi|^2$ can be replaced by $\psi_0^2$, i.e.:
\begin{equation}
 \vec{j}=\frac{e\hslash}{im}(\psi^*\nabla\psi-\psi\nabla\psi^*)-\frac{4e^2}{mc}\psi_0^2\vec{A}
\end{equation}
Taking the curl of the current, one obtains:
\begin{equation}
 \vec{\nabla}\times\vec{j}=-\frac{4e^2}{mc}\psi_0^2\vec{h}
\end{equation}
which is equivalent to the London equation, with penetration depth:
\begin{equation}
 \lambda(T)=\sqrt{\frac{mc^2}{16\pi e^2\psi_0^2}}
\end{equation}
The above expression for the penetration depth is equal to the London ones, where the number of superconducting electrons $n_s$ is replaced by $4\psi_0^2$. As in the London theory, $\lambda$ determines the range of variation of the magnetic field.

\subsection{The Ginzburg-Landau parameter $\kappa(T)$}
The ratio between the two characteristic lengths $\xi(T)$ and $\lambda(T)$ defines the so-called Ginzburg-Landau parameter $\kappa$:
\begin{equation}
 \kappa(T)=\frac{\lambda(T)}{\xi(T)}
\end{equation}

This parameter $\kappa$ is useful to distinguish Type $I$ superconductors from Type $II$ superconductors; for the former ones we have $\kappa\ll1$, while for the latter $\kappa\gg1$. The value $\kappa=1/\sqrt{2}$ \footnote{$\kappa=1/\sqrt{2}$ is the exact value for which the surface energy between a superconductor and a normal metal goes from negative (Type $II$ superconductors) to positive (Type $I$ superconductors)  values.} is special because it is the boundary line between the two families fo superconductors. 

\section{The BCS theory}
In $1957$ three American physicist, J. Bardeen, L. Cooper and J. R. Schrieffer \cite{Bardeen_1957} discovered the mechanism of the superconductivity and nowadays it is often called \emph{Cooper Pairing}. It is a milestone in the history not only of the condesed matter physics, but also of the entire physics. Here we will show something about the theory developed by them, the so-called BCS theory.

\subsection{Attractive interaction and Cooper's argument}
If we consider the ground state of a free electron gas, we have to fill every energetic single electron level untill the Fermi level, everyone with a momentum $\vec{k}$ and an energy $\hslash^2k^2/2m$. Cooper showed with a simple argument that a very small attractive interaction into the system was able to get the  ground state unstable.  

If we take two electrons at the positions $\vec{r}_1$ and $\vec{r}_2$ interacting each other, and we treat the other ones as a free electron gas, the first two electrons because of the Pauli's exclusion principle will stay into states with momentum $k>k_F$. Choosing states for which the centre of mass of the two electrons is fixed, their wave function $\Psi(\vec{r}_1,\vec{r}_2)$ will be only a function of the difference $\vec{r}_1-\vec{r}_2$. Expanding $\Psi$ as plane waves, we have:
\begin{equation}
\label{eq:wave}
 \Psi(\vec{r}_1,\vec{r}_2)=\sum_kg(\vec{k})e^{i\vec{k}\cdot(\vec{r}_1,\vec{r}_2)}
\end{equation}
where $g(\vec{k})$ is the probability amplitude to find one electron in a plane wave state $\vec{k}$ and the other one into the state $-\vec{k}$ (We have to point out also that $g(\vec{k})=0$ for $k<k_F$ because of the occupied states by the other free electrons gas). The Schr\"odinger equation for the two electrons is:
\begin{equation}
 -\frac{\hslash^2}{2m}(\vec{\nabla}_1^2+\vec{\nabla}_2^2)\Psi+V(\vec{r}_1,\vec{r}_2)\Psi=\Big(E+\frac{\hslash^2k_F^2}{m}\Big)\Psi
\end{equation}
where $E$ is the energy of the two electron with respect to the Fermi level, and $V(\vec{r}_1,\vec{r}_2)$ is the interacting potential. Putting the explicit expression of the wave function (Eq: (\ref{eq:wave})) into the Sch\"odinger equation, we obtain:
\begin{eqnarray}
 \frac{\hslash^2}{m}k^2g(\vec{k})+\sum_{k'}g(k')V_{kk'}=(E+2E_F)g(\vec{k})\\
V_{kk'}=\frac{1}{Vol}\int V(\vec{r})e^{-i\vec{r}\cdot(\vec{k}-\vec{k'})}\,d\vec{r}
\end{eqnarray}
This equation within the Pauli principle, represents the so-called Bethe-Goldstone equation for the two electrons problem. If the interaction $V$ is attractive, it is possible to show that a binding state exists. A way to see this is to take a constant attractive potential different from zero only over the Fermi level at an energy $\hslash\omega_D$\footnote{This will be justified in the next subsection.}. If $\hslash\omega_D\ll E_F$ and if we are in the weak interaction limit ($N(0)V\ll1$), we have:
\begin{equation}
 E=-2\hslash\omega_De^{-2/N(0)V}
\end{equation}
where $N(0)$ is the density of states at the Fermi level. We can observe $E<0\,$, so the binding state exists and thus the normal state is unstable with respect to the formation of electrons pairs. 
 
\subsection{The electron-phonon interaction}
Now a good question could be: why into a simple electron gas two electrons have to attract each other? In order to achieve this attraction, the electrons have to couple with other particles or excitations, such as phonons, spin waves $\ldots$ Here we will consider only the electron-phonon interaction, just to give an example of how it is possible to have an attractive interaction.

We want to know the matrix element of the electron-electron interaction between a starting state $|I\rangle$ and a final state $|II\rangle$, for which the electrons are described respectively by plane wave $\vec{k}$,  $-\vec{k}$ for $|I\rangle$ and $\vec{k'}$, $-\vec{k'}$ for $|II\rangle$. This electron-electron matrix element  will have a Coulomb repulsion term $U_q$ and a phononic term; in the latter case the electrons can interact with the lattice via two ways: either the electron $\vec{k}$ emits a phonon with momentum $-\vec{q}$ adsorbed by the electron $-\vec{k}$, or the electron $-\vec{k}$ emits a phonon with momentum $\vec{q}$ adsorbed by the electron $\vec{k}$. Up to the second order of the perturbation theory we can write the matrix element as:
\begin{equation}
\label{eq:Veff}
 \langle I|H|II\rangle=U_q+2\frac{|W_q|^2}{\hslash}\frac{\omega_q}{\omega^2-\omega_q^2}
\end{equation}
where $\omega$ is the energy difference between the starting state $|I\rangle$ and the final state  $|II\rangle$; and $\omega_q$ is the phonon energy. When $\omega<\omega_q$ the phononic term will be negative (i.e. attractive), thus if the Coulomb repulsion is not so big, the total interaction is attractive\footnote{This justifies the assumption of the previous subsection.}.

\subsection{The BCS ground state}
Starting from the above osservations Bardeen, Cooper and Schrieffer proposed their microscopic theory; simplifying the expression of the effective potential of Eq: (\ref{eq:Veff}) to a small square well potential around the Fermi surface, they wrote down the following Hamiltoian in second quantization:
\begin{equation}
 \label{eq:HamBCS}
H=\sum_{\vec{k},\sigma}\epsilon_kc^\dagger_{k,\sigma}c_{k,\sigma}+\sum_{k,\,k',\sigma}V_{kk'}c^\dagger_{k,\sigma}c^\dagger_{-k,\overline{\sigma}}c_{-k,\overline{\sigma}'}c_{k',\sigma}
\end{equation}
where $V_{kk'}$ is the effective potential, $\epsilon_k$ is the kinetic energy of the electrons and $c^\dagger_{k,\sigma}$ ($c_{k,\sigma}$) is the creation (annihilation) operator for an electron with momentum $k$ and spin $\sigma$. 

Because in the real space the electron-phonon coupling is expected to be short range, in the $k$-space it will be long range, so a mean field approach is justified. We can define:
\begin{eqnarray}
 b_k=\langle c_{-k,\downarrow}c_{k,\uparrow} \rangle\\
b_k^*=\langle c^\dagger_{k,\uparrow}c^\dagger_{-k,\downarrow} \rangle
\end{eqnarray}
 then we have:
\begin{equation}
 c_{-k,\downarrow}c_{k,\uparrow}=b_k+\delta b_k\equiv b_k+(c_{-k,\downarrow}c_{k,\uparrow}-b_k)
\end{equation}
substituting these expressions into the Hamiltonian (\ref{eq:HamBCS}) neglecting the square terms $(\delta b_k)^2$, we obtain:
\begin{equation}
\label{eq:Ham1BCS}
H=\sum_{\vec{k},\sigma}\epsilon_kc^\dagger_{k,\sigma}c_{k,\sigma}-\sum_{k}\Big(\Delta^*_kc_{-k,\downarrow}c_{k,\uparrow}+\Delta_kc^\dagger_{k,\uparrow}c^\dagger_{-k,\downarrow}-b_k^*\Delta_k\Big)
\end{equation}
where:
\begin{eqnarray}
 \Delta_{k'}^*=-\sum_kV_{kk'}b^*_k\\
 \Delta_{k}=-\sum_{k'}V_{kk'}b_{k'}
\end{eqnarray}
In order to diagonalize the Hamiltonian (\ref{eq:Ham1BCS}) we need to define two new fermionic operators, $\eta_k$ and $\gamma_k$, by the following unitary transformation:
\begin{eqnarray}
 c_{\scriptscriptstyle k,\uparrow}=\cos\theta\eta_k-\sin\theta\gamma_k^\dagger\\
c_{\scriptscriptstyle -k,\downarrow}^{\scriptscriptstyle \dagger}=\sin\theta\eta_k+\cos\theta\gamma_k^\dagger
\end{eqnarray}
Substituting these operators into the Hamiltonian \ref{eq:Ham1BCS} and putting:
\begin{equation}
 \tan(2\theta)=-\frac{\Delta_k}{\epsilon_k}
\end{equation}
it is possible to diagonalize the Hamiltonian itself\footnote{This means that there are only terms such as $\eta^\dagger\eta$ or $\gamma^\dagger\gamma$, and also every $\Delta_k$ need to have the same phase, i.e. the global order parameter has a global phase.} obtaining:
\begin{equation}
 H=\sum_{k,\sigma}\Big[\epsilon_k+\Delta_kb^\dagger_k\Big]+\sum_kE_k(\eta_k^\dagger\eta_k-\gamma_k^\dagger\gamma_k)
\end{equation}
with:
\begin{equation}
 E_k=\sqrt{\epsilon_k^2+|\Delta_k|^2}
\end{equation}
The expression of $E_k$ shows that $|\Delta_k|$ is a gap into the spectrum, and it represents the superconducting order parameter of the system. 

\subsection{The gap $\Delta$ and the critical temperature $T_c$}
In the BCS framework it is also simple to find a self-consistent equation that defines the gap $\Delta_k$ because the Hamiltonian is written as a free fermion gas. Thus minimizing the free energy of the system with respect to the gap, we obtain:
\begin{equation}
 \Delta_k=-\sum_{k'}V_{kk'}\Delta_{k'}\frac{\tanh(\beta E_{k'}/2)}{2E_{k'}}
\end{equation}
If we also consider the following simple expression of the interaction potential:
\begin{equation}
 V_{kk'}=-V \qquad {\mathrm{for}}\qquad |\omega-\omega_F|<\omega_0
\end{equation}
i.e. an attractive constant interaction around the Fermi level with an amplitude $\omega_0$ (if the nature of the attraction is phononic, the energy $\omega_0$ corresponds to the Debye energy $\omega_D$), the gap is momentum-indipendent, and so the self-consistent equation becomes:
\begin{equation}
 1=V\sum_{k'}\frac{\tanh(\beta E_{k'}/2)}{2E_{k'}}
\end{equation}

Now two simple expansion can be done: one around the critical temperature $T_c$, and the other one around zero temperature.
\vspace{1cm}
\begin{description}
 \item[${\mathbf {T\longrightarrow T_c}}$]
\end{description}

In this limit $\Delta_k\rightarrow0$, so $E_k\simeq\epsilon_k$, and approximating the density of states with its value at the Fermi level $D(0)$ (we have to remember that we are into an energy band around the Fermi level), the self-consistent equation writes as:
\begin{equation}
 1=\lambda\int_0^{\omega_D}\frac{\tanh(\beta\epsilon/2)}{\epsilon}\,d\epsilon
\end{equation}
where $\lambda\equiv VD(0)$. In the weak coupling limit ($\lambda\ll1$) it is possible to find:
\begin{equation}
 K_BT_c=\frac{2e^\gamma}{\pi}\omega_De^{-1/\lambda}
\end{equation}
where $\gamma\simeq0.5772\cdots$ is the Euler constant.  
\vspace{1cm}
\begin{description}
 \item[${\mathbf {T\longrightarrow 0}}$]
\end{description}

In this limit the gap equation becomes:
\begin{equation}
 1=\lambda\int_0^{\omega_D}\frac{d\epsilon}{\sqrt{\epsilon^2+\Delta^2}}
\end{equation}
And in the weak coupling limit ($\lambda\ll1$) we have:
\begin{equation}
 \Delta(T=0)=2\omega_De^{-1/\lambda}
\end{equation}
\vspace{1cm}

From the previous results we can also find a universal ratio value between the gap at zero temperature and the critical temperature:
\begin{equation}
 \frac{\Delta(T=0)}{K_BT_c}\approx1.76\cdots
\end{equation}

\subsection{Anderson's theorem}
An important result that is well known for conventional metallic superconductors is the Anderson's theorem \cite{Anderson_1959}. It states that those materials are insensitive to nonmagnetic impurity doping, so that the superconconductor critical temperature $T_c$, and the superconductor density of states are not affected by the nonmagnetic impurity scattering. In contrast to this behaviour of conventional metallic superconductors, for High-Temperature-Superconductors (see next section for them) doping with a small amount of nonmagnetic impurity like $Zn$ destroys superconductivity, as reported for some materials $YBa_2(Cu_{1-x}Zn_x)_3O_7$ \cite{Tokunaga_1997}, $YBa_2(Cu_{1-x}Zn_x)_4O_8$ \cite{Itoh_2001,Itoh_2001_1}, $La_{1.85}Sr_{0.15}Cu_{1-x}Zn_xO_4$ \cite{Xiao_1990}, $Bi_2Sr_2Ca(Cu_{1-x}Zn_x)_2O_8$ \cite{Kluge_1995}.  

Thus understanding the role of the disorder is really important for a better comprehension of the behaviour of these materials. This will be a main point in our research.

\clearpage
\newpage
 
\section{HTS and cuprates}
In $1986$ K. A. M\"{u}ller and G. Bednorz \cite{Bednorz_1986} discovered new classes of the so-called High-Temperature-Superconductors (HTS). After that a new era in superconductivity began.

We want to review something about these materials, focusing our attention on the cuprates superconductors and on some experiments \cite{Bozovic_2004,Iguchi_2001,Iguchi_2002,Panagopoulos_2004,Cabo_2006,Thisted_2003,Thisted_2004,Lascialfari_2003,Bodi_2002,Podolsky_2007,Mukerjee_2004,Ussishkin_2002,Wang_2001,Ong_2004,Xu_2000,Wang_2006,Hoffman_2002,Kohsaka_2007} that up to now have not a clear interpretation, and that are the starting point of our research thesis. 

\begin{figure}[htb!]
\begin{center}
\includegraphics[clip=true,scale=1.2]{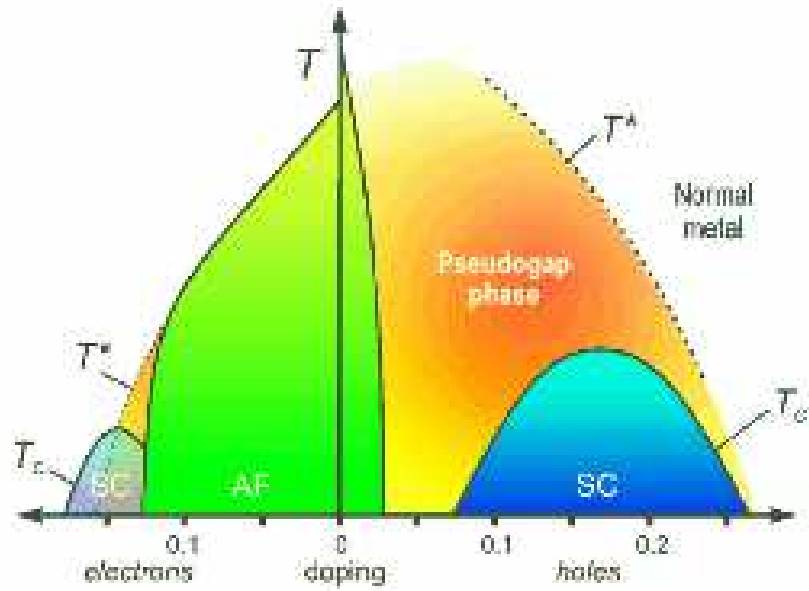}
\caption{Schematic doping phase diagram of electron- and hole-doped HTS (SC indicates the superconducting phase, and AF the antiferromagnetic one). [Figure taken from \cite{Fischer_2007}]}
\label{fig:HTS_phasediag}
\end{center}
\end{figure}

For many years prior to the discovery of HTS, the highest $T_c$ had been stuck at $23$ $K$ ($Nb_3Ge$). This discovery took place in an unexpected material, a transition-metal oxide, so it was clear that some
novel mechanism must be at work. From that time cuprate HTS are studied intensively, both from an experimental point of view and a theoretical point of view. But the superconductivity in these materials is only one aspect of a rich phase diagram which must be understood; many other physical properties are interesting and unclear outside the superconduting region. While there are a lot of HTS materials, they all share a layered structure made up of one or more copper-oxygen planes. But for all of them it is possible to skecth a similar phase diagram (see Fig. \ref{fig:HTS_phasediag})

We start our discussions from a so-called parent compound, the $La_2CuO_4$. There is general agreement that it is an insulator, more precisely a Mott insulator. The last one concept was introduced many years ago \cite{Mott_1949} to describe a situation where a material should be metallic according to band theory, but is insulating due to strong electron-electron repulsion. In our case, in the copper-oxygen layer there is an odd number of electrons per unit cell. More specifically, the copper ion is doubly ionized and is in a $d^{\,9}$ configuration so that there is a single hole in the $d$ shell per unit cell. According to band theory, the band is half-filled and must be metallic. Nevertheless, there is a strong repulsive energy  cost when putting two electrons (or holes) on the same ion, and when this energy (commonly called $U$)  dominates over the hopping energy $t$, the ground state is an insulator due to strong correlation effects. It also follows that the Mott insulator should be an antiferromagnet because when neighboring spins are oppositely aligned one can gain an energy $4t^2/U$ by virtual hopping. This is called the superexchange energy J. The parent compound can be doped by substituting some of the trivalent $La$ by divalent $Sr$. The result is that $x$ holes are added to the $Cu-O$ plane in $La_{2−x}Sr_xCuO_4$, which is called hole doping. In the compound $Nd{2−x}Ce_xCuO_4$ \cite{Tokura_1989}, the reverse happens in that x electrons are added to the $Cu-O$ plane, which is called electron doping. On the hole-doping side the antiferromagnetic order is rapidly suppressed. Almost immediately after the suppression of the antiferromagnet, superconductivity appears. The dome-shaped $T_c$ is characteristic of all hole-doped cuprates. On the electron-doped side, the antiferromagnet is more robust and survives up to higher $x$ concentrations,  beyond which a region of superconductivity arises. We shall focus on the hole-doped materials.

The metallic state above $T_c$ has been under intense study and exhibits many unusual properties not encountered before in any other metal. This region of the phase diagram has been called the pseudogap phase. We will see below that a depletion of the DOS occurs in this region justifying its name. It is not a well-defined phase in that a definite finite-temperature phase boundary has never been found. The region of the normal state above the optimal $T_c$ also exhibits unusual properties. The resistivity is linear in $T$ and the Hall coefficient is temperature dependent \cite{Chien_1991}. Beyond optimal doping (the overdoped region), the standard behaviour returns. A point of view is that the strange metal is characterized by a quantum critical point lying under the superconducting dome \cite{Castellani_1997,Varma_1997,Tallon_2000}. A different point of view is that the physics of HTS is the physics of the doping of a Mott insulator, strong correlation is the driving force behind the phase diagram. The simplest model which captures the strong-correlation physics is the Hubbard model and its strong-coupling limit, the $t-J$ model. 

\subsection{Basic electronic structure of the cuprates}
\begin{figure}[htb!]
\begin{center}
\includegraphics[clip=true,scale=1.2]{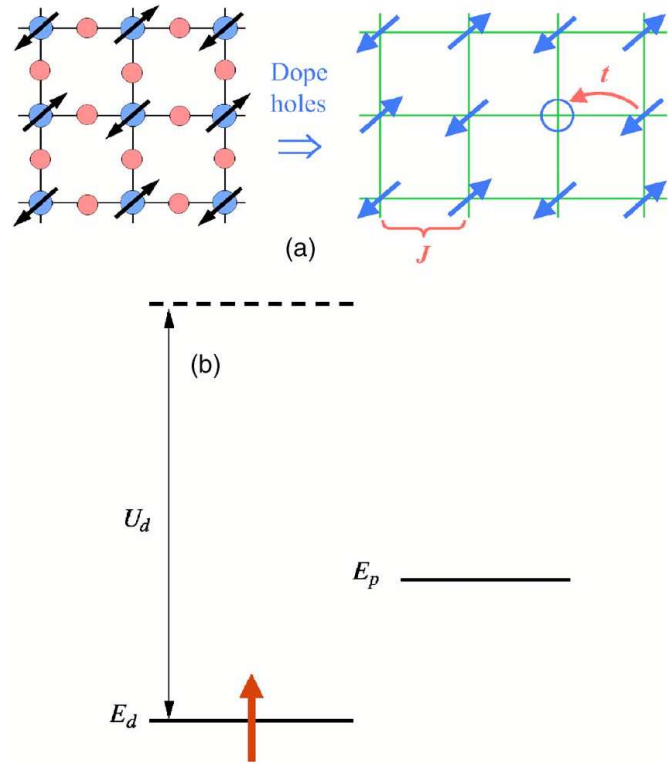}
\caption{Electronic structure of the cuprates. (a) The copper $d$ and the oxygen $p$ orbitals in the hole picture. A single hole with $S=1/2$ occupies the copper $d$ orbital in the insulator. [Figure taken from \cite{Lee_2006}]}
\label{fig:Lee_2006}
\end{center}
\end{figure}

The physics of high-Tc superconductivity is related to the copper-oxygen layer (see Fig. (\ref{fig:Lee_2006})).

In the parent compound $La_2CuO_4$, the formal valence of Cu is $2^+$, which means that its electronic state is in the $d^{\,9}$ configuration. The copper is surrounded by six oxygens in an octahedral environment with the apical oxygen lying above and below $Cu$. The distortion from a perfect octahedron due to the shift of the apical oxygens splits the $e_g$ orbitals so that the highest partially occupied $d$ orbital is $x^2-y^2$. The lobes of this orbital point directly to the $p$ orbital of the neighboring oxygen, forming a strong covalent bond with a large hopping integral $t_{pd}$. Thus the electronic state of the  cuprates can be described by the so-called three-band model, where in each unit cell we have the $Cu$ $d_{x^2-y^2}$ orbital and two oxygen $p$ orbitals \cite{Emery_1987,Varma_1987}. The $Cu$ orbital is singly occupied while the $p$ orbitals are doubly occupied. By substituting divalent $Sr$ for trivalent $La$, the electron count on the $Cu-O$ layer can be changed in a process called doping. For example, in $La_{2−x}Sr_xCuO_4$, $x$ holes per $Cu$ are added to the layer. As seen in Fig. (\ref{fig:Lee_2006}), due to the large $U_d$ the hole will reside on the oxygen $p$ orbital. The hole can hop via $t_{pd}$, and due to translational symmetry the holes are mobile and form a metal, unless localization due to disorder or some other phase transition intervenes. The full description of hole hopping in the three-band model is complicated, on the other hand, there is strong evidence that the low-energy physics (on a scale small compared with $t_{pd}$ and $E_p-E_d$) can be understood in terms of an effective one-band model on the square lattice, with an effective nearest neighbor hopping integral $t$ and with $E_p-E_d$ playing a role analogous to $U$. In the large $U$ limit this maps onto the $t-J$ model plus constraint of non double occupancy:
\begin{equation}
 H=-\sum_{\langle i,j\rangle,\sigma}t_{ij}c^\dagger_{i\sigma}c_{i\sigma}+J\sum_{\langle i,j\rangle}\Big(\vec{S}_i\cdot\vec{S}_j-\frac{1}{4}n_in_j\Big)
\end{equation}
where $c^\dagger_{i\sigma}$ ($c_{i\sigma}$) is the fermion creation (annihilation) operator on site $i$ and spin $\sigma$, $n_i=\sum_\sigma c^\dagger_{i\sigma}c_{i\sigma}$ is the number operator, and $J=4t^2/U$.

\subsection{Underdoped cuprates}
Here we want to show some features of the so-called pseudogap region, that has a rich phenomenology.  Doping an antiferromagnetic Mott insulator with a hole, means that a vacancy is introduced into the antiferromagnetic spin background, and this vacancy would like to hop with an amplitude $t$ to lower its kinetic energy. However, after one hop its neighboring spin finds itself in a ferromagnetic environment.  It is clear that the holes are very effective in destroying the antiferromagnetic background. This is particularly so at $t\gg J$ when the hole is strongly delocalized. The basic physics is the competition between the exchange energy $J$ and the kinetic energy, which is of order $t$ per hole or $xt$ per unit area. When $xt\gg J$, we expect the kinetic energy to win and the system would be a standard metal with a weak residual antiferromagnetic correlation. When $xt\ll J$, however, the outcome is much less clear because the system would like to maintain the antiferromagnetic correlation while allowing the hole to move as freely as possible. This region of the phase diagram is referred to as the pseudogap region.
Below we review some experiments done in this region, focusing our attention on the following ones: 
\begin{description}
\item[$\bullet$ - Knight-shift]
\item[$\bullet$ - Giant proximity effect]
\item[$\bullet$ - STM]
\item[$\bullet$ - Specific heat]
\item[$\bullet$ - Nerst effect]
\item[$\bullet$ - Diamagnetic effects above $T_c$]
\end{description}

\subsection*{Knight-shift}
The Knight shift is a shift in the nuclear magnetic resonance frequency of a paramagnetic substance. It is due to the conduction electrons in metals. They introduce an ``extra'' effective field at the nuclear site, due to the spin orientations of the conduction electrons in the presence of an external field. This is responsible for the shift observed in the nuclear magnetic resonance. Depending on the electronic structure, Knight shift may be temperature dependent. However, in metals which normally have a broad featureless electronic density of states, Knight shifts are temperature independent.

\begin{figure}[htb!]
\begin{center}
\includegraphics[clip=true,scale=1.0]{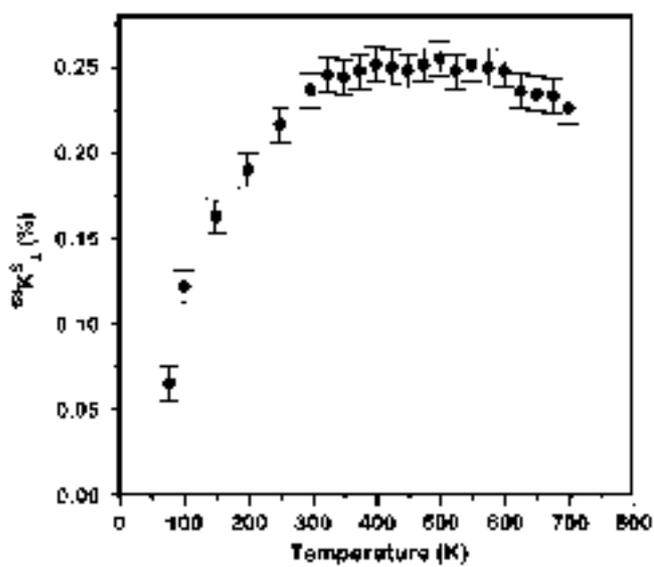}
\caption{Knight shift of the planar $^{63}Cu$ for an underdoped $YBa_2Cu_4O_8$ with $T_c=79\,K$. [Figure taken from \cite{Curro_1997}]}
\label{fig:knightshift}
\end{center}
\end{figure}

But as we can see in Fig. (\ref{fig:knightshift}), the Knight-shift measurement in the underdoped $YB_2Cu_4O_8$ shows that while the spin susceptibility is almost temperature independent between
$700\,K$ and $300\,K$, as in an ordinary metal, it decreases below $300\,K$ and by the time the $T_c$ of $80\,K$ is reached, the system has lost $80\%$ of the spin susceptibility \cite{Curro_1997} indicating the presence of a pseudogap. This one is seen by several other probes, and it may indicate the tendency to form some kind of ordered state, or the presence of preformed pairs, or both.

\clearpage
\newpage

\subsection*{Giant proximity effect}
It is well known that putting a normal metal $(N)$ in close contact with a superconductor $(S)$, it is possible to observe a penetration of the superconducting wave function into $N$ over some characteristic distance $\xi_n$, the coherence length in $N$. This is the standard proximity effect, by means it is possible to build $SNS$ Josephson junctions throughout a current is carried out. This can be easily understood from Eq: (\ref{eq:land}), indeed if we think for simplicity to have a contact between a strong superconductor and a normal metal, at the interface the superconducting order parameter has its maximum value and solving the Eq: (\ref{eq:land}) (neglecting the non linear term) we see that this order parameter decays exponentially over a distance of the order of $\xi$ into the normal metal. So if we have a SNS junction whose barrier has a length $d$ of the order of $\xi$, it is possible to observe a superconducting current through the barrier itself.

In recent experiments \cite{Decca_2000,Bozovic_2004}, $SS'S$ Josephson junctions were considered, where $S'$ is a parent compound HTS. In spite of the really small value of the coherence length of the barrier (roughly $4$ \AA), a current was observed in the junction even if the barrier amplitude $d$ was bigger than the coherence length. For this reason this phenomenon was called Giant Proximity Effect and maybe it is possible to observe it because of the pseudogap nature of the $S'$ barrier as claimed by some authors \cite{Kresin_2003}; the giant scale of the phenomenon is provided by the presence of ``superconducting'' islands that percolating allow the transfer of the current. However this mechanism requires a fine tuning of the spacing between the superconducting islands so that they are of the order of $\xi$. We will propose a different explanation involving the competition with a CDW phase.

\subsection*{STM}
\begin{figure}[htb!]
\begin{center}
\includegraphics[clip=true,scale=1.2]{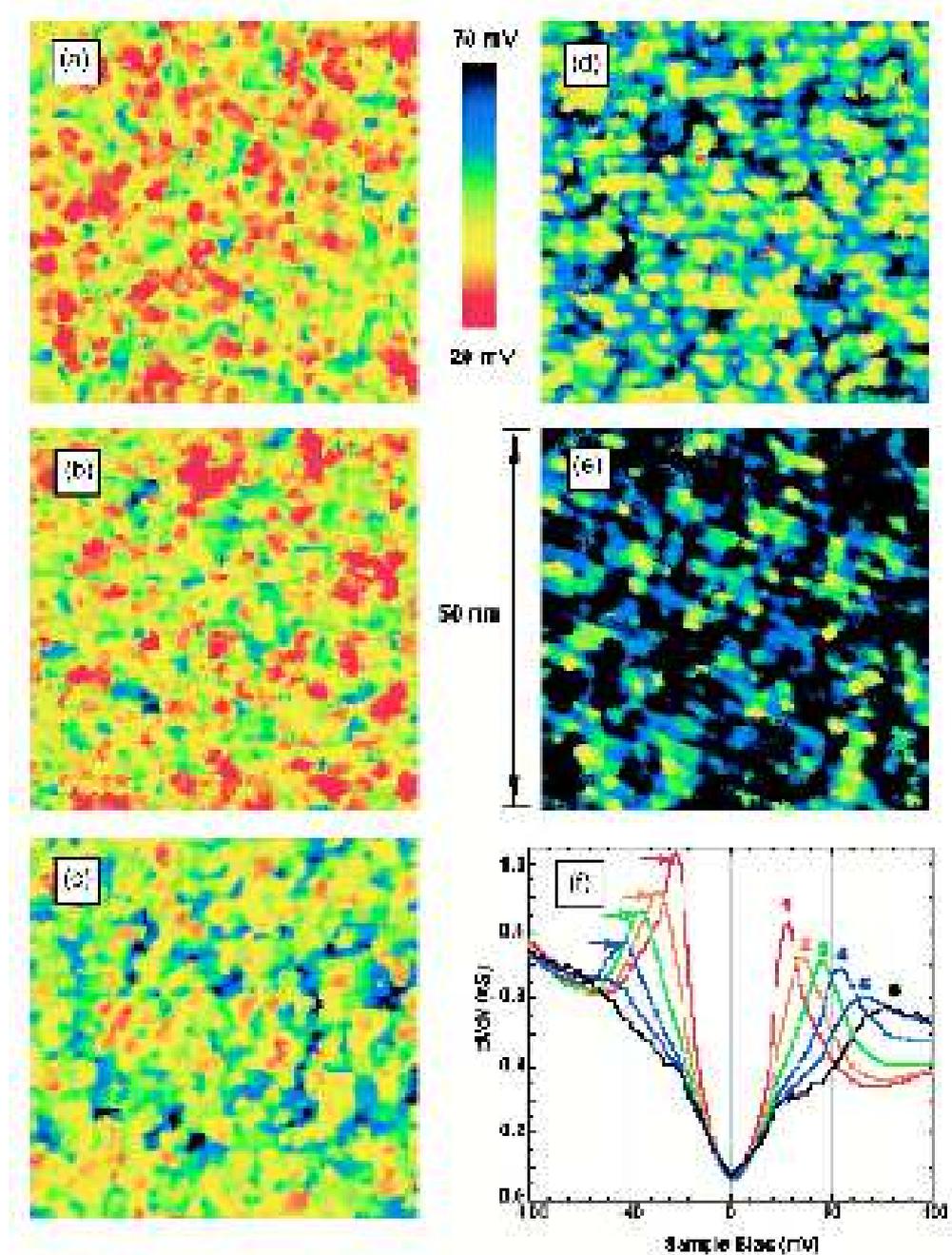}
\caption{ STM images showing the spatial distribution of energy gaps for a variety of samples which are progressively more underdoped from (a) to (e). (f) Th eaverage spectrum for a given energy gap[Figure taken from \cite{McElroy_2005}]}
\label{fig:STM}
\end{center}
\end{figure}

In the past few years, low-temperature STM data have become available, mainly on $Bi-2212$ samples. STM provides a measurement of the local density of states with atomic resolution. It is complementary to ARPES
 in that it provides real-space information but no direct momentum-space information. One important outcome
 is that STM reveals the spatial inhomogeneity of $Bi-2212$ on roughly a $50-100$ \AA length scale, which  becomes more significant with underdoping. As shown in Fig. \ref{fig:STM}, spectra with different energy gaps are associated with different patches and with progressively more underdoping; patches with large gaps become more predominant. There is much excitement concerning the discovery of a static $4\times4$ pattern in this material and its relation to the incommensurate pattern seen in the vortex core of $Bi-2212$ \cite{Hoffman_2002}. How this spatial modulation is related to the pseudogap spectrum is a topic of current debate. In the next chapter we will give more informations about the STM and the charge modulation.

\clearpage
\newpage

\subsection*{Specific heat}
\begin{figure}[htb!]
\begin{center}
\includegraphics[clip=true,scale=1.1]{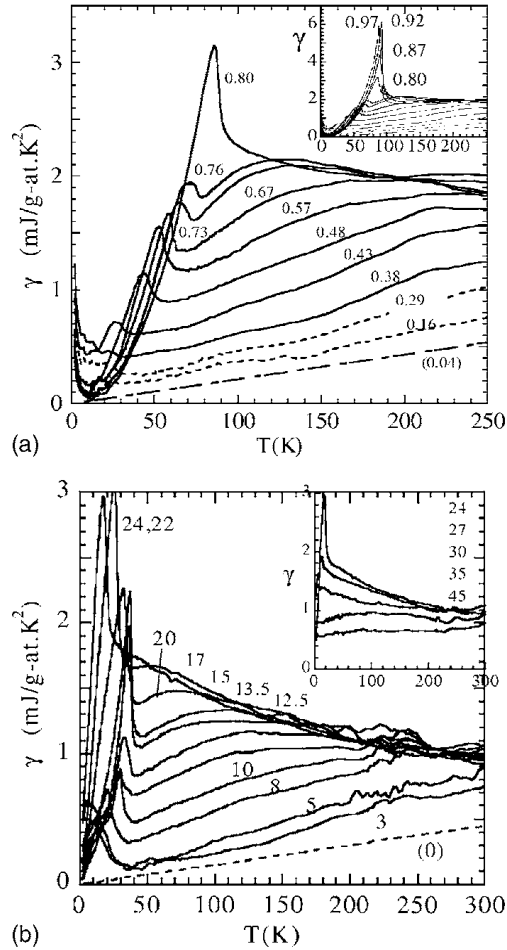}
\caption{The specific-heat coefficient $\gamma$. (a) $YBa_2Cu_3O_{6+y}$. (b) $La_{2−x}Sr_xCuO_4$. Curves are labeled by the oxygen content $y$ in the top figure and by the hole concentration $x$ in the bottom
figure. Optimal and overdoped samples are shown in the inset. The jump in $\gamma$ indicates the  superconducting transition. Note the reduction of the jump size with underdoping. [Figure taken from \cite{Loram_1993,Loram_2001}]}
\label{fig:specheat}
\end{center}
\end{figure}

 A second indication of the pseudogap comes from the linear $T$ coefficient $\gamma$ of the specific heat, which shows a marked decrease below room temperature (see Fig. (\ref{fig:specheat})). Furthermore, the specific-heat jump at $T_c$ is greatly reduced with decreasing doping.

\clearpage
\newpage

\subsection*{Nerst effect}
The Nernst effect in a solid is the detection of an electric field $\vec{E}$ (along $\pm y$, say) when a temperature gradient $-\vec{\nabla}T\parallel\vec{x}$ is applied in the presence of a magnetic field $\vec{H}\parallel\vec{z}$. The Nernst signal, defined as $E$ per unit gradient ($e_N(H,T)=E/|\vec{\nabla} T|$) is generally much larger in ferromagnets and superconductors than in nonmagnetic normal metals. Where $e_N$ is linear in $H$ (conventional metals), it is customary to define the Nernst coefficient $\nu=e_N/B$  with $\vec{B}=\mu_0\vec{H}$. Our focus here, however, is on the Nernst effect in Type-$II$ superconductors, where $e_N$ is intrinsically strongly nonlinear in $H$ \cite{Wang_2006,Ong_2004,Wang_2001,Mukerjee_2004,Ussishkin_2002}.
\vspace{1cm}
\begin{figure}[htb!]
\begin{center}
\includegraphics[clip=true,scale=1.3]{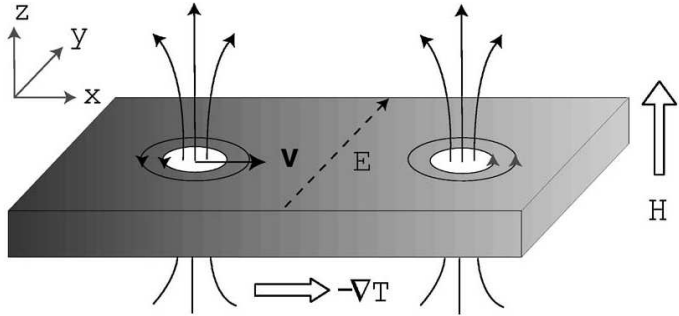}
\caption{The vortex-Nerst effect in a Type $II$ superconductor. Concentric circles represent vortices. [Figure taken from \cite{Wang_2006}]}
\label{fig:nerst}
\end{center}
\end{figure}

The observation of a large Nernst signal $e_N$ in an extended region above the critical temperature $T_c$  in hole-doped cuprates provides evidence that vortex excitations survive above $T_c$. The results support the scenario that superfluidity vanishes because long-range phase coherence is destroyed by thermally created vortices and that the pair condensate extends high into the pseudogap state in the underdoped regime. Nonetheless, acceptance of a vortex origin for $e_N$ above $T_c$ is by no means unanimous; several
models interpreting the Nernst results strictly in terms of quasiparticles have appeared \cite{Kontani_2002,Dora_2003,Tan_2004,Alexandrov_2004}.

\begin{figure}[htb!]
\begin{center}
\includegraphics[clip=true,scale=1.7]{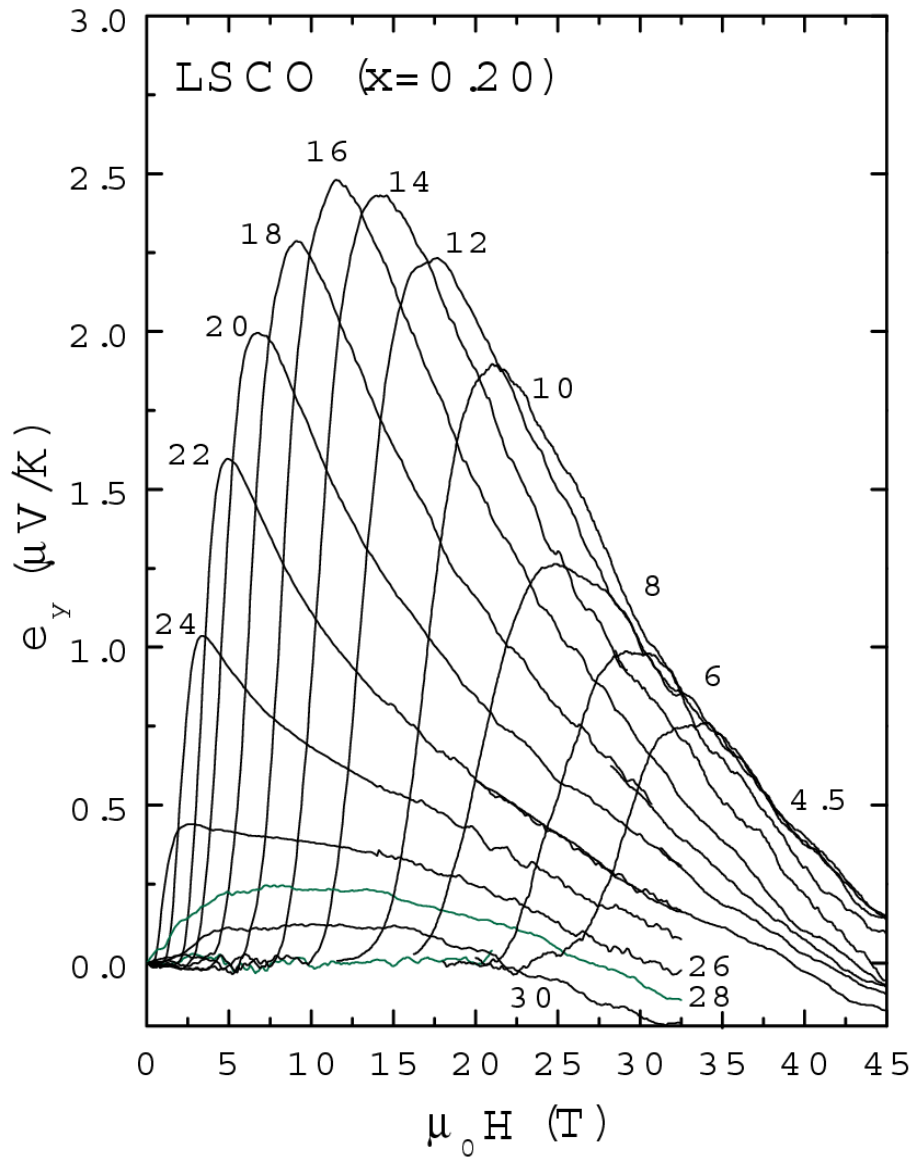}
\caption{The observed Nernst signal $e_N$ vs field $H$ up to $45\,T$ in overdoped $La_{2−x}Sr_xCuO_4$ at selected $T$ $(x=0.20$, $T_{c0}=28.9\,K$). The prominent peak and decrease at high fields are the vortex Nernst signal. At $T>18\,K$, the Nernst signal of the holes which is negative causes $e_N$ to become slightly negative at high fields. [Figure taken from \cite{Ong_2004}]}
\label{fig:nerst2}
\end{center}
\end{figure}

In cuprates, $e_N(T,H)$ exists as a strong signal over a rather large region in the $T-H$ (temperature-field) plane. Fig. (\ref{fig:nerst2}) shows plots of $e_N$ vs $H$ in overdoped $La_{2−x}Sr_xCuO_4$ in which $x=0.20$ and $T_{c0} = 28\,K$ ($T_{c0}$ is the critical temperature in zero-field). The characteristic profile of the curve of $e_N$ vs $H$ below $T_{c0}$ becomes apparent only in very high fields. Starting at the lowest $T$ $(4.5\, K)$, we see that $e_N$ is zero until a characteristic field, where the vortex lattice is known to melt. The solid-liquid melting transition occurs at $H_m$ ($\sim 25\,T$). In the liquid state, $e_N$ rises to a maximum value before decreasing monotonically towards zero at a field that we identify with the upper critical field $H_{c2} \sim 50\, T$ (the field at which the pairing amplitude is completely suppressed). As $T$ increases, both $H_m$ and the peak field  move to lower field values. A complication in overdoped $La_{2−x}Sr_xCuO_4$ is that the hole carriers contribute a moderately large, negative Nernst signal. Close to $T_{c0}$, this carrier contribution pulls the vortex signal to negative values in high fields. The hole
contribution complicates the task of isolating the vortex signal at high $T$ in overdoped samples, but is
negligible for $x\leqslant0.17$.

\begin{figure}[htb!]
\begin{center}
\includegraphics[clip=true,scale=1.7]{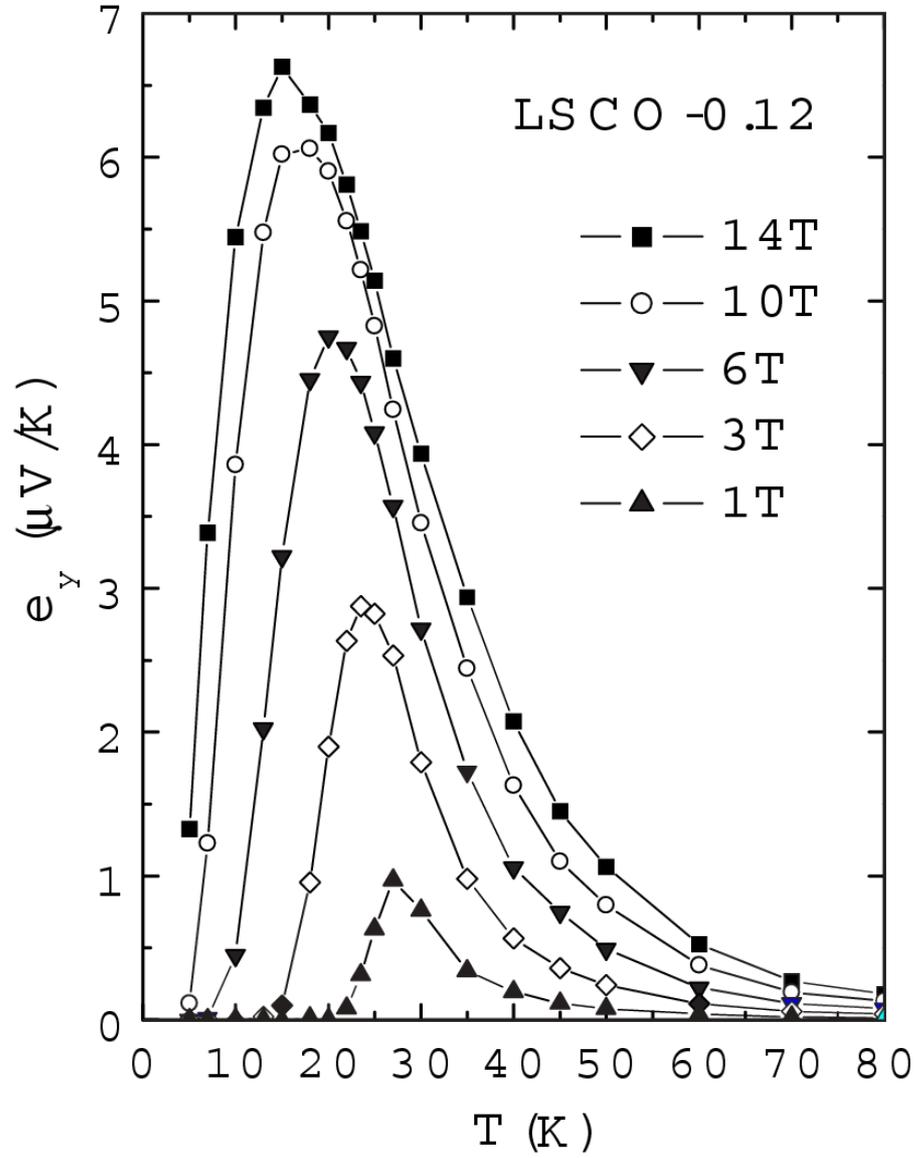}
\caption{The $T$ dependence of $e_N$ at fixed $H$ in underdoped $La_{2−x}Sr_xCuO_4$ ($x = 0.12$, $T_{c0}=28.9\,K$). [Figure taken from \cite{Ong_2004}]}
\label{fig:nerst3}
\end{center}
\end{figure}

\begin{figure}[htb!]
\begin{center}
\includegraphics[clip=true,scale=1.7]{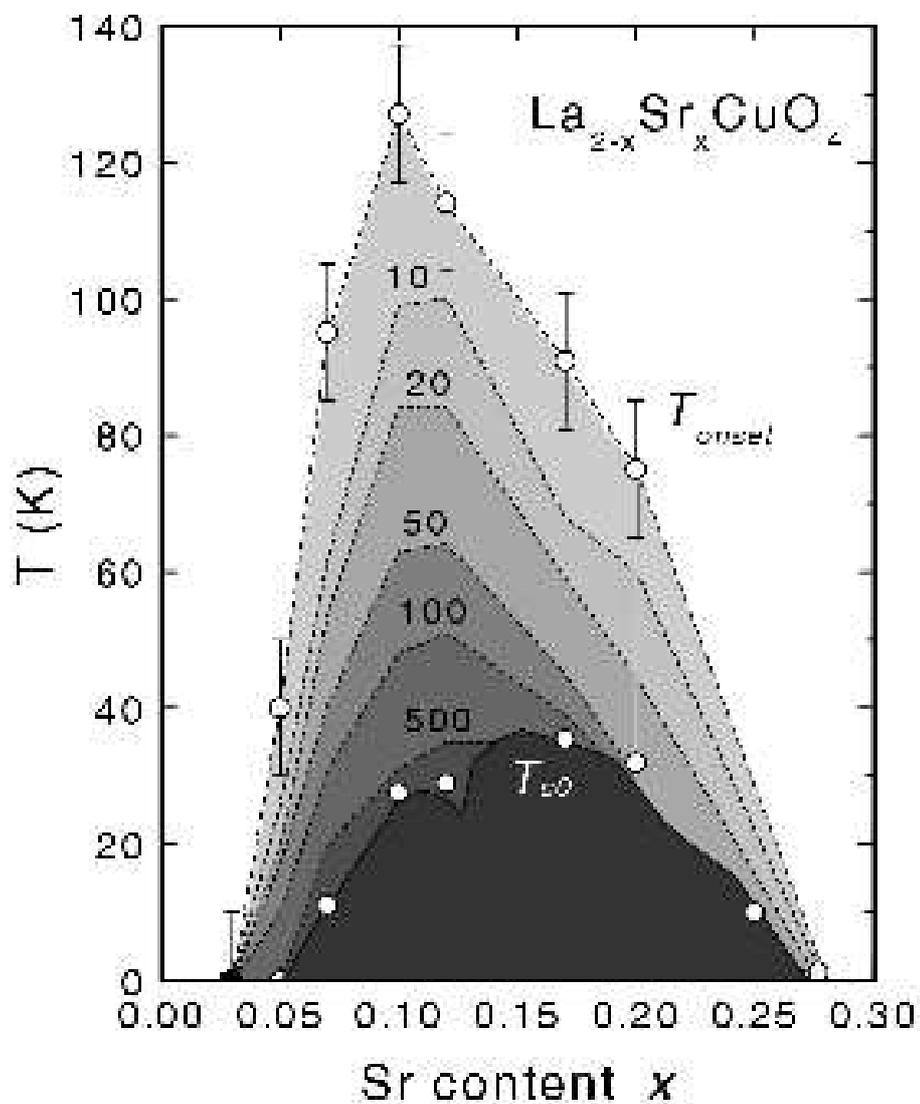}
\caption{Phase diagram of $La_{2−x}Sr_xCuO_4$ showing contour lines of the vortex Nernst signal observed above $T_{c0}$. The vortex-Nernst signal is not observed in the samples at $x=0.26$ and $0.03$. [Figure taken from \cite{Ong_2004}]}
\label{fig:nerst4}
\end{center}
\end{figure}

A different perspective on $e_N(T,H)$ is shown in Fig. (\ref{fig:nerst3}) in underdoped $La_{2−x}Sr_xCuO_4$ ($x=0.12$, $T_{c0}=28.9\,K$). Each curve represents the profile of $e_N$ vs $T$ at fixed field. At this doping, the hole contribution to $e_N$ is negligibly small compared with the vortex signal. The important feature here is that they extends continuously to $T$ high above $T_{c0}$. There is no sign of a sharp boundary separating the vortex liquid state at low $T$ and $H$ from a high-T ``normal state''. Displaying the data in this way brings out clearly the smooth continuity of the vortex signal above and below $T_{c0}$. This continuity is also apparent in contour plots of $e_N(T,H)$ in the $T-H$ plane (see Fig. (\ref{fig:nerst4})).

\clearpage
\newpage

\clearpage
\newpage

\subsection*{Diamagnetic effects above $T_c$}
Above the critical temperature, into the pseudogap state, a rich phenomenology can be observed; for example diamagnetic effects are really important because they are a mark that something related to superconductivity is happening. Iguchi \cite{Iguchi_2001} studied these effects in the pseudogap region of the $La_{2-x}Sr_xCuO_4$.

\begin{figure}[h!]
\centering
\includegraphics[clip=True,scale=1.2]{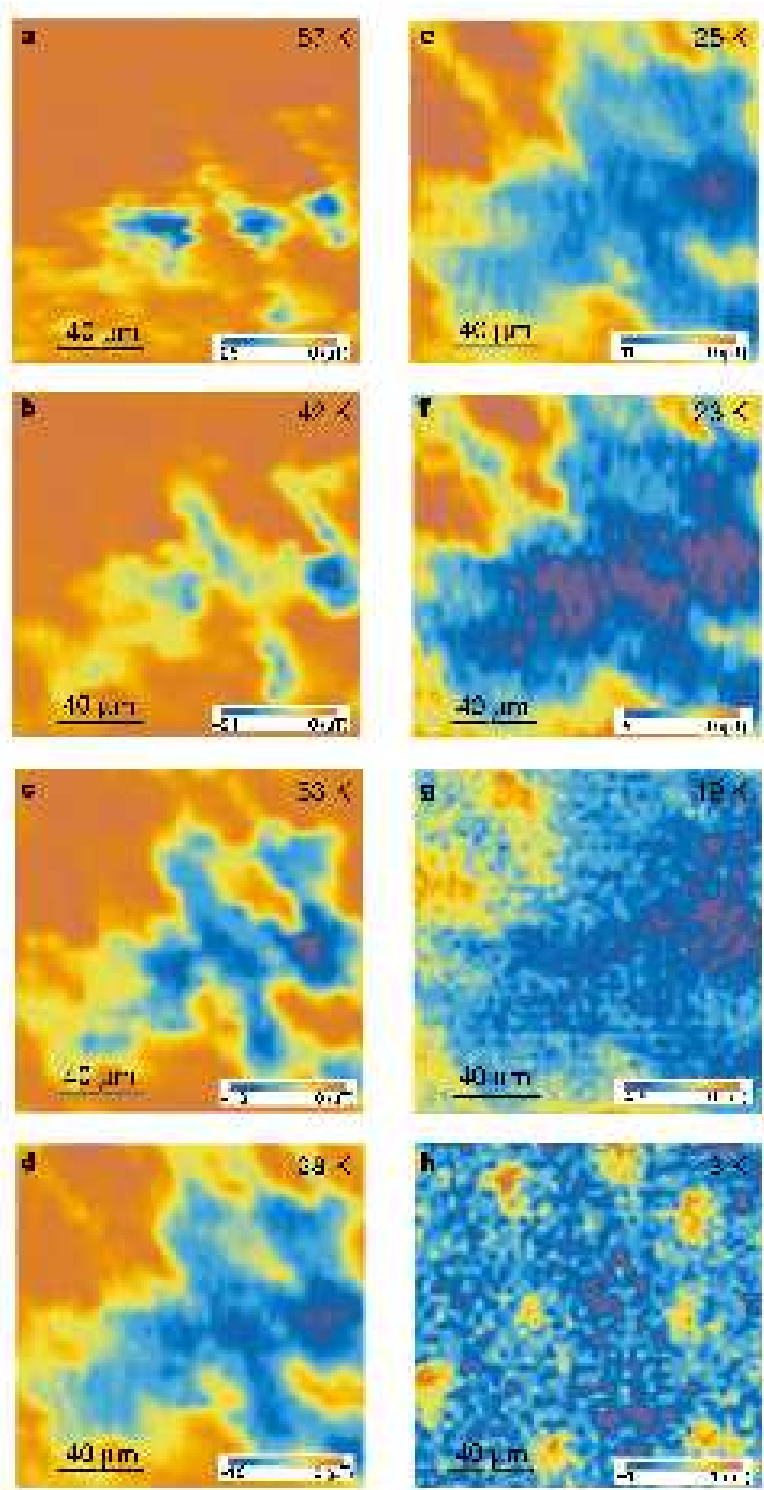}
\caption{Development of magnetic domains with temperature, as observed by scanning SQUID microscopy. $T_c$ was $18\;K$. [Figure taken from \cite{Iguchi_2001}].
}
\label{fig:iguchi}
\end{figure}

In Fig. (\ref{fig:iguchi}) three or more diamagnetic precursor domains, a few tens of micrometres in size, are already present at $57$ $K$. The dark-orange region corresponds to the background-field level. A similar but less developed structure ($<5\mu m$) was also observable at $80$ $K$, which indicates that the
nucleation of diamagnetic domains would start at temperatures significantly higher than $80$ $K$. With reducing temperature, these domains developed to form one big domain, which developed further. The growth of domains became remarkable from temperatures approximately $10$ $K$ above $T_c$. Just after the coverage of most area by the big-domain region, the superconducting transition occurred. In contrast to the development of the domain area, the diamagnetic amplitude of high-temperature domains was significantly large in the first stage and then decreased with reducing temperature.

\clearpage
\newpage

\section{Two-dimensional Superconductivity}
Because of our interest in cuprates HTS and in their properties in the pseudogap region, it is useful to discuss the two-dimensional superconductivity because, as we have seen in the previous section, these materials are layered and the superconductivity takes place in the two-dimensional $CuO$-planes, and also because the pseudogap features could be interpreted in a picture for which there exist preformed Cooper-pairs above the critical temperature $T_c$ that do not have global phase coherence yet. 

Our starting point is the Ginzburg-Landau model, whose effective Hamiltonian is given by:
\begin{equation}
\label{eq:LGXY}
 H=\int\Big[\alpha|\psi|^2+\frac{\beta}{2}|\psi|^4+\frac{1}{2m}\Big|\Big(-i\hslash\vec{\nabla}-\frac{2e\vec{A}}{c}\Big)\psi\Big|^2+\frac{|\vec{\nabla}\times\vec{A}|^2}{8\pi}\Big]d\,\vec{r}
\end{equation}
where $\psi(\vec{r})=|\psi(\vec{r})|e^{i\phi(\vec{r})}$ is the complex scalar order parameter of the superconducting phase.

Now two simplifications have to be made: in Type $II$ superconductors the zero temperature mean-field penetration depth is much greater then the zero temperature coherence length ($\lambda(T=0)\gg\xi(T=0)$), thus fluctuations of the field represented by the last term in Eq: (\ref{eq:LGXY}) around the external field configuration are strongly suppressed and can therefore be neglected; also amplitude fluctuations of the order parameter are neglected too, because in the preformed Cooper-pairs picture the only degree of freedom of the problem ($\sqrt{\rho_s}$ where $\rho_s$ is the superfluid density) is represented by the phase of the order parameter. So we put $|\psi|=1$, simplifying the effective Hamiltonian into:
\begin{equation}
 H=\int\Big[\frac{1}{2m}\Big|\Big(-i\hslash\vec{\nabla}-\frac{2e\vec{A}}{c}\Big)\psi\Big|^2\Big]d\,\vec{r}
\end{equation}
We can also simply further the model because the field $\vec{A}$, as observed before, is frozen, so we can put it equal to zero, obtaining\footnote{We are assuming also that external fields are zero.}:
\begin{equation}
\label{eq:XY}
H=-J_0\int|\vec{\nabla}\psi|^2 d\,\vec{r}=-J_0\int|\vec{\nabla}\phi|^2 d\,\vec{r}
\end{equation}
where $J_0=\hslash^2/2m$. Here we can introduce easily the concept of the superfluidity density $\rho_s$, indeed we can rewrite the Eq: (\ref{eq:XY}) as a kinetic energy:
\begin{equation}
\label{eq:XY3}
H=-\frac{1}{2}\rho_sv_s^2V
\end{equation}
where $V$ is the volume and $v_s=(\hslash/m)(\int|\vec{\nabla}\phi| d\,\vec{r}/V)$ is the superfluid velocity. 

The last modification that we perform is to transform this continuum theory into a lattice one, introducing a short distance cutoff in the problem, which in physical term is taken to be of the order of the coherence length $\xi$ (we remember that in Type $II$ superconductors $\xi\sim4$ \AA, so it is of the order of the real lattice spacing $a$ of the underlying lattice). First of all we change the name of the order parameter from $\psi$ to $\vec{S}$ (we can think to the order parameter as a spin variable with modulus equal to one, and defined on lattice sites of our two dimensional space), and we observe that:
\label{latt-cont}
\begin{equation}
\label{eq:S}
 \vec{S}_i\cdot\vec{S}_j=1-\frac{1}{2}(\vec{S}_i-\vec{S}_j)^2
\end{equation}
so the square gradient of the order parameter of Eq: (\ref{eq:XY}) can be written as a sum of nearest neighbour (along $x$ and $y$ directions) spin variables square differences, and by means of Eq: (\ref{eq:S}) we have:
\begin{equation}
 H=-J\sum_{\langle ij\rangle}(1-\vec{S}_i\cdot\vec{S}_j)
\end{equation}
where $J=J_0/a^{2-d}$, being $a$ the lattice spacing (we take $a_x\equiv a_y$ for simplicity) and $d$ the dimension over which is defined the system. Using the condition $|\vec{S}|\equiv 1$, we can also write:
\begin{equation}
\label{eq:XY2}
 H=-J\sum_{\langle ij\rangle}(1-cos(\phi_i-\phi_j))
\end{equation}
This is the well known $XY$ model, that we will use to describe some important features of the two-dimensional superconductivity.

\subsection{Quasi-long-range-order}
If we consider the model (\ref{eq:XY2}) at low temperature, we expect that thermal fluctuations in $\phi$ are small, this means that we can expand the cosine-term to leading order arriving at:
\begin{equation}
 H=-\frac{J}{2}\sum_{\langle ij\rangle}(\phi_i-\phi_j)^2
\end{equation}
The correlation function that probes the phase coherence in the system is:
\begin{equation}
\label{eq:gammas}
 \Gamma(\vec{r})=\langle e^{i(\phi_i-\phi_j)}\rangle
\end{equation}
where $\vec{r}=\vec{i}-\vec{j}$ is the lattice vector between the points $i$ and $j$. Eq: (\ref{eq:gammas}) can be evalueted easily because we have a gaussin integral, so we can write:
\begin{equation}
\label{eq:gamma}
 \Gamma(\vec{r})=e^{-\langle(\phi(\vec{r})-\phi(\vec{0}))\phi(\vec{0})\rangle}
\end{equation}
Thus the phase coherence is reflected on the correlation function:
\begin{equation}
 G(\vec{r})=\langle(\phi(\vec{r})-\phi(\vec{0}))\phi(\vec{0})\rangle
\end{equation}
This correlation function can be computed passing in the Fourier space and using the equipartition theorem applied to the Fourier-modes $\phi(\vec{k})$:
\begin{equation}
 \frac{J}{2}k^2\langle\phi(\vec{k})\phi(-\vec{k})\rangle=\frac{K_BT}{2}
\end{equation}
 Thus we get:
\begin{eqnarray}
\label{eq:Gr}
 G(\vec{r})&=&\frac{K_BT}{J}\int \frac{d^2k}{(2\pi)^2}\frac{e^{i\vec{k}\cdot\vec{r}}-1}{q^2}=\nonumber\\
&=&\frac{K_BT}{2\pi J}\ln(r/a)
\end{eqnarray}
where $a$ is some short-distance cutoff ($a=e^{-\gamma}/2\sqrt{2}$, with $\gamma$ the Euler constant). We can note that the phase fluctuations vanish when $T\rightarrow0$, this means that phase-correlations become truly long-ranged. Inserting Eq: (\ref{eq:Gr}) into Eq: (\ref{eq:gamma}), we have:
\begin{equation}
 \label{eq:gamma2}
\Gamma(\vec{r})=\Big(\frac{r}{a}\Big)^\eta
\end{equation}
where $\eta=K_BT/2\pi J$. An important feature of $\Gamma(\vec{r})$ is that it goes to zero for $\vec{r}\rightarrow0$ at every temperature different from zero. Hence, there is never true long-range order at finite temperature in a two dimensional superconductor. This is a specific example of the Hohenberg-Mermin-Wagner theorem that states that at finite $T$ no continous symmetry can be spontaneously broken in dimensions $d\leqslant2$ \cite{Hohenberg_1967,Mermin_1966}. What we at most can have is power-law decay (which is slower than exponential decay characteristic of short-range order).
 Even if long-range order does not exist, this does not mean that there is not any energy cost to ``twist'' the phase of the superconducting order parameter at low temperature (we shall call this energy cost \emph{phase stiffness}). On the other hand, at very high temperatures, phases are randomly oriented relative to each other even on short length scales, and a local twist is expected to come at no cost in the free energy. Hence somewhere in between low and high temperatures a phase transition must occur. Clearly it will be a peculiar transition from a quasi-ordered state to a disordered one, and not from an ordered state to a disordered one. 
 
\subsection{Vortex-antivortex pairs}
In order to understand the nature of the phase transition that occur in the $XY$ model, we have first of all to define the vorticity $q$ of a phase field:
\begin{equation}
\oint d\vec{l}\cdot\nabla\phi=2\pi q
\end{equation}
where $q=\{0,\pm1\pm2,\cdots\}$. Phase fields for which $q=0$ are topologically different from those where $q\neq0$; it is impossible to continually deform a phase field with $q=0$ into one with $q\neq0$. Vortices will be generated spontaneously at high temperatures since this will increase the configurational entropy of the system, lowering the free energy. We can note that for a superconductor $\nabla\phi$ gives rise to an electric current, and the curl of this current is a magnetic field, thus $q$ can be seen as the quantized magnetic field penetrating through the area enclosed by the contour over which the integral is taken. Because no net magnetic field can be generated throughout the system by thermal fluctuations, vortices must be always generated in pairs of opposite vorticity, a vortex-antivortex pair. At low temperature where vortices are expected to be unimportant, they are tightly bound; while at high temperature there is an unbinding of these pairs, responsible for destroying the phase stiffness of the system.

\subsection{Stiffness or Helicity modulus and the BKT transition}
The pioneering work of Berezinsky, Kosterlitz and Thouless \cite{Berezinsky_1971,Kosterlitz_1973,Kosterlitz_1974} allowed the better comprehension of this peculiar transition, after them called Berezinsky-Kosterlitz-Thouless (BKT) transition. They showed that it is possible to define a stiffness parameter that has a jump of $2/\pi$ at a transition temperature; this stiffness parameter is anything else that the phase stiffness of the $XY$ model, or in other way the superfluid density $\rho_s$ of the system. The phase-stiffness of the $XY$ model represents the energy cost of introducing twists in the phase of the superconducting order parameter. The phase stiffness, often called the Helicity modulus $\Upsilon$, of the $XY$ model is defined as the cost in free energy of an initial twist $\Delta\Phi$ in the phase of the order parameter across the system:
\begin{equation}
\label{eq:helicity}
 \Upsilon\sim\frac{\partial^2F}{\partial\Delta\Phi^2}\Big|_{\Delta\Phi=0}
\end{equation}
 This twist may be viewd as adding a vector potential $\vec{A}$ to the argument of the cosine in the $XY$ model (by minimal coupling). By standard quantum mechanics the current is given by the first derivative of the free energy with respect to the added vector potential, so the second derivative of the free energy with respect to the vector potential is equal to the first derivative of the current with respect to the vector potential, that is the superfluid density $\rho_s$. Thus we have $\Upsilon(T)\sim\rho_s(T)$\footnote{The Helicity modulus could be defined in the same way of the superfluidity density of the equation \ref{eq:XY3}, i.e. the cost in free energy of an initial twist $\Delta\Phi$ in the phase of the order parameter across the system could be written as $F=(1/2)\Upsilon(\Delta\Phi)^2V$. Thus comparing this expression with Eq: (\ref{eq:XY3}) we have: $\Upsilon=(\hslash/m)^2\rho_s$.}, and thanks to Kosterlitz and Thouless we know that at the transition we have the following jump:
\begin{equation}
\lim_{T\rightarrow T_c^-}\,\frac{\hslash^2}{m^2}\frac{\rho_s(T)}{K_BT}=\frac{2}{\pi}
\end{equation}

\subsection{An historical note}
Historically the concept of the Helicity modulus was rigorously defined for a $d-$dimensional isotropic system with an $n-$vector $(n\geqslant2)$ order parameter by M. E. Fisher et al. \cite{Fisher_1973}. This definition is phrased in terms of well-defined equilibrium free energies, and it is given by:
\begin{equation}
\label{eq:fisher_stiff}
 \Upsilon(T) = \lim_{A(\Omega),L(\Omega)\rightarrow\infty} \Big(\frac{L(\Omega)}{2\theta^2A(\Omega)}\Big)[F(T;\Omega,W_\theta^{+-})-F(T;\Omega,W_\theta^{++})]
\end{equation}
where $\Omega$ is the domain over which is defined the order parameter, $L(\Omega)$ is the direction along which we want to calculate the Helicity Modulus, $A(\Omega)$ is the cross-sectional area orthogonal to $L(\Omega)$, $W_\theta$  are wall potentials which establish a definite phase angle $\theta$ for the order parameter. Finally the superscript symbols $+$ or $-$ over the potential barriers indicate that the angle $\theta$ at the ends of the domain are choosen either positively or negatively with respect to a fixed axis. So the $(+,+)$ or $(-,-)$ wall combinations will yeld uniform bulk phases in which the order parameter has a constant phase angle indipendent of positions; while the mixed $(+,-)$ barrier potentials impose some sort of ``twist'' on the system. The simplest choice for wall potentials consists in taking periodic and anti-periodic boundary conditions for the system, but nonetheless the definition (\ref{eq:fisher_stiff}) is difficult to employ, particularly for non homogeneus system we cope with.

A different but equivalent and most powerful definition of the Helicity Modulus was given later by Jasnow et al. \cite{Rudnick_1977}, using the idea that $\Upsilon$ can be considered as a response function of the system to a distorsion $\Delta\phi$ of the order parameter; so we can write down:
\begin{equation}
\Upsilon(T) = \lim_{L\rightarrow\infty}\frac{\partial^2f}{\partial\epsilon^2}\Big|_{\epsilon\,{\scriptscriptstyle=\,0}} 
\end{equation}
where $L$ is the length of the system along the direction we are calculating the Stiffness, $f$ is the free energy density, and $\Delta\phi=\epsilon L$. This is exactly the definition introduced above with Eq:  (\ref{eq:helicity}). If we want to write the Stiffness directly using the distorsion parameter $\Delta\phi$, we have for a cubic system:
\begin{equation}
\label{eq:jasnow_stiff}
\Upsilon(T) = \lim_{L\rightarrow\infty}\frac{1}{L^{d-2}}\frac{\partial^2F}{\partial\Delta\phi^2}\Big|_{\Delta\phi\,{\scriptscriptstyle=\,0}} 
\end{equation}
where $d$ is the dimension over wich the system is defined, and $F$ is the extensive free energy.
\begin{figure}[htb!]
\centering
\includegraphics[clip=True,scale=1.0]{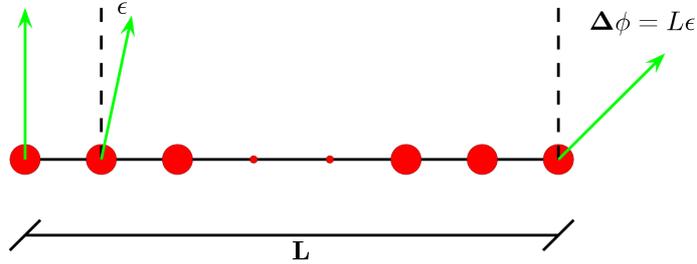}
\caption{A cartoon to show the meaning of the variables that are into Eq: (\ref{eq:jasnow_stiff}).}
\label{fig4-cap5}
\end{figure}

\subsection{A toy model}
Now we want to calculate the Helicity Modulus for a really simple toy model, first of all in order to see immediately an application of the just introduced definition, but more than anything because the zero temperature Stiffness value for this toy model will give us an insight about another different way to calculate the Helicity Modulus at zero temperature. 

\begin{figure}[htb!]
\centering
\includegraphics[clip=True,scale=1.0]{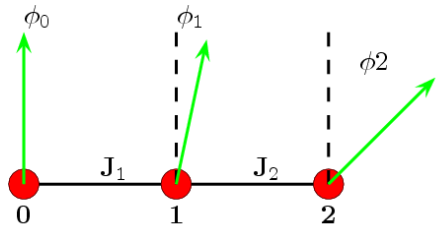}
\caption{A sketch of the toy model defined by Eq: (\ref{eq:toy}).}
\label{fig5-cap5}
\end{figure}

The toy model is a simple ferromagnetic $1-$dimensional $XY$ model defined on only three sites; its Hamiltonian is:
\begin{equation}
\label{eq:toy}
 H = -J_1\cos(\phi_0-\phi_1)-J_2\cos(\phi_1-\phi_2)
\end{equation}
and the positions of the variables are showed in the Fig: \ref{fig5-cap5}.

In order to find the Helicity Modulus $\Upsilon$, first we have to fix arbitrarily the value of $\phi_0$ and then we have to ``twist'' the variable $\phi_2$; for simplicity we put:
\begin{eqnarray}
 \left \{
\begin{array}{l}
\phi_0=0\\\\
\phi_2=\phi_0+\Delta\phi\equiv\Delta\phi
\end{array}
\right.
\end{eqnarray}  
To evaluate the Stiffness it is necessary to find the free energy of the system and then to take its second derivative with respect $\Delta\phi$ and evaluate this one for $\Delta\phi=0$. We know that the free energy is given by:
\begin{equation}
 F = -\frac{1}{\beta}\ln Z
\end{equation}
so we have:
\begin{equation}
 \frac{\partial^2F}{\partial\Delta\phi^2}\Big|_{\Delta\phi\,=\,0}=-\frac{1}{\beta}\Big[\frac{1}{Z}\frac{\partial^2Z}{\partial\Delta\phi^2}-\frac{1}{Z^2}\Big(\frac{\partial Z}{\partial\Delta\phi}\Big)^2\Big]_{\Delta\phi\,=\,0}
\end{equation}
Before calulating the partition function $Z$ and its $1^{st}$ and $2^{nd}$ derivatives, we perform a change of variables in order to write a more symmetric expression for the Hamiltonian; the variable transformation is defined by:
\begin{equation}
 \phi'_i=\phi_i-\frac{\Delta\phi}{2}i \qquad {\mathrm{with}} \qquad i\in\{0,1,2\}
\end{equation}
and then the Hamiltonian Eq: (\ref{eq:toy}) becomes:
\begin{equation}
\label{eq:toy2}
 H = -J_1\cos(\phi+\Delta\phi/2)-J_2\cos(\phi-\Delta\phi/2)
\end{equation}
where we also renamed the variable $\phi'_1$ into $\phi$ for simplicity. At this point we can write down:
\begin{eqnarray}
Z|_{\Delta\phi=0} & = & \int_0^{2\pi} d\phi \; e^{\beta(J_1+J_2)\cos\phi}\equiv Z_0\\
\frac{\partial Z}{\partial\Delta\phi}\Big|_{\Delta\phi=0} & = & \frac{\beta}{2}(J_1-J_2)\int_0^{2\pi} d\phi\;e^{\beta(J_1+J_2)\cos\phi}\sin\phi\equiv0\\
\frac{\partial^2 Z}{\partial\Delta\phi^2}\Big|_{\Delta\phi=0} & = & \int_0^{2\pi} d\phi\;e^{\beta(J_1+J_2)\cos\phi}\Big[\frac{\beta^2}{4}(J_2-J_1)^2\sin^2\phi+\nonumber\\
& &-\frac{\beta}{4}(J_2+J_1)\cos\phi\Big]
\end{eqnarray}
and using the modified Bessel functions of first kind, $I_0(z)$ and $I_1(z)$, we have the following expression for the Stiffness of the toy model:
\begin{equation}
 \Upsilon = 2\Big(\frac{1}{J_1}+\frac{1}{J_2}\Big)^{-1}\;\frac{I_1\big[\beta(J_1+J_2)\big]}{I_0\big[\beta(J_1+J_2)\big]}
\end{equation}
We can underline that if $J_1=0$ or $J_2=0$ the Stiffness is equal to zero; this can be simply understood because if one link ($J$) is missing, then the variables $\phi_0$ and $\phi_2$ are not connected and so a change into one of them doesn't influence the other one. But more important is the observation of the zero temperature value of the Stiffness:
\begin{equation}
 \Upsilon(T=0)= 2\Big(\frac{1}{J_1}+\frac{1}{J_2}\Big)^{-1}
\end{equation}
As it is possible to note, if we think to $J_1$ and $J_2$ as two conductances, the expression of the zero temperature Stiffness is proportional to the global conductance of the system. This result is really interesting because, as we are going to show in the chapter (\ref{chapter5}), it is valid also for a  two-dimensional lattice system in a complete general way. Indeed, as we will see in more detail in the chapter (\ref{chapter5}), we shall interested in evaluating the stiffness of a random $XY$ model:
\begin{equation}
 H=-\sum_{\langle i,j\rangle}J_{ij}\cos(\phi_i-\phi_j)
\end{equation}
where $J_{ij}$ are quenched random bonds.

  \clearpage{\pagestyle{empty}\cleardoublepage}
  \chapter{Charge Density Waves: a Glance}
  \label{chapter2}
\vspace{3cm}
In this chapter we want to describe the ideas underlying the Charge Density Waves state, giving also an overview to Scanning Tunnelling Microscopy technique, and then illustrating some recents experiments performed with this technique on High-Temperature-Superconductors, especially in the so-called pseudogap state.

\section{Charge Density Waves: Basic Concepts}
Density waves are broken symmetry states of metals, due to electron-phonon or electron-electron interactions. The CDW is an electronic-lattice instability (while the electron-electron intraction generates the so-called Spin Density Waves (SDW)), and the driving force behind the CDW instability is the reduction in the energy of electrons in the material as a consequence of establishing a spontaneous periodic modulation of the crystalline lattice with an appropriate wave vector. 

CDW were first discussed by Fr\"{o}hlich in $1954$ and by Peierls in $1955$; the highly anisotropic band structure is really important to observe this ground state in metals. Indeed the experimental evidence of these ground state was found much later their theoretical prediction, when the so-called low-dimensional materials were discovered and investigated.
\subsection{The 1-dimensional electron gas}
Because of the importance of the low-dimensionality for observing CDW, it could be useful to review some results regarding the one-dimensional electron gas.

Considering the 1D free electron gas, its energy dispersion is given by:
\begin{equation}
 \epsilon(k)=\frac{\hslash^2k^2}{2m}
\end{equation}
and the Fermi energy is equal to:
\begin{equation}
 \epsilon(k_F)=\frac{\hslash^2k_F^2}{2m}
\end{equation}
where the Fermi wavevector is:
\begin{equation}
 k_F=\frac{N_0\pi}{2L}
\end{equation}
and $N_0$ is the total number of electrons, while $L$ is the length of the 1D chain. The topology of the Fermi surface in 1D is really peculiar, indeed it has only two points, $\pm k_F$. This kind of Fermi surface gives a response to an external perturbation completely different from that in higher dimensions. If we consider a time indipendent potential $\phi(\vec{r})$ acting on an electron gas, the rearrangement of the charge density $\rho(\vec{r})$, expressed in the Fourier space and within the framework of linear response theory, is given by:
\begin{equation}
 \rho(\vec{q})=\chi(\vec{q})\phi(\vec{q})
\end{equation}
where $\chi(\vec{q})$ is the so-called Lindhard response function, that in $d$ dimensions is equal to:
\begin{equation}
 \chi(\vec{q})=\int\frac{d\vec{k}}{(2\pi)^d} \frac{f_k-f_{k+q}}{\epsilon_k-\epsilon_{k+q}}
\end{equation}
and $f_k=f(\epsilon_k)$ is the Fermi function. For the one-dimensional case, assuming a linear dispersion relation around the Fermi energy, the Lindhard response function becomes:
\begin{equation}
 \chi(q)=-e^2n(\epsilon_F)\ln\Big|\frac{q+2k_F}{q-2k_F}\Big|
\end{equation}
where $n(\epsilon_F)$ is the density of states at the Fermi level; $\chi(q)$ diverges for $q=2k_F$, and this implies that at $T=0$ the electron gas is unstable with respect to the formation of a periodically varying electron charge. 

\subsection{The mean-field CDW ground state}
The electron-phonon interaction and the divergence of the electronic response at $q=2k_F$ in one dimension give a strongly renormalized phonon dispersion spectrum $\Omega(q)$, generally referred to as the Kohn anomaly. This renormalization is deeply temperature-dependent. At some temperature and for $q=2k_F$, $\Omega(q)$ becomes zero, thus identifying a phase transition to a state where a periodic static lattice distortion and a periodically varying charge modulation develop. So we have the CDW state. This transition is generally called Peierls transition, but also Kuper (1955) and Fr\"{o}hlich (1954) studied it.

Now we will skecth how it is possible to obtain the transition to a CDW state in a one-dimensional electron gas coupled to the underlying chain of ions through electron-phonon interaction, in the framework of the mean field theory using the so-called Fr\"{o}hlich Hamiltonian:
\begin{equation}
\label{eq:frohlich}
H = \sum_k\epsilon_kc^\dagger_kc_k+\sum_q\hslash\omega_qb^\dagger_qb_q+\sum_{k,q}g_qc^\dagger_{k+q}c_k(b^\dagger_{-q}+b_q) 
\end{equation}
the first term is the Hamiltonian of the electron gas where $c^\dagger_k$ and $c_k$ are the fermionic creation and annihilation operators for the electron states with energy $\epsilon_k=\hslash^2k^2/2m$; the second term is the Hamiltonian describing the lattice ions vibrations, where $b^\dagger_q$ and $b_q$ are the bosonic creation and annihilation operators for the phonons with a wavevector $q$, and $\omega_q$ being the normal mode frequencies; the third term is the interaction Hamiltonian, where $g_q$ is the electron-phonon coupling constant.

Writing down the equation of motion of the normal coordinates of the ions, and using the linear response theory it is possible to find the renormalized phonon frequency:
\begin{equation}
 \omega^2_{ren,\,q}=\omega^2_q+\frac{2g^2\omega_q}{\hslash}\chi(q,T)
\end{equation}
The phonon frequency for $q=2k_F$ becomes:
\begin{equation}
 \omega^2_{ren,2k_F}=\omega^2_{2k_F}-\frac{2g^2n(\epsilon_{\scriptscriptstyle F})\omega_{2k_F}}{\hslash}\ln(1.14\epsilon_{\scriptscriptstyle F}/k_BT)
\end{equation}
With decresing temperature the renormalized phonon frequency goes to zero and this defines the mean field CDW transition temperature:
\begin{equation}
 k_BT_{CDW}^{mf}=1.14\epsilon_{\scriptscriptstyle F}\exp(-1/\lambda)
\end{equation}
where $\lambda$ is the dimensionless electron-phonon coupling constant
\begin{equation}
 \lambda = \frac{g^2n(\epsilon_{\scriptscriptstyle F})}{\hslash\omega_{2k_F}}
\end{equation}
Below the CDW transition temperature there is a ``frozen-in'' lattice distorsion and then the mean lattice ionic displacement $\langle u(x)\rangle$ is different from zero:
\begin{equation}
 \langle u(x)\rangle = \Delta u\cos(2k_{\scriptscriptstyle F}x)
\end{equation}
with
\begin{equation}
 \Delta u = \Big(\frac{2\hslash}{NM\omega_{k_{\scriptscriptstyle F}}}\Big)\frac{|\Delta|}{g}
\end{equation}
and $N$ is the number of lattice sites for unit length, $M$ is the ionic mass, and $\Delta$ is the CDW gap energy opened in the energy dispersion at the Fermi level. It is also possible to calculate the modulation of the electronic density:
\begin{equation}
 \rho(x) = \rho_{\scriptscriptstyle 0}\Big[1+\frac{\Delta}{\hslash v_{\scriptscriptstyle F}k_{\scriptscriptstyle F}\lambda}\cos(2k_{\scriptscriptstyle F}x)\Big]
\end{equation}
where $\rho_{\scriptscriptstyle 0}$ is the constant electronic density in the metallic state.

To summarize we have seen that the ground state has a periodic modulation both of the charge density and lattice distorsion, and also the gap opening in the energy dispersion at the Fermi level turns the material into an insulator.

\subsection{Variants and (In-)Commensurate CDW}
\begin{figure}[htb!]
\centering
\includegraphics[clip=True,scale=1.0]{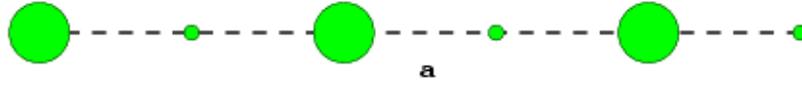}
\caption{A 1-dimensional charge density wave; it represents a kind of variant. The radius of the circles is proportional to the charged density.}
\label{fig:1dCDW1}
\end{figure}

\begin{figure}[htb!]
\centering
\includegraphics[clip=True,scale=1.0]{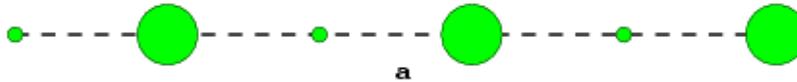}
\caption{A 1-dimensional charge density wave; it represents a kind of variant. The radius of the circles is proportional to the charged density.}
\label{fig:1dCDW2}
\end{figure}

As seen above the electron-phonon coupling allows a periodic modulation both of the charge density and lattice distorsion. For one-dimensional systems we can imagine the CDW in a simple way: if we have a chain for which the ions are placed at a distance $a$ (the lattice spacing) each other, we can think to have more charge density on a lattice site and less on the neighbour site and so on, as skecthed in Fig. (\ref{fig:1dCDW1}). In this case we have created a charge modulation with a period equal to twice the lattice spacing, but it is clear that this modulation breaks the translational symmetry of the system, indeed we can easily think to have a charge modulation that is equal to the previous one but shifted by one lattice spacing, as shown in Fig. (\ref{fig:1dCDW2}). The cartoons showed in Fig. (\ref{fig:1dCDW1})  and (\ref{fig:1dCDW2}) are valid if we have a number of electron for lattice site ($n$) equal to one (i.e. $n=1$). This is consistent with the periodicity obtained in weak coupling: $\lambda=2\pi/2k_F=2a/n$, so for $n=1$ we have $\lambda=2a$ as skecthed in the figures.

\begin{figure}
\centering
\includegraphics[clip=True,scale=3.0]{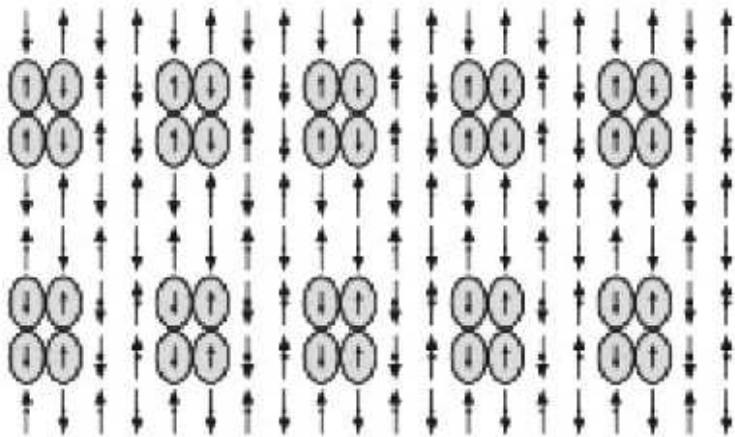}
\caption{A system of two half-filled stripes. The radius of the circles gives the hole density; the length of the arrow gives the local magnetization. Notice that every cluster of four big circles contains in total approximately two holes with a singlet orientation for the spins: real-space Cooper pairs. \newline [Figure taken from \cite{Bosch_2001}]}
\label{fig:stripe}
\end{figure}

These two kinds of modulation will be called ``variants'' of the CDW, borrowing the term from  crystallography. The number of these variants can increase if we have a charge modulation with a larger period. In our one-dimensional case where $\lambda/a$ is an integer, the CDW is said commensurate (below we will explain better this concept) and $\lambda/a$ gives the numbers of ``variants''. In general this number is given by the number of atoms in the unit cell. In Fig. (\ref{fig:stripe}) we show a charge and spin density wave of purely electronic origin; in this case there are $64$ variants because the unit cell is given by a rectangle of $8\times4$ atoms that gives $32$ different trnslations, and we have also to multiply this value for $2$, that represents the $90^\circ$ rotational symmetry breaking. Indeed the building block of a two-dimensional charge ordered  pattern is a two-dimensional unit cell, that could breaks translational and rotational symmetries in many ways. If we imagine an experiment in which the system is quenched from a paramagnetic state to a CDW, it can nucleates a CDW variant into a region and another variant into a different region. Even more if we have quenched impurities into the sample, these will favour a variant in one region and a different variant in another region. The resulting state for this case could be a policristal charged ordered state, where more ordered patterns, corresponding to different variants, mismatch each others don't allowing for the observation of a single ordered state. As we shall point out later, in our research we will study the simplest situation for which we have only two variants of charge ordering.

We can also point out another important feature that emerges from the pictures Fig. (\ref{fig:1dCDW1}) and (\ref{fig:1dCDW2}); these one represent a strong coupling limit behaviour, indeed in our chain we have  sites with a really big charge density and others sites with a really poor charge density. This situation can be viewed as a preformed electrons pairs scenario, where in the sites with a huge charge density we have two electrons (with opposite spins in order to satisfy the Pauli's exclusion principle), and the other sites are almost empty. In this strong coupling framework we can think to introduce an Ising like pseudospin variable that has an up value corresponding to the sites where there are electrons pairs, and a down value corresponding to the sites where there aren't electrons. This picture with an Ising like pseudospin variable will be really useful to formulate our model, as we shall see in the next chapter.

Another important concept has to be introduced at this level: the difference between a commensurate and an incommensurate CDW. The former is a CDW for which the charge modulation has a period equal to a rational number of the underlying lattice spacing $a$, while the latter is a CDW for which the ratio between the period of the charge modulation and the lattice spacing is equal to an irrational number. In other words if we think to get a maximum of the charge modulation corresponding to a lattice site, for the commensurate CDW we will always find another lattice site over which the charge modulation is maximum, while for the incommensurate CDW this does not happen.  


\section{Experimental techniques and CDW}
Hereafter we will focus our attention to hole-doped cuprates. In these materials doped holes tend to aggregate into one-dimensional domain walls (stripes) separating regions of antiferromagnetically ordered spin domains (see Fig. (\ref{fig:stripe})); but also a different kind of order can be observed, the checkboard one.

Stripes are characterized by modulations of the charge density at a single ordering vector $\vec{Q}$ and its harmonics $\vec{Q}_n=n\vec{Q}$ with $n$ an integer. In a crystal, we can distinguish different stripe states not only by the magnitude of $\vec{Q}$, but also by whether the order is commensurate [when
$|\vec{Q}|a=2\pi(m/n)$ where $a$ is the lattice constant and $n$ is the order of the commensurability] or incommensurate with the underlying crystal, and on the basis of whether $\vec{Q}$ lies along a symmetry axis or not. In the cuprates, stripes that lie along or nearly along the Cu-O bond direction are called ``vertical'' and those at roughly $45^\circ$ to this axis are called ``diagonal''. 

Checkerboards are a form of charge order that is characterized by bidirectional charge density modulations, with a pair of ordering vectors $\vec{Q}_1$ and $\vec{Q}_2$ (where typically $|\vec{Q}_1|=|\vec{Q}_2|$). Checkerboard order generally preserves the point group symmetry of the underlying crystal if both ordering vectors lie along the crystal axes. In the case in which they do not, the order is rhombohedral checkerboard and the point group symmetry is not preserved. As with stripe order, the wave vectors can be incommensurate or commensurate, and in the latter case $\vec{Q}_ja=2\pi(m/n,m'/n')$. Commensurate order, as with stripes, can be site centered or bond centered.

\begin{figure}[htb!]
\centering
\includegraphics[clip=True,scale=.67]{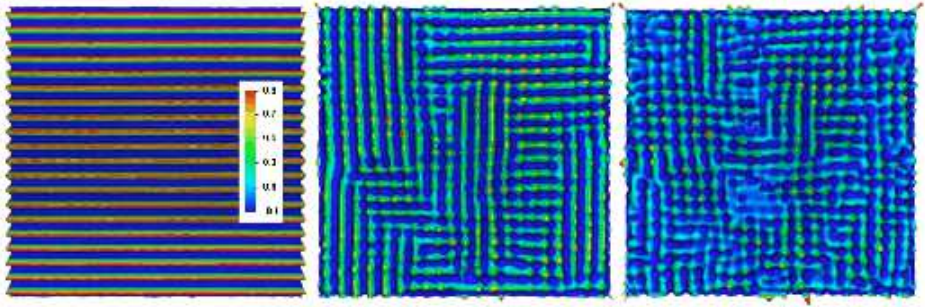}
\caption{Left panel: Highly stripe-ordered system, with weak impurities. Center panel: Otherwise identical to the first system (including the spatial distribution and concentration of impurities), but the
strength each impurities has increased. Right panel: Identical to the left panel, except for a more strong disorder intensity. [Figure taken from \cite{Robertson_2006}] }
\label{fig:stripe2}
\end{figure}

\begin{figure}[htb!]
\centering
\includegraphics[clip=True,scale=0.67]{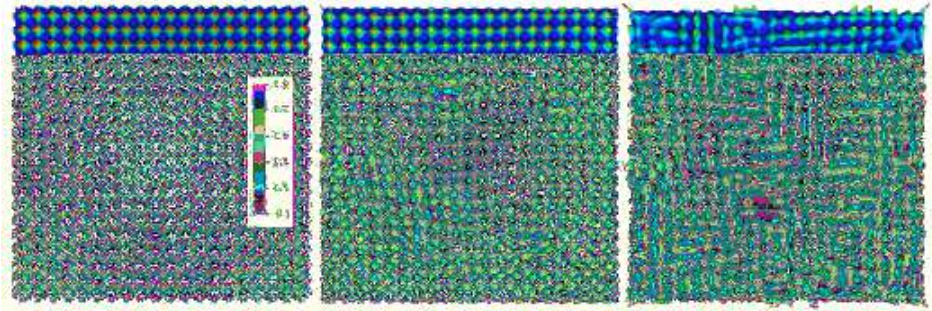}
\caption{The parameters entering the effective Hamiltonian and the impurity realizations are identical here to the panels of Fig. (\ref{fig:stripe2}), with the exception of the symmetry breaking term. In the center panel, because the checkerboard state is more stable than the analogous stripe state. Unlike the stripe ordered system, the checkerboard system does not break into domains, but rather develops pair wise dislocations (visible in the central panel). Note the similarity between the right panel of each set of Fig. (\ref{fig:stripe2}) and (\ref{fig:check}). [Figure taken from \cite{Robertson_2006}] }
\label{fig:check}
\end{figure}

If we have a material with CDW modulation, it is not always simple to distinguish a stripe order from a checkboard order, expecially when the disorder effect is strong. For example if we see the right panels of the Fig. (\ref{fig:stripe2}) and (\ref{fig:check}), they appear really similar even if they cames from different order modulation (stripe and checkboard ones) \cite{Robertson_2006}. 
\vspace{1cm}
\clearpage
\newpage

\clearpage
\newpage

\subsection{Neutron and X-ray scattering}
Despite extensive experimental work on the incommensurate spin fluctuations and ordering in cuprate superconductors, experimental studies on the charge counterpart have been relatively scarse. In recent years, the Scanning Tunneling Microscopy (STM) technique has attracted much attention due to its ability to provide real space image of charge distribution (we will discuss this technique better in the next subsection). Although these STM studies provide unprecedented information on the inhomogeneous distribution of charge density and superconducting gap, due to the surface sensitive nature of the technique, its application has been so far limited to a subset of cuprate samples. In contrast, neutron and X-ray scattering investigations have been performed in many materials \cite{Zimmermann_1998,Abbamonte_2002,Abbamonte_2005,Tranquada_1999,Tranquada_1995,Tranquada_1995_1,Cheong_1991,Thurston_1992,Mason_1992,Yamada_1998,Tranquada_1996,Tranquada_1997,Stock_2002,Stock_2004}. 

We have to stress that neutron scattering gives only indirect evidences of charge modulations through its coupling with the underlying lattice, thus a lattice distrosion due to charge modulations can be seen by neutron scattering. On the other hand, X-ray scattering can couple directly to the charge degree of freedom\footnote{Except for when the incident photon energy is near the absorption edges, the largest contribution to the X-ray scattering comes from the structural modulation accompanying charge order. In this case, X-ray scattering is similar to the neutron one.}. The first X-ray study of the charge stripes was done with very high energy intensity X-rays by Zimmermann \cite{Zimmermann_1998}, in order to obtain a good momentum resolution. Recent avability of the LBCO crystals have made it possible to carry out more detailed investigations using soft X-ray resonant scattering \cite{Abbamonte_2005}.

In these kinds of experiments the signature of charge ordering is new peaks in the static structure function corresponding to a spontaneous breaking of symmetry, leading to a new periodicity longer than the lattice constant of the host crystal. It is also interesting to understand if the charge order eventually observed is commensurate or incommensurate; one way to determine this is to observe the position of the charge order Bragg peak as function of the temperature or of the pressure. If this position is locked then the CDW is commensurate. 

\subsection{The STM technique}
First of all we review some concepts about one of the most important experimental technique used to study HTS cuprates nowadays: the Scanning Tunneling Microscopy.

This kind of spectroscopy was invented by Binnig and Rohrer (1982) \cite{Binning_1982,Binning_1982_1}, and it gives the possibility to study a material with a spatial resolution down to the atomic scale. A really nice and historical demonstration of the possibility for studying superconductors with STM was achieved by Hess (1989) \cite{Hess_1989}, showing the electronic structure of the vortex core of the vortex lattice in $NbSe_2$. Thanks to its highly spatial and energy resolution, the STM is complementary to other techniques like optical spectroscopy and Angle-Resolved-PhotoEmission-Spectroscopy (ARPES), which offer $k-$space resolution.

The STM measures the tunneling current that flows between a metallic tip and a conducting sample separated by a thin insulating layer, generally vacuum. It allows one to obtain not only the surface topography with atomic scale resolution, but also the Local Electron Density of States (LDOS). The phenomenon behind the STM is the well-known quantum tunneling of electrons between two electrodes separated by a thin potential barrier. The core of this microscopy is the metallic tip, which is free to move in the $x-y$ plane above the sample scanning its surface, and in the $z$ direction. The tunneling regime is defined by three interdependent parameters: the electrode spacing $d$, the tunneling current $I$, and the bias voltage $V$ between the tip and the sample.

\begin{figure}[htb!]
\centering
\includegraphics[clip=True,scale=1.35]{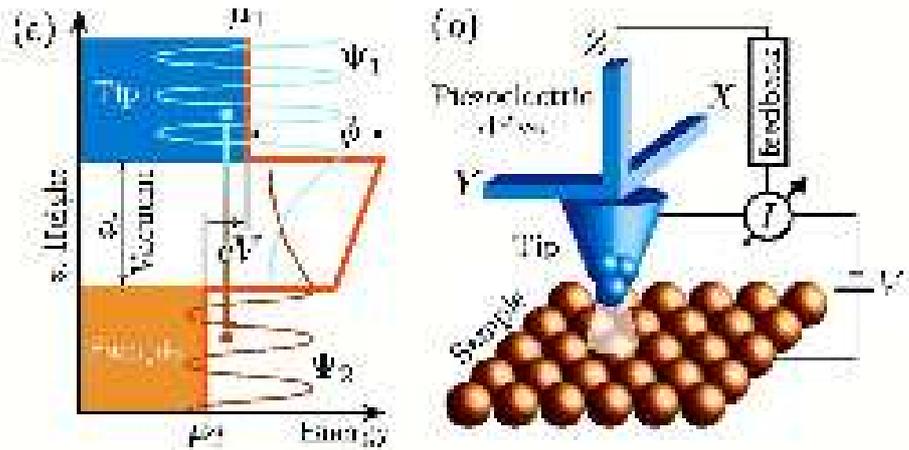}
\caption{(a) Tunneling process between the tip across a vacuum barrier of width $d$ and height $\phi$. The electron wave functions $\Psi$ decay exponentially into vacuum with a small overlap, allowing electrons to tunnel from one electrode to the other. (b) Schematic view of the scanning tunneling microscope. \newline [Figure taken from \cite{Fischer_2007}] }
\end{figure}

This kind of microscopy can be used in two different operating modes: a constant-current imaging mode, and a constant-height imaging mode. In the first case the tunneling current $I$ is kept constant by a feedback adjustment of the tip during the scan, so recording the height of the tip as a function of position it is possible to obtain an image, $z(x,y)$, of the surface of the sample. In the second operating mode the tip is scanned over the sample at a constant absolute height, recording the tunneling current $I(x,y)$ as a function of position. The last mode is the most used operating mode, and it can be showed theoretically that the tunneling conductance $dI/dV$ provides a measurement of the sample LDOS.

\begin{figure}[htb!]
\centering
\includegraphics[clip=True,scale=1.3]{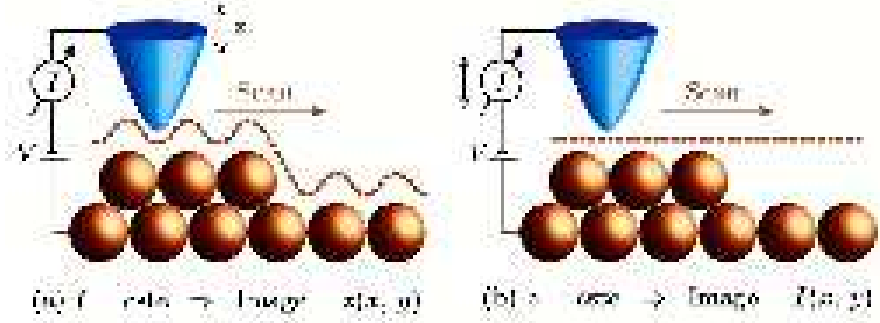}
\caption{Generic STM operating mode: (a) constant-current and (b) constant-height imaging. \newline [Figure taken from \cite{Fischer_2007}]}
\end{figure}

The starting point to intepret the STM measurements is the tunneling Hamiltonian formalism, that it is able to provide understanding both of the single-particle and pair-tunneling phenomena. The basic idea is to describe the transfer of particles across the barrier by a phenomenological tunneling Hamiltonian:
\begin{equation}
 H_{\scriptscriptstyle T} = \sum_{\lambda,\,\rho}T_{\lambda\rho}c^\dagger_\rho c_\lambda + h.c.
\end{equation}
where $\lambda$ is the label of single-particle states on the left side of the junction, and $\rho$ is the label for the analogous states on the right side. The operator $c_\lambda$ destroys a state in the left side, while the operator $c^\dagger_\rho$ creates a state in the right side of the barrier. $T_{\lambda\rho}$ is the tunneling matrix that depends upon the geometry of the tuneling junction and on the electronic states on both states. If a low bias voltage $V$ is applied across the junction, the total current can be calculated using linear-response theory. Neglecting the current of the electron pairs, the single-particle current is given by:
\begin{equation}
\label{eq:current}
 I_s = \frac{2\pi e}{\hslash}\int d\omega [f(\omega-eV)-f(\omega)] \sum_{\lambda,\,\rho}|T_{\lambda\rho}|^2 A_\lambda(\omega-eV)A_\rho(\omega)
\end{equation}
where $A_\lambda(\omega)$ and $A_\rho(\omega)$ are the single-particle spectral function of the tip and sample materials, and $f(\omega)$ is the Fermi function.

At zero temperature and assuming constant tunneling matrix elements Eq. (\ref{eq:current}) leads to a simple formula, that shows how the bias dependence of the conductance directly probes the DOS of the sample:
\begin{equation}
 \sigma(V) = \frac{dI_s}{dV} = \frac{2\pi e}{\hslash}|T|^2 N_T(0)N_S(eV)
\end{equation}
where ``T'' and ``S'' indicate ``tip'' and ``sample'' and $N(\omega)$ is the DOS at the Fermi level. It is necessary to underline also that for understanding local probes like STM, it is better to write down expression related to sample LDOS $N_{sample}(\vec{x},\omega)$:
\begin{equation}
 I_s \propto \int d\omega [f(\omega-eV)-f(\omega)] N_{tip}(\omega-eV)N_{sample}(\vec{x},\omega)
\end{equation}
Assuming a constant tip DOS the tunneling conductance becomes:
\begin{equation}
 \sigma(\vec{x},V) \propto \int d\omega [-f'(\omega-eV)]N_{sample}(\vec{x},\omega)
\end{equation}
where $f'$ is the derivative of the Fermi function.

\subsubsection{Some experimental observations}
Before giving some experimental evidences of charge modulations in HTS thanks to STM spectroscopic measurements, we can show some other findings about HTS obtained with STM. 

An important application of STM consists in surface characterization of a sample; the most widely studied HTS using STM is $Bi_2Sr_2CaCu_2O_{8+\delta}$ ($Bi2212$), because of the simplicity to prepare atomically flat and clean surfaces by cleaving, but $YBa_2Cu_3O_7$ ($Y123$) is also widely investigated.  

\begin{figure}[htb!]
\centering
\includegraphics[clip=True,scale=0.8]{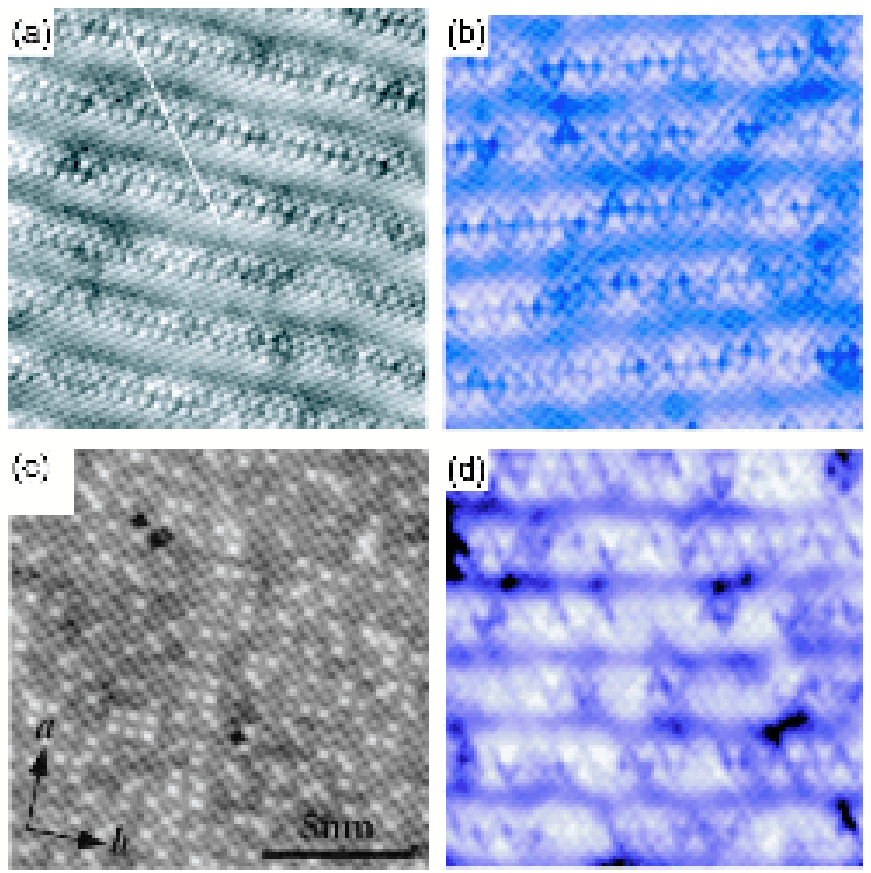}
\caption{STM images of the BiO surface of Bi-based HTS cuprates. (a) $Bi_2Sr_2CuO_6$ at $4.6$ $K$ \cite{Shan_2003}; (b) $Bi_2Sr_2CaCu_2O_6$ at $4.2$ $K$ \cite{Pan_2001}; (c) $Bi_2Sr_2CaCu_2O_6$ at $4.3$ $K$ \cite{Kinoda_2003}; (d) $Bi_2Sr_2CaCu_2O_6$ at $4.3$ $K$ \cite{Pan_2000}.}
\end{figure}

\begin{figure}[htb!]
\centering
\includegraphics[clip=True,scale=1.15]{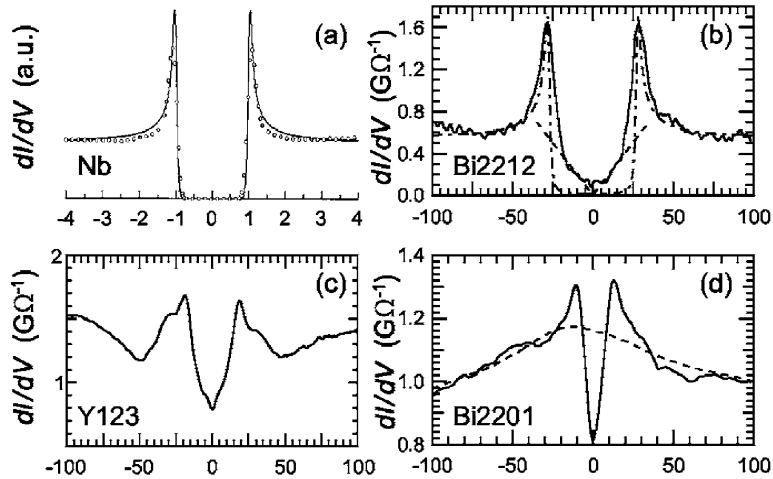}
\caption{(a) Nb at $335$ $mK$ (BCS-like) \cite{Pan_1998}; (b) Bi2212 at $4.8$ $K$ \cite{Renner_1995}; (c) Y123 at $4.2$ $K$ \cite{Maggio-Aprile_1995}; (d) Bi2201 at $2.5$ $K$ (solid line) and $82$ $K$ (dashed line) \cite{Kugler_2001}.}
\label{fig:4}
\end{figure}

At low temperature (and zero magnetic external field) the STM performed on HTS gives the possibility to identify generic features in the electronic DOS of the superconducting phase. For example observing conductance spectra it is possible to see: linear V-shaped energy dependence near the Fermi level, that is consistent with the d-wave symmetry; a large finite conductance at $V=0$; multiple broad coherence peaks at the gap edges (e.g. see Fig. (\ref{fig:4})).

\begin{figure}[htb!]
\centering
\includegraphics[clip=True,scale=1.3]{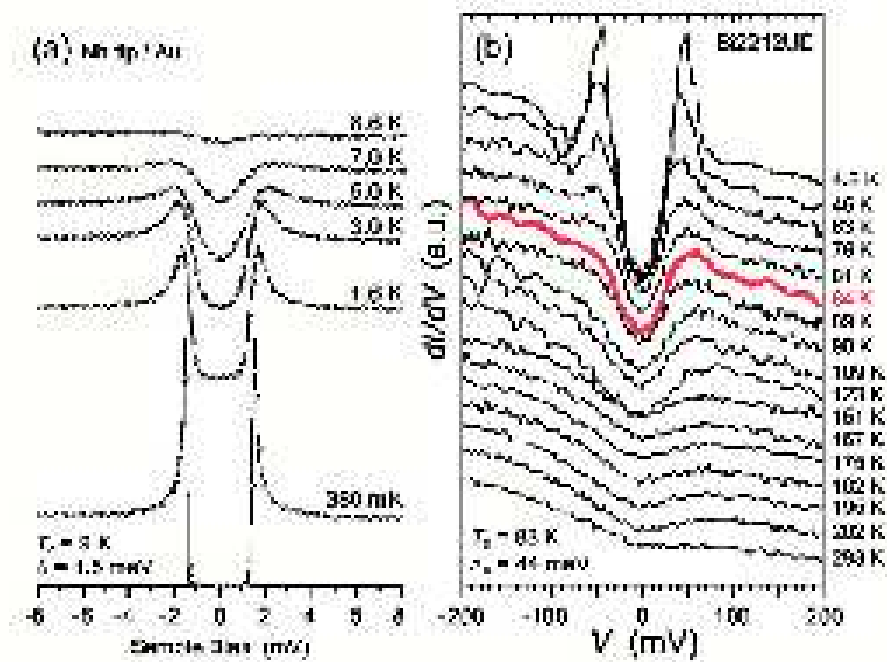}
\caption{T dependence of DOS measured by STM. (a) Nb (BCS-like) with $T_c\approx 9$ $K$ \cite{Pan_1998}; (b) Bi2212 (HTS) with $T_c\approx 83$ $K$ \cite{Renner_1998}.}
\label{fig:5}
\end{figure}

The STM is also used to investigate the so-called pseudogap region in HTS; many experimental techniques provide evidence for an unconventional normal state characterized by the opening of a gap in the electronic spectrum at a temperature $T^*$ above the critical temperature. What is the origin of the pseudogap phase is not clear yet, and there are essentially two possible theoretical interpretations: the pseudogap is the manifestation of some order unrelated to and/or in competition with the superconducting order; the pseudogap is the precursor of the superconducting gap, and reflects pairs fluctuations above $T_c$. In this thesis we will explore the possibility that the pseudogap is due to a disordered CDW of preformed Cooper pairs analogous to Fig. (\ref{fig:stripe}). The study of the pseudogap phase by STM is not simple because measuring the $T$ dependence of the LDOS is difficult due to tip shifts, that induce current variations caused not only by temperature variations but also by different tunneling locations. However observing the T-dependence of quasiparticle DOS in $Nb$ (a conventional BCS superconductor) and of Bi2212, it is clear the existence of a pseudogap into the HTS above $T_c$(see Fig. (\ref{fig:5})).

\vspace{0.5cm}

Another beautiful application of STM consists obtaining real-space imaging of vortices. The first observation of them was achieved using Bitter decoration ten years later (1967) the theoretical prediction by Abrikosov (1957); while the first STM image of vortex lattice was obtained by Hess roughly thirty years later (1989) on $NbSe_2$. 

\begin{figure}[htb!]
\centering
\includegraphics[clip=True,scale=1.8]{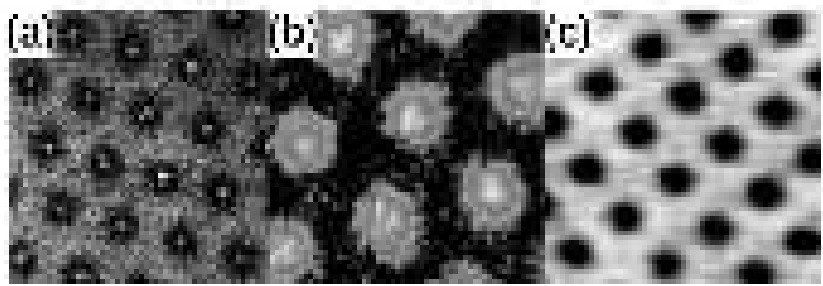}
\caption{STM images of vortex latices in conventional superconductors. (a) Hexagonal lattice in $NbSe_2$ \cite{Renner_1991}; (b) Hexagonal lattice in $MgB_2$ \cite{Eskildsen_2002}; (c) Square lattice in $LuNi_2B_2C$ \cite{DeWilde_1997}.}
\label{fig:6}
\end{figure}

\section{Charge Modulations in HTS}
Another important feature of HTS that can be studied by STM is the existence of spatial charge modulations like stripes, or charge-density-waves. 
\begin{figure}[htb!]
\centering
\includegraphics[clip=True,scale=1.4]{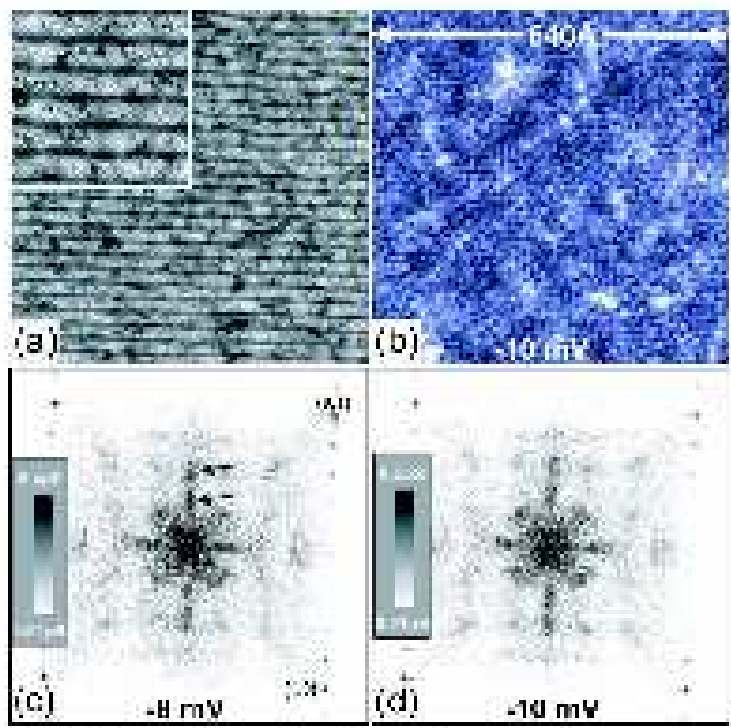}
\caption{Quasiparticle interference in $Bi2212$. Real-space (a-b) and Fourier-spaces (c-d) images of conductance maps at $-8$ $mV$ and $-10$ $mV$ \cite{McElroy_2003}.}
\label{fig:7}
\end{figure}

The presence of periodic spatial modulations of the DOS (not necessarily of the density) in HTS was observed for the first time by Hoffman \cite{Hoffman_2002}, who detected a modulation of the LDOS with a periodicity of about $4\,a_{\scriptscriptstyle 0}$ around the center of a vortex. Then many others experiments were able to see DOS modulations either in the superconducting state, or in the pseudogap state, or in vortex cores. Initially there was a problem interpreting DOS modulations in the superconducting state as due to charge modulations because in some experiments these modulations were dispersed in energy and not in others; the puzzle was solved interpreting these findings in terms of quasiparticle interference due to scattering from impurities and other inhomogeneities, and hence not due to a real CDW. This explains why in some experiments the DOS modulations into the superconducting state were not observed, because of the relatively good homogeneity of samples.

A square pattern modulation was observed in vortex cores, and a very similar order was found also in the pseudogap state: both are nondispersive (this suggests a CDW state), and the overall spectra are similar. This open a new question about the precise relation between the vortex core state and the pseudogap state.
\vspace{0.5cm}
\begin{figure}[htb!]
\centering
\includegraphics[clip=True,scale=1.3]{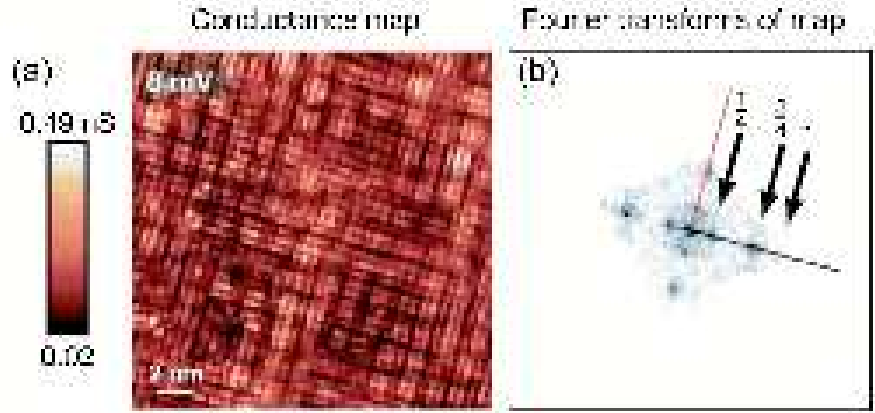}
\caption{Conductance map on $Ca_{1-x}Na_xCuO_2Cl_2$ at $8$ $mV$ and the corresponding Fourier transfrom \cite{Hanaguri_2004}.}
\label{fig:8}
\end{figure}

\begin{figure}[t!]
\centering
\includegraphics[clip=True,scale=1.5]{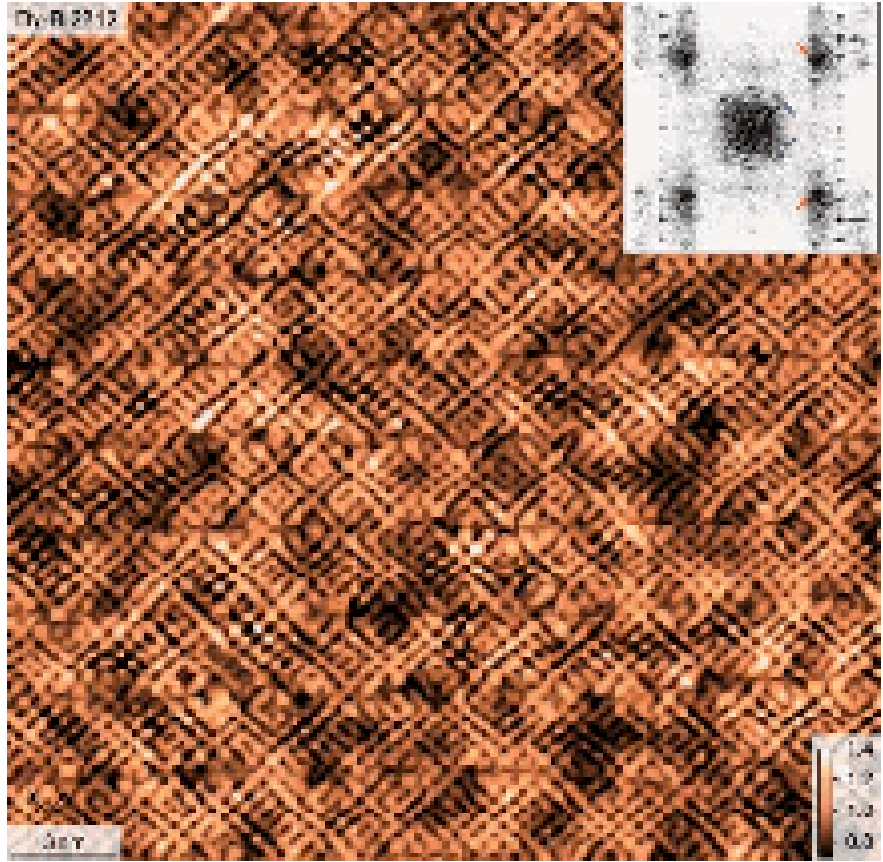}
\caption{A $25-nm^2$ $R$ map; no long-range order is apparent. Instead, we see randomly distributed electronic variations of the Cu-O-Cu bond state with equal probability of orientation along the two Cu-O axes. The Cu-O bond directions are shown as pairs of orthogonal black arrows. The inset shows its Fourier transform. The predominant peaks occur at wave vectors $\vec{q}\sim(3/4,0)$ and $(0,3/4)$ in units $2\pi/a_0$ (orange arrows), and the peaks at $\vec{q}\sim(1/4,0)$ and $(0,1/4)$ (blue arrows) are weaker. Atomic peaks $\vec{q}\sim(1,0)$ and $(0,1)$ are shown by black arrows. [Figure taken from \cite{Kohsaka_2007}].}
\label{fig:Kohsaka}
\end{figure}

Kohsaka et al. \cite{Kohsaka_2007} proposed a different technique to detect real charge modulations. Their proposal is based on theoretical studies that suggest that doping-induced correlation changes might be found directly as an asymmetry of electron tunneling currents with bias voltage\footnote{Electron extraction at negative sample bias being
strongly favored over electron injection at positive sample bias.} as observed by Anderson et al. \cite{Anderson_2006} and Randeira et al. \cite{Randeira_2005}. We remember that The STM tip-sample tunneling current is given by:
\begin{equation}
\label{eq:cur}
 I(\vec(r),z,V)=f(\vec{r},z)\int_0^{eV}N(\vec{r},E)dE
\end{equation}
where $z$ is the tip's surface-normal coordinate, $V$ is the relative sample-tip bias, and $N(\vec{r},E)$ is the sample's LDOS at lateral locations $\vec{r}$ and energy $E$. Unmeasurable effects due to the tunneling matrix elements, the tunnel-barrier height, and $z$ variations from electronic heterogeneity are contained in $f(\vec{r},z)$. For a simple metallic system where $f(\vec{r},z)$ is a featureless constant, Eq. (\ref{eq:cur}) shows that spatial mapping of the differential tunneling conductance $dI/dV(\vec{r},V)$ yields $N(\vec{r},E=eV)$. However, for the strongly correlated electronic states in a lightly hole doped
cuprate, the situation is much more complex. According to the theoretical studies, the correlations cause the ratio $Z(V)$ of the average density-of-states for empty states $\overline{N}(E=+eV)$ to that of filled states $\overline{N}(E=-eV)$ to become asymmetric by an amount:
\begin{equation}
\label{eq:zeta}
Z(V)=\frac{\overline{N}(E=+eV)}{\overline{N}(E=-eV)}\approx\frac{2n}{1+n} 
\end{equation}
where $n$ indicates the number of holes per unit cell. Spectral-weight sum rules \cite{Randeira_2005} also indicate that the ratio $R(\vec{r})$ of the energy-integrated $N(\vec{r},E)$ for empty states $E>0$ to  that of filled states $E<0$ is related to $n$ by:
\begin{equation}
\label{eq:erre}
 R(\vec{r})\approx\frac{2n(\vec{r})}{1-n(\vec{r})}
\end{equation}
Eq. (\ref{eq:zeta}) and (\ref{eq:erre}) have also a practical advantage; if we define the ratios $Z(\vec{r},V)$ and $R(\vec{r},V)$ in terms of the tunneling current:
\begin{eqnarray}
Z(\vec{r},V) &=& \frac{dI/dV(\vec{r},z,+V)}{dI/dV(\vec{r},z,-V)}\\
R(\vec{r},V) &=& \frac{I(\vec{r},z,+V)}{I(\vec{r},z,-V)}
\end{eqnarray}
we see immediately from Eq. (\ref{eq:cur}) that the unknown effects in $f(\vec{r},z)$ are all canceled out
by the division process. Thus, $Z(\vec{r},V)$ and $R(\vec{r},V)$ contain important physical information and are also expressible in terms of measurable quantities only. An example of the power of the application of this prescription is given in the Fig. (\ref{fig:Kohsaka}).

Recent systematic doping and temperature dependent STM studies \cite{Wise_2008} show CDW modulations in
the HTS $Bi_{2-y}Pb_ySr_{2-z}La_zCuO_{6+x}$. Authors find that a static and  non-dispersive, checkerboard-like electronic lattice exists over a wide range of doping, and that its wavelength increases with increasing hole density, supporting the physical picture of CDW formation. 

\subsection{The Random Field Ising Model}
As we have seen up to now, it is possible to find in underdoped cuprates CDW configurations, where on short length scales there is an ordered charge modulation, while not on long length scales, where a glassy like CDW behaviour emerges. This disordered CDW pattern could be due to the pinning of differents CDW ``variants'' caused by local impurities, as ions located out of the Cu-O plane acting as pinning fields with respect to a CDW variant. So at a coarse grained level, keeping in mind the strong coupling picture previously described using an Ising like pseudospin variable, we can think local impurities as quenched random fields coupled wiht the pseudospin Ising variable. Clearly as we already underlined, the real systems present more than two variants\footnote{A better schematization of this situation could be obtained using a Potts variable.} but for simplicity we shall consider only the situation with two variants, because it is able to catch the most important features that are in the pseudogap state. So a natural ingredient of our model will be the presence of quenched random fileds that couple with the Ising like part of the order parameter.

Now we are going to describe how a glassy CDW pattern is a natural consequence of the presence of impurities. Indeed we are handling a Random Field Ising Model (RFIM), and it is known that the RFIM in two dimensions has not an ordered state also at zero temperature \cite{Imry_1975}. Here we want to show briefly the so called \emph{Imry-Ma argument}, which asserts that quenched random fields induce the lack of long range order into the system if it is defined on a $d$-dimensional space lower than a critical dimension $d_l$ (the so called lower critical dimension) depending on the symmetry of the order parameter\footnote{The Imry-Ma argument is not so stringent to assure that for the RFIM the lower critical dimension is two; the rigorous demonstration of this fact is due to Imbrie \cite{Imbrie_1984}.}. First of all we have to underline that the interesting case to analyze is the weak disorder limit, otherwise the result is obvious: in the strong limit the disorder acts as a strong pinning centre settling the variable to a fix value independently from the interaction from the others variables. 

We take a spin system $\vec{S}_i$ interacting with quenched random fields $\{h_i\}$ which are coupled with our variables and have $\langle h \rangle=0$ and $\langle h^2 \rangle=\sigma^2$. Now we take at zero temperature a state for which every spin point along the same direction and we will call it ``positive'' for brevity; now we choose a domain of linear size $L$ and we reverse all spins in the opposite direction, we we will call this situation ``negative'' for brevity. Clearly whithout quenched random fields this reversing operation is not possible at zero temperature because we have to pay an energy cost to form the interface between the positive and negative state. But if there are random fields into the system, the situation is different because we can gain energy from these random fields. While the lack of energy due to the interface is proportional to the length of the interface itself, the gain is proportional to the square root of the number of spins inside the domain (this is a direct consequence of the central limit theorem, because every fields into the domain give an energy contribution proportional to $\sigma$, so assuming these fields indipendent each others their global contribution will be proportional to $\sqrt{N}$, where $N$ is the number of spins into the domain). If we indicate with $J$ the coupling constant between two spins, the interface energy is given by $JL^{d-1}$, while the volume energy gain is given by $\sigma L^{d/2}$, because $N\propto L^d$. 

We can summerize this energetic balance as follow:
\begin{equation}
 \Delta E\propto JL^{d-1}-\sigma L^{d/2}\qquad\mathrm{Ising\;like}
\end{equation}
\begin{equation}
 \Delta E\propto JL^{d-2}-\sigma L^{d/2}\qquad\mathrm{Heisenberg\; like}
\end{equation}
The system will break itself into domains if $\Delta E<0$, thus:
\begin{eqnarray}
\left \{
\begin{array}{l}
d<2\qquad\mathrm{Ising\;like}\\
d<4\qquad\mathrm{Heisenberg\;like}
\end{array}
\right.
\end{eqnarray}
and these domains have a characteristic sizes $L^*$ of the order of:
\begin{eqnarray}
\label{eq:scaledomain}
\left \{
\begin{array}{l}
L^*\approx\Big(\frac{J}{\sigma}\Big)^{2/(2-d)}\qquad\mathrm{Ising\;like}\\
L^*\approx\Big(\frac{J}{\sigma}\Big)^{2/(4-d)}\qquad\mathrm{Heisenberg\;like}
\end{array}
\right.
\end{eqnarray}
We can observe from Eq. (\ref{eq:scaledomain}) that the cases $d=2$ for the Ising situation, and $d=4$ for the Heisenberg situation are marginal ones, and Eq. (\ref{eq:scaledomain}) lacks its meaning. 

For the two dimensional RFIM Binder gave the right expression of $L^*$ taking into account roughness of the domain interface \cite{Binder_1983}:
\begin{equation}
 L^*\propto e^{(J/\sigma)^2}
\end{equation}
$L^*$ is also called Binder scale. 

  \clearpage{\pagestyle{empty}\cleardoublepage}
  \chapter{A Model for CDW-SC Competition}
  \label{chapter3}
\vspace{3cm}
 Here we shall introduce the model that will be studied in the next chapters. The model is an oversimplified and coarse-grained model for the description of the competition between SC and CDW. Even if it is possible to introduce on a phenomenlogical basis the model using only some arguments based on the symmetry properties of the competing orders, here we will show how to reach the same result starting from a microscopic Hamiltonian, giving to the model a more stable ground. Nevertheless it is important to stress that this microscopic derivation does not pretend to be a demonstration of the validity of our model, but it is simply a way to better understand the reasons in studying it and to build a bridge between the experiments and the features of the model that we will analyze.
  
\section{The repulsive Hubbard model}
The Hubbard model, an interacting fermion model independently proposed in $1963$ by Gutzwiller \cite{Gutzwiller_1963}, Hubbard \cite{Hubbard_1963} and Kanamori \cite{Kanamori_1963}, has played an extremely relevant role in condensed matter physics. Originally designed to describe, in a simplified way, the effects of the competition between electronic  delocalization and correlations within the narrow $d$-band of transition metal-oxides, the model has then been shown to be the ideal tool to describe the relevant collective features of these materials. 

The simplest expression of the Hubbard model relevant to the physics of cuprates is the single-band two-dimensional Hubbard model:
\begin{equation}
H= \sum_{i,j,\sigma}t_{ij}c_{i\sigma}^\dagger c_{j\sigma}+U\sum_in_{i\uparrow}n_{i\downarrow}+{\mathrm h.c.}
\end{equation}
where $c_{i\sigma}^\dagger$ ($c_{i\sigma}$) creates (annihilates) an electron at site $i$ with spin $\sigma$, $t_{ij}$ is the inter-site hopping, $n_{i\uparrow}=c_{i\uparrow}^\dagger c_{i\uparrow}$ ($n_{i\downarrow}=c_{i\downarrow}^\dagger c_{i\downarrow}$) is the number operator for a spin up (down) on the site $i$, $U>0$ is the Hubbard constant corresponding to the on-site Coloumb repulsion and $h.c.$ stands for ``hermitian conjugate''. For many purposes it is sufficient to assume that $t_{ij}$ is non-zero only when $i$ and $j$ are nearest neighbors, and we will assume:
\begin{eqnarray}
 t_{ij} =\left \{
\begin{array}{l}
-t\qquad {\mathrm if} \qquad \langle i,j\rangle\\\\
0\qquad {\mathrm otherwise}
\end{array}
\right.
\end{eqnarray} 
Despite the simplicity of the model, its solution turns out to be a very difficult task from a theoretical point of view. Only in one-dimension an exact solution is known. An important approach is the mean field theory solution of the model, even if it is not accurate in lower spatial dimensions due to neglecting spatial and temporal fluctuations; an improvement of this approach is the so called dynamical mean field theory, which neglects only spatial fluctuations. Another class of approximate analytic methods is known as the slave particle methods, closely related to the so called Gutzwiller projection variational approach, which searches for an approximate ground state of the model. Finally two others ways to attack  the model are the purely numerical Monte Carlo and Exact Diagonalization methods; the first one is computationally exspensive in terms of computer time, while the second one is expensive in terms of computer memory.

\section{The Hubbard to Heisenberg mapping}
\label{section2}
Now we want to review how and when it is possible to map a Hubbard model into a quantum antiferromagnetic  Heisenberg model (more details can be found in the appendix (\ref{appendix3})). 

 The repulsive Hubbard model ($U>0$) can be transformed in the strong coupling limit ($U\gg t_{ij}$) into an effective spin Heisenberg model with antiferromagnetic coupling. This can be done either using the  standard methods of perturbation theory or performing a canonical transformation. We shall describe the second way as showed in \cite{Chao_1978}. The idea of this technique is to divide the Fock space of the system into two subspaces: one containing only single occupied states (plus the vacuum), and the other one with only double occupied states. In the strong coupling limit, these two subspaces are well separated in energy each other, so that in the lowest order approximation the dynamic of the system will be described only by the dynamic within the two subspaces, neglecting the dynamics associated to the transition from one subspace to the other one. The idea is then to divide the starting Hamiltonian into two pieces, one ($H_0$) associated to the intra-band dynamics, and the other one ($H_1$) associated to the inter-band dynamics; performing a canonical transformation on $H$ it is possible to obtain an effective Hamiltonian for the low-energy subspace as a perturbative expansion in $H_1$ .

The decomposition of the starting Hamiltonian will be:
\begin{equation}
 H=H_0+\epsilon H_1
\end{equation}
where we introduced $\epsilon$ for bookeeping the perturbative orders, but at the end it will be put equal to one. The target now is to find an operator ${\mathcal S}$ for which the following canonical transformation:
\begin{equation}
\tilde H = e^{-i\epsilon\mathcal S}He^{i\epsilon\mathcal S}
\end{equation}
does not have linear $\epsilon$-terms. For this reason at the $2^{nd}$ order in $\epsilon$ the effective Hamiltonian reads:
\begin{equation}
 \tilde H=H_0+\frac{i}{2}[H_1,\mathcal S]\epsilon^2
\end{equation} 
Now it is necessary to find $H_0$, $H_1$ and $\mathcal S$ in order to write down the explicit expression of the effective Hamiltonian. To achieve this target it is useful to define two projector operators $P_1$ and $P_2$ respectively for the subspace with double occupied states and for the subspace with empty or single occupied states. Thanks to these projectors it is possible to rewrite the starting Hamiltonian $H$ using the following operators: $P_1HP_1$, $P_2HP_2$ and $P_1HP_2$, $P_2HP_1$; the former two represent the intra-band dynamics, while the latter two represent the inter-band dynamics. Schematically we can write the effects of these operators in the following way:
\begin{eqnarray}
\label{eq:transitions}
 \left \{
\begin{tabular}{llll}
$|0\,,\uparrow\rangle$ &$\Longrightarrow$& $|\uparrow\,,0\rangle$& $P_1HP_1$\\
$|\downarrow\,,\uparrow\downarrow\rangle$&$ \Longrightarrow$&$|\uparrow\downarrow\,,\downarrow\rangle$&$P_2HP_2$\\
$|0\,,\uparrow\downarrow\rangle$&$\Longrightarrow$&$|\downarrow\,,\uparrow\rangle$&$ P_1HP_2$\\
$|\uparrow\,,\downarrow\rangle$&$\Longrightarrow$&$|\uparrow\downarrow\,,0\rangle$&$P_2HP_1$
\end{tabular}
\right.
\end{eqnarray} 
where the meaning of the above symbols is the following: considering for simplicity a two site system (the generalization to more sites is trivial and more details can be found into the appendix (\ref{appendix3})), we indicate with $|\cdot,\cdot\rangle$ a generic state of our Fock space, where at the first position we write the state of the first site, and at the second position we write the state of the second site. A single state site can be empty ($0$), or single occupied with a spin up ($\uparrow$) or down ($\downarrow$), or double occupied with two opposite spins ($\uparrow\downarrow$). Thus for this two sites system the global Fock space is given by the following Fock states:
\begin{equation}
\begin{tabular}{llll}
$|0,0\rangle$ & $|\uparrow,0\rangle$ & $|\downarrow,0\rangle$& $|0,\uparrow\rangle$\\
$|0,\downarrow\rangle$ & $|\uparrow,\uparrow\rangle$ & $|\uparrow,\downarrow\rangle$ & $|\downarrow,\uparrow\rangle\nonumber$\\
$|\downarrow,\downarrow\rangle$ & $|\uparrow\downarrow,0\rangle$ & $|0,\uparrow\downarrow\rangle$ & $
|\uparrow,\uparrow\downarrow\rangle$\\
$|\downarrow,\uparrow\downarrow\rangle$ & $|\uparrow\downarrow,\uparrow\rangle$ & $|\uparrow\downarrow,\downarrow\rangle$ & $|\uparrow\downarrow,\uparrow\downarrow\rangle$
\end{tabular}
\end{equation}
So with the symbol:
\begin{equation}
|0\,,\uparrow\rangle\qquad \Longrightarrow\qquad |\uparrow\,,0\rangle
\nonumber
\end{equation}
we intend a transition from the state $|0\,,\uparrow\rangle$ to the state $|\uparrow\,,0\rangle$, and because they are clearly both in the subspace selected by the projector $P_1$, the operator that act this transformation is $P_1HP_1$. In the same way we can read the scheme reported by Eq. (\ref{eq:transitions})  (Obviously in that scheme there are not all the possible transformations, but only some ones as examples).

Now $H_0$ and $H_1$ can be easily identified:
\begin{eqnarray}
 H_0&=&P_1HP_1+P_2HP_2\\
H_1&=&P_1HP_2+P_2HP_1
\end{eqnarray}
After some algebra it is possible to find the relations among $\mathcal S$ and the projectors $P_1$ and $P_2$, and then to write the expression of the effective Hamiltonian:
\begin{equation}
 \tilde H=P_1HP_1+P_2HP_2-\frac{1}{U}[P_1HP_2HP_1-P_2HP_1HP_2]
\end{equation}
Considering both two-sites and three-sites interactions, the most general form of the effective Hamiltonian is:
\begin{eqnarray}
 \tilde H &=& -t\sum_{\langle ij\rangle,\sigma}\big(\widehat{c}_{i\sigma}^\dagger \widehat{c}_{j\sigma}+h.c\big)+\frac{4t^2}{U}\sum_{\langle ij\rangle}\Big(\vec{S}_i\cdot\vec{S}_j-\frac{\widehat{n}_i\widehat{n}_j}{4}\Big)+\nonumber\\
&&-\frac{t^2}{U}\sum_{\langle ijm\rangle\sigma}\big(\widehat{c}^\dagger_{i\sigma}\widehat{n}_{j\overline{\sigma}}\widehat{c}_{m\sigma}-\widehat{c}^\dagger_{i\sigma}\widehat{c}^\dagger_{j\overline{\sigma}}\widehat{c}_{j\sigma}\widehat{c}_{m\overline{\sigma}}+h.c.\big)
\end{eqnarray}
where
\begin{eqnarray}
 \left \{
\begin{array}{l}
\widehat{c}^\dagger_{i\sigma}=c^\dagger_{i\sigma}(1-n_{i\overline{\sigma}})\\
\widehat{c}_{i\sigma}=c_{i\sigma}(1-n_{i\overline{\sigma}})\\
\widehat{n}_{i\sigma}=\widehat{c}^\dagger_{i\sigma}\widehat{c}_{i\sigma}
\end{array}
\right.
\end{eqnarray}
and $\vec{S}$ is the quantum spin operator. If this effective Hamiltonian is studied for the special half-filling $(n_{i\uparrow}+n_{i\downarrow}=1)$ case, it reduces to a quantum antiferromagnet Heisenberg model:
\begin{equation}
 \tilde H = J\sum_{\langle ij\rangle}\Big(\vec{S}_i\cdot\vec{S}_j-\frac{1}{4}\Big)
\end{equation}
where $J=4t^2/U$. We'll focus our attention to this special situation.

\section{The attractive Hubbard model}
Now we shall describe another really important model, the attarctive ($U<0$) Hubbard model. It is the simplest model able to describe the SC-CDW competition. It can be written as:
\begin{eqnarray}\label{eq:Hm}
H=\sum_{<i,j>}\sum_{\sigma}t_{ij}c_{i\sigma}^\dag c_{j\sigma}-\frac{1}{2}|U|\sum_{i,\sigma}n_{i\sigma}n_{i,\overline{\sigma}}+h.c.
\end{eqnarray}
With this model it is possible to study in a simple and systematic way both the weak coupling limit ($U\ll t_{ij}$) corresponding to the standard BCS superconductivity, and the strong coupling limit ($U\gg t_{ij}$) corresponding to the Bose-Eistein Condensation (BEC) driven superconductivity. In HTS the overdoped regime is close to the BCS theory, while in the underdoped regime probably the BEC scenario dominates. So a natural question is how to relate them and what is the nature of the BCS-BEC crossover. This problem deals with the ``bosonic'' nature of the Cooper pairs and the possibility of interpreting  superconductivity itself as a superfluid of charged bosons. Even if this interpretation is not correct in the framework of the BCS theory, i.e. when the attractive coupling between the electrons is weak: indeed as a result of the small value of the binding energy, the Cooper pairs have a large size and are strongly overlapped, which make not possible to neglect the fermionic nature of their components. The situation is radically different in the opposite limit, when the attractive interaction between the electrons are strong. In this case the size of the Cooper pairs is really small and, as a consequence, they can be actually regarded as charged bosons. In this situation the onset of superconductivity, and hence the value of the critical temerature, is no longer controlled by the formation of Cooper pairs as in the BCS case, but by the superfluid transition associated to the Bose condensation of electron pairs. In this framework the pseudogap state observed in underdoped cuprates can be understood as a landmark of the presence of preformed pairs not yet condensed into a unique quantum state. In this region of the phase diagram the destruction of the superconducting ordering, and the behaviour of $T_c$ would not be determined by the breaking of the Cooper pairs, whose binding energy scale is given by the superconducting gap $\Delta$, but by phase fluctuations associated to the condensate wave function , whose energy scale is governed by the superconducting stiffness.  

Because of our interest on the pseudogap region of underdoped cuprates, a good starting point for the formulation of our model is the attractive Hubbard model. Moreover this model presents further advantage for us because it has both a superconducting order and a charge density wave ordering competing between each other, and while away from half-filling the ground state of this model is always a SC, exactly at half-filling the SC and the CDW are degenerate (for a review on the attractive Hubbard model there is \cite{Micnas_1990}). A lot of informations about the attractive Hubbard model can be easily obtained transforming this model into the more familiar repulsive Hubbard model, for which there exists a more extensive literature.  This can be achieved by a canonical transformation known as the ``attraction-repulsion'' transformation. In this way it is possible to obtain a one-to-one correspondence between the states of the repulsive Hubbard model and the states of the attractive Hubbard model, and between the charge and spin operators of the two models. In the next subsection we will see at work the ``attraction-repulsion'' transformation.

\subsection{The ``Attraction-Repulsion'' transformation}
In order to map the negative Hubbard model into a positive Hubbard model, we have to perform a canonical transformation known as ``attraction-repulsion'' transformation, defined by:
\begin{eqnarray}
\label{eq:att-rep}
\left\{ \begin{array}{l}
c_{i\uparrow}^\dag=b_{i\uparrow}^\dag\\
c_{i\uparrow}=b_{i\uparrow}\\
c_{i\downarrow}^\dag=e^{i\vec{Q}\cdot\vec{R}_i}b_{i\downarrow}\\
c_{i\downarrow}=e^{-i\vec{Q}\cdot\vec{R}_i}b_{i\downarrow}^\dag
\end{array}
 \right.
\end{eqnarray}
where $Q_i$ satisfies the relation $e^{i\vec{Q}\cdot\vec{R}}=-1$ for every traslational vector $\vec{R}_i$ that transforms one sub-lattice into the other one (I'm assuming the model defined on a bipartite lattice). Also the operators $b_{i\sigma}$ and $b_{i\sigma}^\dag$ satisfy the same anticommutating rules of the operators $c_{i\sigma}$ e $c_{i\sigma}^\dag$. 

If we indicate the single site states of the attractive Hubbard model with:
\begin{equation}
\label{eq:att}
 |0\rangle_a\qquad |\uparrow\rangle_a\qquad |\downarrow\rangle_a\qquad |\uparrow\downarrow\rangle_a
\end{equation}
and the states of the repulsive Hubbard model with:
\begin{equation}
\label{eq:rep}
 |0\rangle_r\qquad |\uparrow\rangle_r\qquad |\downarrow\rangle_r\qquad |\uparrow\downarrow\rangle_r
\end{equation}
where $a$ stays for ``attractive'' and $r$ for ``repulsive'', it is possible to show using Eq.  (\ref{eq:att-rep}) and the anticommutantig rules, that the following correspondence among the states (\ref{eq:att}) and (\ref{eq:rep}) is valid:
\begin{eqnarray}
\left\{ \begin{tabular}{lcl}
$|0\rangle_a$ & $\Longleftrightarrow$ & $|\downarrow\rangle_r$\\
$|\uparrow\rangle_a$ & $\Longleftrightarrow$ & $|\uparrow\downarrow\rangle_r$\\
$|\downarrow\rangle_a$ & $\Longleftrightarrow$ & $|0\rangle_r$\\
$|\uparrow\downarrow\rangle_a$ & $\Longleftrightarrow$ & $|\uparrow\rangle_r$
\end{tabular}
\right.
\end{eqnarray}
We can also define the number, spin and charge density operators for the attractive Hubbard model by: 
\begin{eqnarray}
\left\{ \begin{array}{l}
n_{i\sigma}=c_{i\sigma}^\dag c_{i\sigma}\\
\sigma_i^+=(\sigma_i^-)^\dag=c_{i\uparrow}^\dag c_{i\downarrow}\\
\sigma_i^z=\frac{1}{2}(n_{i\uparrow}-n_{i\downarrow})\\
\rho_i^+=(\rho_i^-)^\dag=c_{i\uparrow}^\dag c_{i\downarrow}^\dag\\
\rho_i^z=\frac{1}{2}(n_{i\uparrow}+n_{i\downarrow}-1)
\end{array}
 \right.
\end{eqnarray} 
These quantities after the transformation become:
\begin{eqnarray}
\left\{ \begin{array}{l}
\sigma_i^\dag=e^{-i\vec{Q}\cdot\vec{R}_i} b_{i\uparrow}^\dag b_{i\downarrow}^\dag\\
\sigma_i^z=\frac{1}{2}(b_{i\uparrow}^\dag b_{i\uparrow}+b_{i\downarrow}^\dag b_{i\downarrow}-1)\\
\rho_i^+=e^{i\vec{Q}\cdot\vec{R}_i} b_{i\uparrow}^\dag b_{i\downarrow}\\
\rho_i^z=\frac{1}{2}(b_{i\uparrow}^\dag b_{i\uparrow}-b_{i\downarrow}^\dag b_{i\downarrow})
\end{array}
 \right.
\end{eqnarray}
So defining the analogous number, spin and charge density operators for the repulsive Hubbard model:
\begin{eqnarray}
\left\{ \begin{array}{l}
\bar{n}_{i\sigma}=b_{i\sigma}^\dag b_{i\sigma}\\
\bar{\sigma}_i^+=(\bar{\sigma}_i^-)^\dag=b_{i\uparrow}^\dag b_{i\downarrow}\\
\bar{\sigma}_i^z=\frac{1}{2}(\bar{n}_{i\uparrow}-\bar{n}_{i\downarrow})\\
\bar{\rho}_i^+=(\bar{\rho}_i^-)^\dag=b_{i\uparrow}^\dag b_{i\downarrow}^\dag\\
\bar{\rho}_i^z=\frac{1}{2}(\bar{n}_{i\uparrow}+\bar{n}_{i\downarrow}-1)
\end{array}
 \right.
\end{eqnarray}
we will have the following correspondence:
\begin{eqnarray}
\label{eq:HpHn}
\left\{ \begin{array}{l}
\sigma_i^+=(\sigma_i^-)^\dag=e^{-i\vec{Q}\cdot\vec{R}_i}\bar{\rho}_i^+\\
\sigma_i^z=\bar{\rho}_i^z\\
\rho_i^+=(\rho_i^-)^\dag=e^{i\vec{Q}\cdot\vec{R}_i}\bar{\sigma}_i^+\\
\rho_i^z=\bar{\sigma}_i^z
\end{array}
 \right.
\end{eqnarray} 
The equation (\ref{eq:HpHn}) shows that the density operators in the new (repulsive) representation play the same role of the spin operators in the old (attractive) representation and viceversa. Also the Hamiltonian after the attraction-repulsion transformation reads as:
\begin{eqnarray}
\tilde{H} &=&\sum_{<i,j>}\sum_\sigma t_{ij} b_{i\sigma}^\dag b_{j\sigma}
+\frac{1}{2}|U|\sum_{i\sigma}\bar{n}_{i\sigma}\bar{n}_{i-\sigma}
\end{eqnarray}

This shows that the negative (attractive) Hubbard model can be transformed easily into a positive (repulsive) Hubbard model. 

\section{Our model for the CDW-SC competition}
At this level we have all ingredients to define our model. We have to take in mind that we want to focus ourselves on the pseudogap region of HTS cuprates where it is possible to have both precursor superconductivity effects and CDW ordering. Thus a good candidate for the description of this region is the attractive Hubbard model which has at half-filling a ground state degeneracy between the SC and the CDW. But as we have seen into the chapter (\ref{chapter2}), glassy like CDW configurations are found in the pseudogap region, so we need to break the degeneracy of the half-filled attractive Hubbard model in order to favour the CDW state. This can be done introducing into the Hamiltonian (\ref{eq:Hm}) an interaction term $W_{ij}$ in the following way:
\begin{eqnarray}
\label{eq:Hm1}
H=\sum_{\langle i,j\rangle}\sum_{\sigma}t_{ij}c_{i\sigma}^\dag c_{j\sigma}-\frac{1}{2}|U|\sum_{i,\sigma}n_{i\sigma}n_{i,\overline{\sigma}}+\nonumber\\
+\frac{1}{2}\sum_{\langle i,j\rangle}\sum_{\sigma,\sigma'}W_{ij}n_{i\sigma}n_{j\sigma'}+h.c.
\end{eqnarray}
and after the ``attraction-repulsion'' transformation this Hamiltonian reads as:
\begin{eqnarray}
\label{eq:Hm2}
\tilde{H} &=&\sum_{\langle i,j\rangle}\sum_\sigma t_{ij} b_{i\sigma}^\dag b_{j\sigma}
+\frac{1}{2}|U|\sum_{i\sigma}\bar{n}_{i\sigma}\bar{n}_{i-\sigma}+2\sum_{\langle i,j\rangle}W_{ij}\bar{\sigma}_{i}^z\bar{\sigma}_{j}^z
\end{eqnarray}
A more clear way to see that $W_{ij}$ introduces an anisotropy term into the Hamiltonian consist to take the strong coupling limit of Eq. (\ref{eq:Hm2}). This limit it is also really important for our study because we would describe the system in the framework of preformed Cooper pairs. Taking this strong coupling limit of Eq. (\ref{eq:Hm2}) we obtain an antiferromagnetic quantum Heisenberg model as we have seen in the section $2$:
\begin{equation}
\label{eq:JJ}
 H=\sum_{\langle i,j\rangle}\Big[J_\parallel(S^x_iS^x_j+S^y_iS^y_j)+J_\perp S^z_iS^z_j\Big]
\end{equation}
where $J_\parallel=J$, $J_\perp=J+W$ with $J=4t^2/U$ and $W_{\langle ij\rangle}\equiv W$. Hereafter we shall consider a slightly different anysotropy term which has essentially the same symmetries but that is easier to analyze:
\begin{equation}
\label{eq:JG}
 H=J\sum_{\langle i,j\rangle}\vec{S}_i\cdot\vec{S}_j-G\sum_i(S^2_i)^2
\end{equation}
We have substituted an exchange anisotropy term with a ionic anisotropy term. $G$ breaks the rotational spin symmetry and because we want to favour the CDW state it is necessary that $G>0$. 

At long wave lengths the antiferromagnetic  order parameter behaves classically in the sense that the only effect of quantum fluctuations is to renormalize the original parameters \cite{Halperin_1989}  (renormalized classical regime). Thus the spin is treated as a classical variable and at this point the antiferromagnetic model can be mapped trivially on the ferromagnetic one just by using the staggered magnetization as a variable. 

Treating the model classically there are two possible phases. Either the magnetization is into the $XY$ plane (we will refer to this one as SC), or it is along the $z-$axis (we will refer to this one as CDW). In Eq. (\ref{eq:JJ}) this is controlled by the ratio $J_\perp/J_\parallel$; however this ratio also controls the stiffness of the two phases. On the other hand in Eq. (\ref{eq:JG}) the ratio $G/J$ controls only the relative stability of the two phases. For simplicity we assume that the two phases have the same stiffness.

We have now to introduce the last ingredient to our model; as we said in the chapter (\ref{chapter2}), in cuprate HTS there are random impurities out of the superconducting Cu-O plane that act as pinning fields with respect the CDW. So in the pseudogap region it is possible to observe glassy like CDW configurations where the different CDW ``variants'' are pinned by the quenched impurities. We can model this effect adding to the Hamiltonian a term that couple the $z-$component of the spin variable with a quenched random field. Then the model that we will analyze reads:
\begin{equation}\label{eq:Heis}
 H=-J\sum_{\langle i,j\rangle}\vec{S}_i\cdot\vec{S}_j-G\sum_i(S_i^z)^2+\frac{W}{2}\sum_ih_iS_i^z
\end{equation} 
where $\vec{S}_i=\{S_i^x,S_i^y,S_i^z\}$ is a classical Heisenberg spin with $|\vec{S}|=1$, $J$ is a positive coupling constant. The first term represents the nearest neighbor interaction of the order parameter. The second term breaks the symmetry in spin space with $G>0$ favoring a CDW state; $h_i$ are statistical independent quenched random variables with a flat probability distribution between $-1$ and $+1$; also $W>0$. These random fields would mimic impurities always present in the real samples. We have to note also that the length of our spins is fixed to one because we are thinking to stay in the framework of preformed Cooper pairs, for which it is important the phase of the order parameter and not its amplitude.
\begin{figure}[htb!]
\centering
\includegraphics[clip=true, scale=.7]{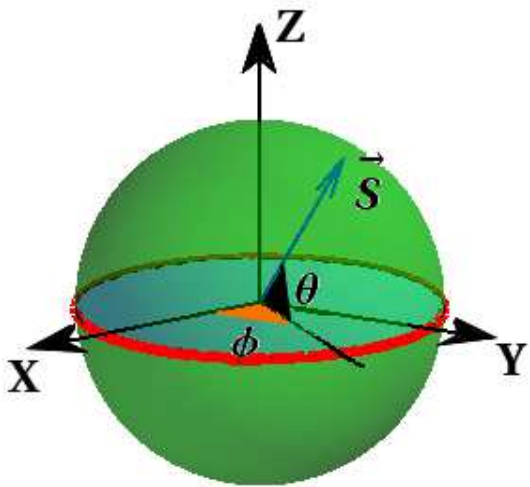}
\caption{A sketch of the order parameter.}
\end{figure}

\section{The magnetic field and the Peierls substitution}
Understanding how a system respond to an external applied magnetic field is important from both the  theoretical and the experimental points of view, so we shall show how to introduce the magnetic field into the effective model that we'll study.

First of all we want to underline that the magnetic field can be coupled either with the orbital degrees of freedom of electrons or with the spin degrees of freedom of electrons; in the last case we have a Zeeman coupling term $-\mu \vec{B}\cdot\vec{S}$, but now we are interested in the orbital coupling. The standard way to introduce this magnetic coupling is the so called Peierls Substitution:
\begin{equation}
 c_{i\sigma}\longrightarrow c_{i\sigma}e^{-i\frac{e}{\hslash}\int_{\vec{R}_0}^{\vec{R}_i}\vec{A}(\vec{r})\cdot d\vec{r}}
\end{equation}
where $\vec{R}_0$ is an arbitrary fixed vector, while $\vec{A}(\vec{r})$ is the potential vector related to the magnetic field $\vec{B}$. Performing this transformation on the repulsive Hubbard model we have:
\begin{equation}
H= -\sum_{\langle ij\rangle\sigma}te^{-i\frac{e}{\hslash}\int_{\vec{R}_i}^{\vec{R}_j}\vec{A}(\vec{r})\cdot d\vec{r}}c_{i\sigma}^\dagger c_{j\sigma}+U\sum_in_{i\uparrow}n_{i\downarrow}+h.c.
\end{equation}
We want to stress now that the half-filling repulsive Hubbard model with magnetic field and in the strong coupling limit is equal to a Heisenberg model too, so the magnetic field doesn't play any role unless either more than two-body terms are considered or more then the $2^{nd}$ order terms into the perturbation series is taken. 

For the half-filling attractive ($U < 0$) Hubbard model, the situation in a magnetic field is completely different because we have coupled electrons able to move forming a current. The effective Hamiltonian reads now:
\begin{eqnarray}
 \tilde H&=&-\frac{2t^2}{U}\sum_{\langle ij\rangle\sigma}n_{i\sigma}n_{j\overline{\sigma}}+\nonumber\\
&&+\frac{2t^2}{U}\sum_{\langle ij\rangle}\Big(c_{i\uparrow}^\dagger c_{i\downarrow}^\dagger c_{j\uparrow}c_{j\downarrow}e^{-i\frac{2e}{\hslash}\int_{\vec{R}_i}^{\vec{R}_j}\vec{A}(\vec{r})\cdot d\vec{r}}+c_{j\uparrow}^\dagger c_{j\downarrow}^\dagger c_{i\uparrow}c_{i\downarrow}e^{i\frac{2e}{\hslash}\int_{\vec{R}_i}^{\vec{R}_j}\vec{A}(\vec{r})\cdot d\vec{r}}\;\Big)\nonumber\\
\end{eqnarray}
writing also:
\begin{equation}
 A_{ij}\equiv\frac{2e}{\hslash}\int_{\vec{R}_i}^{\vec{R}_j}\vec{A}(\vec{r})\cdot d\vec{r}
\end{equation}
and
\begin{eqnarray}
 \left \{
\begin{array}{l}
S_i^+=c^\dagger_{i\uparrow}c^\dagger_{i\downarrow}\\
S_i^-=c_{i\uparrow}c_{i\downarrow}
\end{array}
\right.
\end{eqnarray}
We have
\begin{eqnarray}
 \tilde H&=&-\frac{2t^2}{U}\sum_{\langle ij\rangle\sigma}n_{i\sigma}n_{j\overline{\sigma}}+\frac{2t^2}{U}\sum_{\langle ij\rangle}\Big(S_{i}^+S_{j}^-e^{-iA_{ij}}+S_{j}^+S_{i}^-e^{iA_{ij}}\Big)
\end{eqnarray}
and using the relations:
\begin{eqnarray}
 \left \{
\begin{array}{l}
S_i^+=S_{i}^x+iS_{i}^y\\
S_i^-=S_{i}^x-iS_{i}^y\\
S_i^z=\frac{1}{2}(n_{i\uparrow}-n_{i\downarrow})
\end{array}
\right.
\end{eqnarray}
it is possible to write the following effective Hamiltonian:
\begin{eqnarray}
 \tilde H&=& J\sum_{\langle ij\rangle}\Big[(S^x_iS^x_j+S^y_iS^y_j)\cos A_{ij}-(S^x_iS^y_j-S^y_iS^x_j)\sin A_{ij}+S^z_iS^z_j-\frac{1}{4}\Big]\nonumber\\
\end{eqnarray} 
where $J=4t^2/U$, and if $A_{ij}$ is equal to zero the standard Heisenberg Hamiltonian is recovered.

If we want to obtain the physical current $j_{ij}$ through a bond, we have to evaluate the following expression:
\begin{equation}
\label{eq:currentCap3}
 j_{ij}=-\frac{\partial H}{\partial A}
\end{equation}
where $A$ is the potential vector between the points $\vec{R}_i$ and $\vec{R}_j$;  $A$ can be considered a constant over the distance of a lattice spacing $a$, thus we can write:
\begin{equation}
 A_{ij}\equiv\frac{2e}{\hslash}\int_{\vec{R}_i}^{\vec{R}_j}\vec{A}(\vec{r})\cdot d\vec{r}\sim\frac{2e}{\hslash}aA
\end{equation}
So Eq. (\ref{eq:currentCap3}) can be rewritten as:
\begin{equation}
\label{eq:currentCap3_2}
 j_{ij}=-\frac{\partial H}{\partial A}\sim-\frac{2ea}{\hslash}\frac{\partial H}{\partial A_{ij}}
\end{equation}

  \clearpage{\pagestyle{empty}\cleardoublepage}
  \chapter[GPE in Competing CDW-SC Systems]{Giant Proximity Effect in Competing Charge Density Waves - Superconducting Systems}
  \label{chapter4}
\vspace{3cm}
In this chapter we want to study the simple coarse-grained model of the competition between SC and CDW  introduced in the previous chapter. We show that in a Josephson-Junction geometry the system has a Giant Proximity Effect (GPE) as observed in cuprate High-$T_c$ Superconducting junctions that recently caught a  lot of attention both from experimental and theoretical point of view.


\section{Introduction and formulation of the problem}


A striking property of cuprate superconductors is the possibility to observe an unusually large proximity effect as reported by many authors \cite{Decca_2000,Bozovic_2003,Bozovic_2004}. We refer to the case of  $(S-S'-S)$ junctions where the electrodes are High Temperature Superconductors and the barrier $S'$ is made by an underdoped cuprate, with $T_c'<T_c$. Here $T_c'$ is  the critical temperature of the barrier and $T_c$ the one of the electrodes. The GPE consists in a finite superconducting current through the Josephson junction whose barrier thickness $L$ is larger than the proximity coherence length $\xi_{S'}$ $(L\gg\xi_{S'})$. 
We focus on the geometry where the barrier is parallel to electodes so that the Josephson current flows along the c-axes where the coherence length is very short ($\xi_c\approx 4$ \AA). 
It has been proposed that this phenomenon could be explained with the presence of preformed pairs, superconducting fluctuations, or droplets well above $T_c'$. Our idea is that they could be due to the  competion between Superconductivity and Charge-Density-Waves. We will show below that the distance over which a superconductor is converted into a CDW can be much larger than $\xi_{S'}$. Thus a superconductor can propagate into a CDW state for much larger distance than into a normal metal. Of course it is not clear that the state above $T_c'$ can be described by a CDW. Indeed, long range ordered CDW are rarely observed in cuprate HTS. However our proposal is that the system can be viewed as a CDW glass (see chapter (\ref{chapter2})) due to quenched disorder. This however will be not essential for ours arguments, therefore for the moment we will ignore the role of disorder. 

\begin{figure}[htb!]
\centering
\includegraphics[clip=True,scale=1.2]{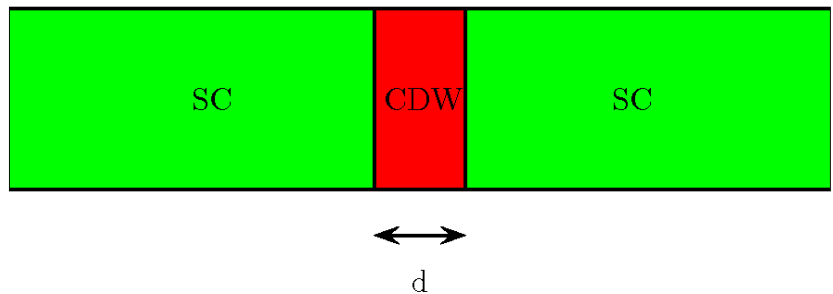}
\caption{A SC-CDW-SC sandwich.}
\label{fig1}
\end{figure}

Now we want to present analytic and numerical results for the properties of a $SS'S$ Josephson Junction system with the geometry showed on Fig. (\ref{fig1})
\vspace{1cm}

\begin{figure}[htb!]
\centering
{
\psfrag{X}{X}
\psfrag{Y}{Y}
\psfrag{Z}{Z}
\psfrag{t}{$\theta$}
\psfrag{f}{$\phi$}
\includegraphics[clip=true,scale=0.7]{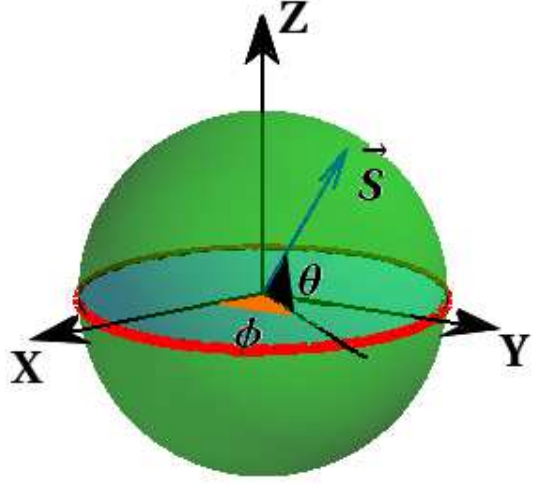}
}
\caption{Order parameter.}
\label{fig2}
\end{figure}

Taking in mind the model introduced in the previous chapter (Eq. (\ref{eq:Heis})), and discarding the contribution of the quenced random fields, we will work in a Ginzburg-Landau framework describing our system using a three dimensional order parameter $\vec{S}\equiv\{S_x,S_y,S_z\}$ (see Fig. (\ref{fig2})), where the first two components ($S_x,S_y$) are the real and imaginary parts of the superconducting order parameter and the third component ($S_z$) represents the order parameter of the CDW state. Also $\vec{S}$ will be subjected to the constrain $\mid\vec{S}\mid \equiv 1$, so that, when $\vec{S}$ is totally in the $xy$ plane, superconductivity could be lost only by phase decoherence and not by amplitude annihilation of the order parameter itself (see chapter (\ref{chapter3}) for a deeper explanation of the model). 
\vspace{0.2cm}

A similar approach was used by Demler \emph{et al.} \cite{Demler_1998} to study the same problem but in the Zhang's $SO(5)$ scenario \cite{Zhang_1997}, where the two competing phases are the Superconductivity and the Anti-Ferromagnetism. Even if many our results are similar to Demler's ones, his perspective to explain GPE was unsuccessful. Indeed, Bozovic \emph{et al.} \cite{Bozovic_2003} reported that one-unit-cell thick antiferromagnetic $La_2CuO_4$ barrier layers remain insulating and completaly block a supercurrent. We have to notice also that others authors have investigated theoretically this problem with different techniques \cite{Alvarez_2005,Covaci_2006,Alexandrov_2007}. 

We will describe the system using the following functional, obtained from the lattice Hamiltonian Eq. (\ref{eq:Heis}) taking its continuum limit\footnote{The continuumm limit of the lattice model is easily get following the reverse steps described in the chapter ($\ref{chapter1}$) when we had to transform the continuum $XY$ model into its lattice expression (see section (\ref{latt-cont})).}:
\begin{eqnarray}
\label{eq:functional}
F\Big[\vec{S}(\vec{x})\Big]=\frac{J}{2a^{d-2}}\int |\vec{\nabla}\vec{S}|^2d\vec{x}-\frac{G}{a^d}\int\Big(S_z(\vec{x})\Big)^2d\vec{x}\,,
\end{eqnarray}
where the order parameter $\vec{S}$ is defined into a $d-$dimensional space and it has to satisfy the constraint $\mid\vec{S}\mid = 1$. We notice also that $a$ represents the lattice spacing of the underlying lattice over which is defined the Hamiltonian descried in the previous chapter (we remember that in this chapter we are not considering the effects of quenched disorder). We put:
\begin{eqnarray}
 \left \{
\begin{array}{l}
\frac{J}{a^{d-2}}\equiv\rho\\\\
\frac{G}{a^d}\equiv g
\end{array}
\right.
\end{eqnarray} 
We have to underline here that the constant $\rho$ has the same meaning of the stiffness introduced in the chapter (\ref{chapter1}) speaking about the $XY$ model. But here we have a three dimensional spin variable, so we can think to have a stiffness into the $XY$ plane (as for the $XY$ model) and another one to describe the energy cost to pass from the $XY$ plane to the $z$-axis. We have taken both stiffnesses  equal for simplicity. 

From here in we will study the functional Eq. (\ref{eq:functional}) relatively to the geometry showed on Fig. (\ref{fig1}), so considering a one dimensional problem. We can rewrite the functional in spherical coordinates in order to satisfy automatically the constraint on the length of the order parameter:
\begin{eqnarray}
 F[\theta(x),\phi(x)] & = & 
\int dx \Big\{\frac{\rho}{2}\Big[\Big(\frac{d \theta}{dx}\Big)^2+\cos^2\theta\Big(\frac{d\phi}{dx}\Big)^2\Big]-g\sin^2\theta\Big\} \,,
\end{eqnarray}
where $\theta\in[-\pi/2,\pi/2]$ and $\phi\in[0,2\pi)$ are the two angles as sketched in Fig. (\ref{fig2}), and obviously related to the Cartesian components of the order parameter via these relations:
\begin{eqnarray}
\left \{
\begin{array}{l}
S_x=\cos\theta\cos\phi\\
S_y=\cos\theta\sin\phi\\
S_z=\sin\theta
\end{array}
\right.
\end{eqnarray}
While the angle $\theta$ is related to the CDW order, the angle $\phi$ represents the phase on the SC order parameter.

If we want to study the proximity effect in a $SS'S$ Josephson junction using the above functional, we have to choose the anisotropy constant $g$ to be negative outside the barrier (i.e. for $x<-d/2$ and $x>d/2$) and positive into the barrier (i.e. for $-d/2<x<d/2$). We put the x-axis origin in the middle point of the barrier, whose thickness is $d$. From a mathematical point of view this problem is well defined if we choose proper boundary conditions; we know that the two superconductors forming the junction have both a well-defined phase. 
So we can choose the following boundary conditions for the superconductors:
\begin{eqnarray}
\left \{
\begin{array}{l}
\theta(-\infty)=\theta(+\infty)=0\\
\phi(-\infty)=0\\
\phi(+\infty)=\Delta\Phi
\end{array}
\right.
\end{eqnarray}
and we have to impose continuity at the interface barrier. For simplicity the previous conditions could be simplifyed considering a junction where the anisotropy constant $g$ outside the barrier is much greater than the one inside the barrier itself; this is the case of rigid superconducting boundary conditions. So we can write:
\begin{eqnarray}
\left \{
\begin{array}{l}
\theta(-d/2)=\theta(+d/2)=0\\
\phi(-d/2)=0\\
\phi(+d/2)=\Delta\Phi
\end{array}
\right.
\end{eqnarray}
and solve the problem more easily only into the barrier. This choice for the boundary condition is clearly a simplification, and its physical meaning is that we are assuming there is not any kind of penetration for the CDW into the SC. 

So we have to minimize our functional, or in other words we have to solve the Eulero-Lagrange equations related to 
\begin{equation}
 F[\theta(x),\phi(x)]=\int dx  f\Big(\theta,\phi,\frac{d\theta}{dx},\frac{d\phi}{dx}\Big)\,,
\end{equation}
so
\begin{eqnarray}
\left \{
\begin{array}{l}
\frac{d}{dx}\Big(\frac{\partial f}{\partial(d\theta/dx)}\Big)-\frac{\partial f}{\partial \theta}=0\\\\
\frac{d}{dx}\Big(\frac{\partial f}{\partial(d\phi/dx)}\Big)-\frac{\partial f}{\partial \phi}=0
\end{array}
\right.
\end{eqnarray}
and explicitly we can write down:
\begin{eqnarray}
& & 
\rho\frac{d^2\theta}{dx^2}+\rho\sin\theta\cos\theta\Big(\frac{d\phi}{dx}\Big)^2+2g\sin\theta\cos\theta=0\\
& & 
\frac{d}{dx}\Big(\cos^2\theta\frac{d\phi}{dx}\Big)=0
\end{eqnarray} 
As it is possible to see, $\phi$ is a cyclic coordinate (i.e. it does not appear explicitly into the expression of the functional), and so the current $I$ related to this coordinate  must to be conserved; this is also clear observing the second Eulero-Lagrange equation:
\begin{equation}
 I=\cos^2\theta\frac{d\phi}{dx}\equiv constant
\label{eq:diff_phi}
\end{equation}
We have to stress also that $I$ is not simply the Noether current related to the coordinate $\phi$, but it represents the real physical current $\mathcal I$ passing through the junction, indeed performing the same calculations seen into the last section of the chapter (\ref{chapter3}) speaking about the Peierls substitution, we obtain for the physical current:
\begin{equation}
 {\mathcal I} = \frac{2ea}{\hslash}I
\end{equation}

Taking the conservation law Eq. (\ref{eq:diff_phi}) we can rewrite the first Eulero-Lagrange equation in the following way:
\begin{eqnarray}
& &
\xi_g^2\frac{d^2\theta}{dx^2}+\xi_g^2I^2\frac{\sin\theta}{\cos^3\theta}+\frac{1}{2}\sin(2\theta)=0
\label{eq:diff_theta}
\end{eqnarray}
where
\begin{equation}
\xi_g^2=\frac{\rho}{2g}
\end{equation}
We want to underline that $\xi_g$ has the physical dimension of a length, and it represents the length scale over which the order parameter of the system pass from the CDW value to the SC value and viceversa. 

\section{Outlines of the solutions}
In this section we show the solutions of Eq. (\ref{eq:diff_theta}) and (\ref{eq:diff_phi}) for our order parameter, but it is also important to stress that two others kind of solutions are possible for our problem. So we have three families of solutions, and we refer to them as:
\begin{description}
 \item[$\bullet$ SC] Superconducting solution;
 \item[$\bullet$ MIXED/CDW] Mixed/Charge-Density-Wave solution;
 \item[$\bullet$ MIXED] Mixed solution.
\end{description}

Here we want to give the profiles of the order parameter in these cases, showing how they change by tuning the parameters of the problem. More details can be found into the Appendix (\ref{appendix4}). 

\subsection{SC solution}
As can be checked easily, there is a trivial solution $\theta(x)\equiv0$ of the equation \ref{eq:diff_theta}, and the corresponding value for $\phi(x)$ is $\phi(x)=(\Delta\Phi/d)x+(\Delta\Phi/2)$. This kind of solution, where our order parameter stays into the $XY$ plane, changing its phase from zero to $\Delta\phi$ linearly, corresponds in our picture to the superconducting state. The energy of this solution (i.e. the value of the functional $F[\theta(x),\phi(x)]$ ) is:
\begin{equation}
 E_{SC} = C \Delta\Phi^2
\label{eq:energy_SC}
\end{equation}
where 
\begin{equation}
 C=\rho/2d=(\frac{g\xi_g}{2})\frac{1}{(d/2\xi_g)}
\end{equation}
is a constant depending only on the parameters of our problem, and remember the definition of the stiffness into the chapter (\ref{chapter1}), it represents the SC stiffness itself.

\subsection{MIXED/CDW solution}
We have analyzed another kind of solution in which we have a perfect Charge Density Wave ($\theta=\pm\pi/2$) in the middle point of the junction (i.e. for $x=0$), or in a symmetric-middle region of the junction (i.e. for $-c/2<x<c/2$ with $c<d$). This kind of solution cannot be found from Eq.  (\ref{eq:diff_theta}) and (\ref{eq:diff_phi}), because of its non-analyticity in $x=0$ or $x=\pm\,c$. In this case the order parameter pass through the North or South pole of its domain space, so $\phi(x)$ remains arbitrary in the perfect CDW region; while $\theta(x)$ must to be obtained solving the following differential equation:
\begin{equation}
 \frac{d^2\theta}{dx^2}+\frac{1}{2}\Big(1-\frac{c}{d}\Big)^2\Big(\frac{1}{\xi_g}\Big)^2\sin(2\theta)=0
\label{eq:diif_thetaCDW}
\end{equation}
In this situation we have to put for our boundary condition: $\theta(\pm d)=0$, $\theta(\pm c)=\pi/2$, $\phi(x)\equiv 0$ for $-d/2<x<c$, $\phi(x)\equiv \Delta\Phi$ for $c<x<d/2$, and $0\leqslant c\leqslant d$. 

 The energy of this solution is given by:
\begin{equation}
 E_{CDW}=C\Big\{-\frac{4c}{d}\Big(\frac{d}{2\xi_g}\Big)^2+\frac{4d}{d-c}\int_0^1dt\Big[\Big(\frac{d\theta}{dt}\Big)^2-\Big(1-\frac{c}{d}\Big)^2\Big(\frac{d}{2\xi_g}\Big)^2\sin^2\theta\Big]\Big\}
\label{eq:energy_CDW}
\end{equation}

This solution is presented only for completeness, indeed we have found, comparing its energy with the SC and MIXED energy, that it is not stable. 

\subsection{MIXED solution}
The solution of the Eq. (\ref{eq:diff_theta}) and (\ref{eq:diff_phi}) in the mixed case, for which the order parameter is an analytic function, can be written down in closed form using elliptic integrals as shown in the following:

\begin{eqnarray}
&&
\frac{2x+d}{2\xi_g}=\frac{\cos\theta_0}{\sqrt{\cos^2\theta_0+\omega^2}}F(\varphi;k^2)\label{eq:theta_Mixed}\\\nonumber\\
&&
\phi(x)=\frac{\Delta\Phi}{2}\frac{\Pi(\varphi;\sin^2\theta_0,k^2)}{\Pi(\pi/2;\sin^2\theta_0,k^2)}\label{eq:phi_Mixed}\\\nonumber\\
&&
\sin\varphi=\Big|\frac{\sin\theta(x)}{\sin\theta_0}\Big|\\\nonumber\\
&&
k^2=\frac{\cos^2\theta_0\sin^2\theta_0}{\cos^2\theta_0+\omega^2}\\\nonumber\\
&&
\omega=I\xi_g\\\nonumber\\
&&
\xi_g^2=\frac{\rho}{2g}
\end{eqnarray} 

Here $F(\varphi;k^2)$ is an elliptic integral of first kind, and $\Pi(\varphi;\sin^2\theta_0,k^2)$ is an  elliptic integral of third kind. So Eq. (\ref{eq:theta_Mixed}) and (\ref{eq:phi_Mixed}) define implicitly our order parameter, where $\theta_0\equiv\theta(0)$ is obtained by solving the following equations:
\begin{equation}
\frac{d}{2\xi_g}=\frac{\cos\theta_0}{\sqrt{\cos^2\theta_0+\omega^2}}K(k^2)
 \label{eq:theta0}
\end{equation}
where $K(k^2)=F(\pi/2,k^2)$ is a complete elliptic integral of first kind. We have to point out also that $\omega$ represents the dimensionless currrent.

 The energy of this solution is given by:
\begin{equation}
 E_{MIXED}=4C\int_0^1dt\Big[\Big(\frac{d\theta}{dt}\Big)^2-\Big(\frac{\omega^2}{\cos^2\theta}-\sin^2\theta\Big)\Big(\frac{d}{2\xi_g}\Big)^2\Big]\Big\}
\label{eq:energy_MIXED}
\end{equation}

\subsection{Summary}
Now we want to show in the following two plots the outlines of the solutions described above.
\vspace{0.7cm}
\begin{figure}[htb!]
\centering
\includegraphics[clip=true,scale=0.65,angle=90]{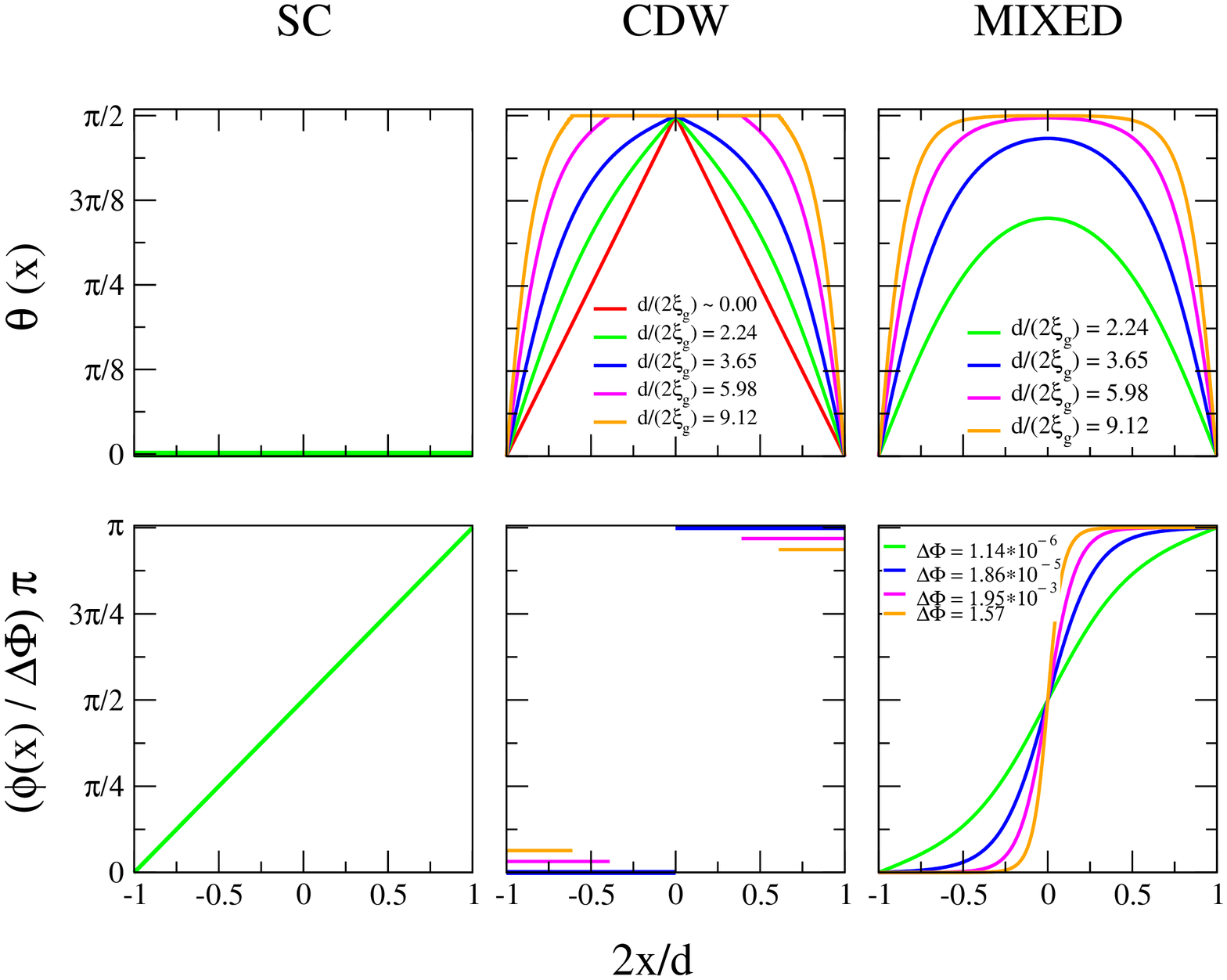}
\caption{Components $\theta(x)$ and $\phi(x)$ of the order parameter for the three different kinds of solutions and for different values of the ratio $d/(2\xi_g)$.}
 \label{fig:panel1}
\end{figure}
\vspace{0.7cm}

\parbox{\textwidth}{~\\}
\vspace{1cm}
\begin{figure}[htb!]
\centering
\includegraphics[clip=true,scale=0.65,angle=90]{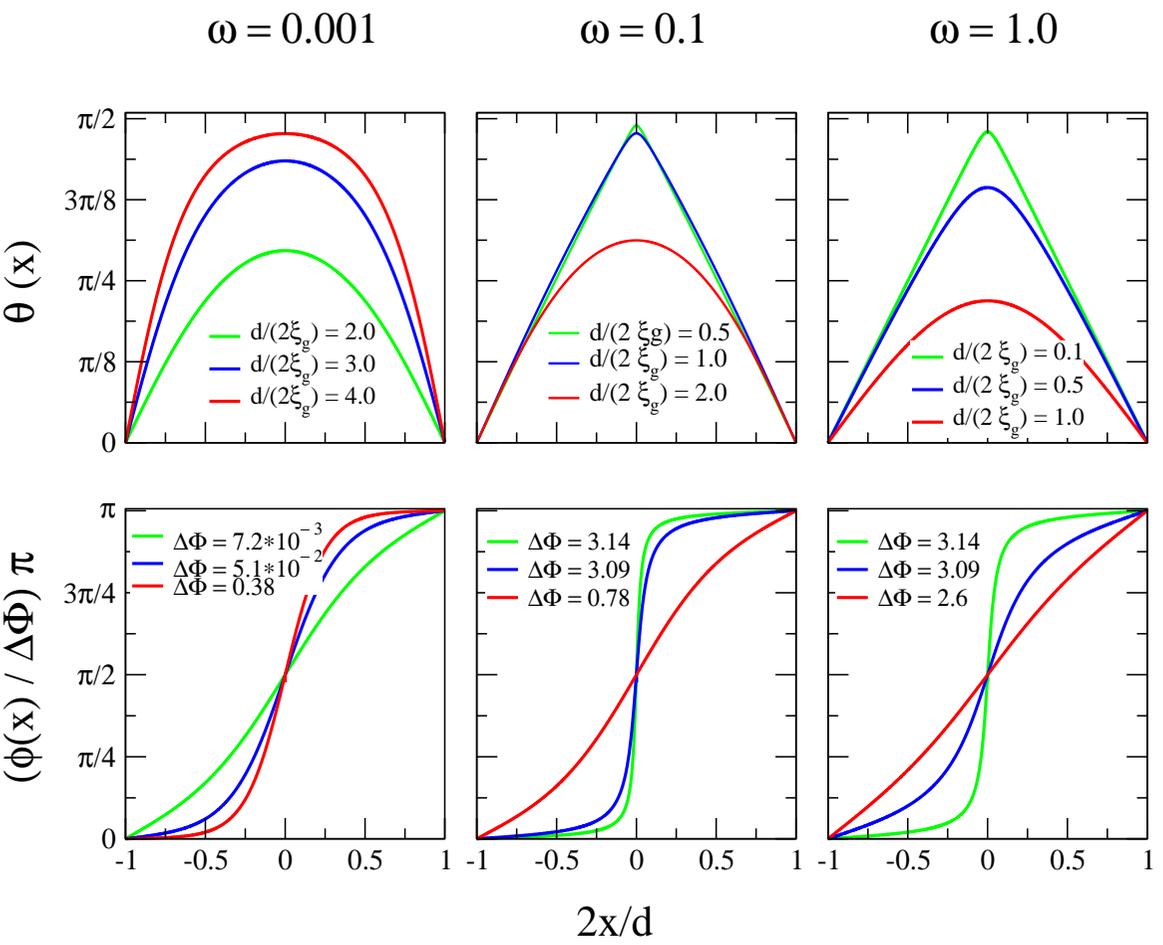}
\caption{Components $\theta(x)$ and $\phi(x)$ of the order parameter for three different current $\omega$  values and for different values of the ratio $d/(2\xi_g)$.}
 \label{fig:panel2}
\end{figure}

\newpage

Into Fig. (\ref{fig:panel1}) it is possible to see on the left side of the panel the outline of the superconducting solution, with $\theta(x)\equiv0$ and the phase angle $\phi(x)$ changing linearly across the junction. In the centre of the picture is shown the solution for the MIXED/CDW case, for which $\theta(x)$ is not differentiable because it has a cusp in the middle point of the junction or a pure CDW segment in the middle region of the junction itself. Finally, on the rigth side of the panel is drawn the solution for the Mixed case; here it is possible to see how increasing the ratio $d/(2\xi_g)$ the solution goes from a more similiar superconducting behaviour to a more similar MIXED/CDW. Into Fig.  (\ref{fig:panel2}) we also show the mixed solution for three different values of the current $\omega$, tuning also the ratio $d/(2\xi_g)$. 

We can better understand the behaviour of these solutions building a ``phase diagram'' for our problem, and commenting the solutions showed above relating them to the ``phase diagram'' itself.  

\section{Phase diagram and behaviour of the current}
Until now we discussed in a complete general way the solutions of our problem whitout saying which is the ``best'' solution for a given set of parameters. We have three different kinds of solutions competing with each other, so we need to understand when one of them is better than others, i.e. when its energy is lower than others ones. To achive this target we have to compare the energies of the three solutions\footnote{These energies are evaluated numerically for the MIXED and MIXED/CDW cases by Eq.  (\ref{eq:energy_MIXED}) and (\ref{eq:energy_CDW}) respectively, while the SC energy (eq. \ref{eq:energy_SC}) is known analitically.}, and then we can build a phase diagram into the space of parameters $\Delta\Phi$ and $d/(2\xi_g)$. These parameters are the natural ones to build up the phase diagram because they are quantities easily controllable from the experimental point of view, indeed we can easily tune the length of the barrier $d$ and the current or the phase difference passing across the junction (as we shall see below the current is related to the phase difference). 

\newpage
\parbox{\textwidth}{~\\}
\begin{figure}[htb!]
\centering
\includegraphics[clip=true,scale=0.5]{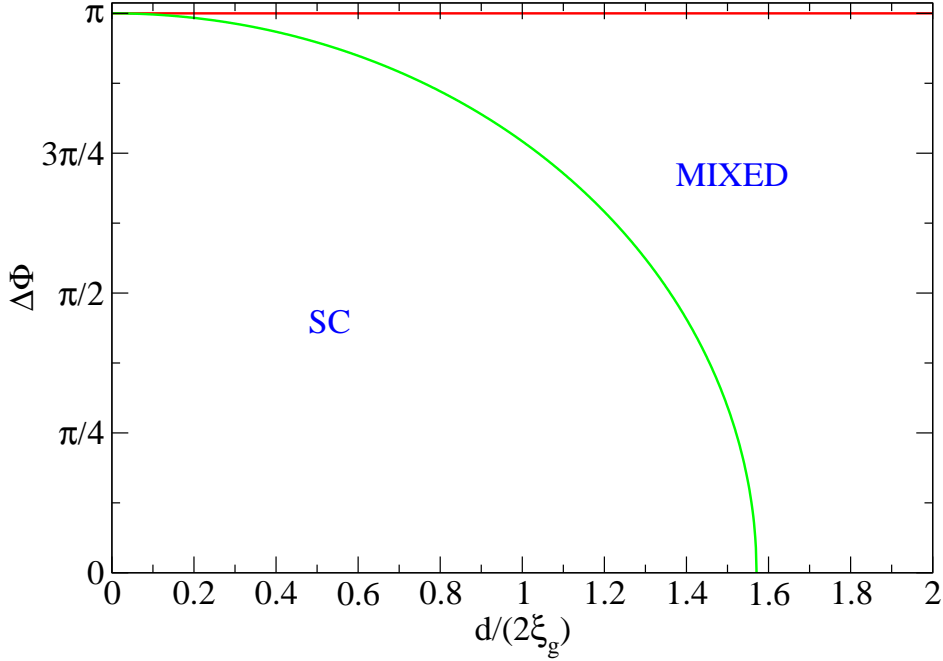}
\caption{Phase diagram with $\Delta\Phi$ \emph{vs} $d/(2\xi_g)$; under the green line the order parameter has only the x and y components, otherwise all components are not zero.}
 \label{fig:phasedia}
\end{figure}
\vspace{0.7cm}

As we can see into Fig. (\ref{fig:phasedia}) the MIXED/CDW solution is never a good one (we anticipated this result previously speaking about of the MIXED/CDW solutions), even if for some values of the parameters it is really similar to the MIXED solution. The line dividing the superconducting region from the mixed one in the phase diagram can be obtained analytically, and it is given by:
\begin{equation}
\Delta\Phi^*=\sqrt{\pi^2-4\Big(\frac{d}{2\xi_g}\Big)^2}
\end{equation}
At this point we can comment the behaviour of the solutions of our problem looking at the ``phase diagram''. First of all we have to remember that we have rigid superconducting boundary conditions ($\phi(-d/2)=0$ and $\phi(+d/2)=\Delta\Phi$); if we fix the thickness of our barrier, we can observe two different behaviours: (i) For $\Delta\Phi=0$, if this thickness is small respect to $\xi_g$ the order parameter prefers to remain into the $XY$ plane forming a pure SC state because the energy cost to leave the $XY$ plane in order to gain a CDW component is higher than the cost in stiffness to bend the order parameter out of the $XY$ plane. Since the width of an interface is $\xi_g$, when $d<\xi_g$ there is not enough space to fit the interface. If the phase difference $\Delta\Phi$ across the junction is increased from zero, at some point the system prefers to form a partial CDW because in this way it reduces the superconducting stiffness. (ii) If the thickness of the barrier is large enough the order parameter is always able to gain a CDW component giving a MIXED solution, because in this situation the barrier is so large that the order parameter can go to the CDW solution. Indeed, we have to remember that over a length of the order of $\xi_g$ the order parameter is able to pass from the SC to the CDW value. 

\begin{figure}[htb!]
\centering
\includegraphics[clip=true,scale=0.65,angle=90]{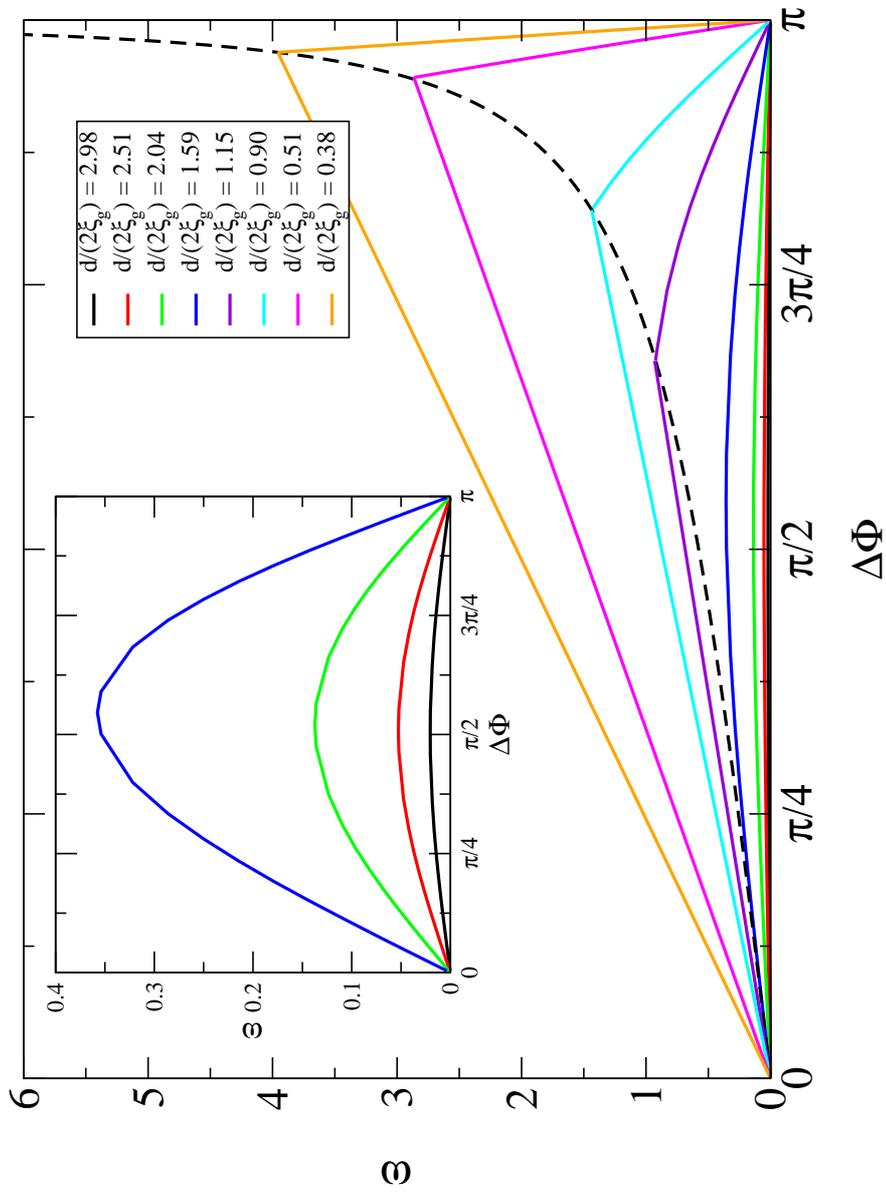}
\caption{The current $\omega$ as a function of the phase difference $\Delta\Phi$ for different values of the ratio $d/2\xi_g$.}
 \label{fig:current}
\end{figure}

Another remarkable behaviour of the system is represented by the relation between the current $\omega$ and the phase difference $\Delta\Phi$ across the junction. As we can see from the caratteristic Current-Phase Difference (see Fig. (\ref{fig:current})) the standard Josephson like behaviour is recovered only for $d/(2\xi_g)\gg1$, otherwise we have a non-analytic current behaviour with a linear dependence for small phase differences (this is the superconducting solution) and a non-linear dependence for greater phase differences (this is the mixed solution). 

We have to point out that the black dashed line in the Fig. (\ref{fig:current}) is the locus of the points where the current goes from the linear behaviour to the non-linear one. This line can be obtained analytically, and it is defined by: 
\begin{equation}
\label{eq:critical}
 \omega=\frac{\Delta\Phi^*}{\sqrt{\pi^2-\Delta{\Phi^*}^2}}
\end{equation}
The Current-Phase Difference plot is a really important result because it can be compared directly with experiments, indeed if our description is exact it would be possible to observe for $SS'S$ junction a non analytical behaviour of the current, and also the maximum current that can be carried out by the junction (the so-called critical current) should be fitted by Eq. (\ref{eq:critical}).

Finally in our framework we could give an explanation of the GPE based on the following observation:  
by means of the coupling between the CDW order parameter and the SC order parameter, the system will have a stiffness associated to the CDW ordering, or in other words there will be a length scale $\xi_g$ related to the capability of the system to pass from the SC state to the CDW state (in our model this means that the spin variable has to go from the $XY$ plane to the $z$-axis). The length scale $\xi_g$ has not anything in common with the coherence length $\xi'$ of the barrier $S'$ (that we remember it is really small, $\xi'\sim 4$\AA), but if $\xi_g$ is big enough it is possible to observe a GPE derived by the presence of the CDW. We can think that there is a latent stiffness stored into the CDW.  

\parbox{\textwidth}{~\\}
\vspace{0.7cm}

Of course this scenario implies that CDW Bragg peaks have to be observed above the superconducting critical temperature $T_c'$ of the barrier, but this is not the case. In the next chapter we will show that in the presence of disorder the CDW can be replaced by glassy CDW, which will not show Bargg peaks.

  \clearpage{\pagestyle{empty}\cleardoublepage}
  \chapter[Disorder and Competition between CDW and SC]{Competition between CDW and SC in the presence of disorder}  
  \label{chapter5}
\vspace{3cm}

In this chapter we are going to show some results regarding zero temperature properties of the model introduced previously. The Random Field Anisotropic Heisenberg Model (RF-AHM), defined on a two-dimensional square lattice, is analyzed numerically looking for features that recall the phenomenology of HTS highligthed in the previous chapters. Mainly our target is to find glassy-like CDW configurations which enable also the establishment of some kind of superconducting order, either local or global in the system. We want to remember that the Hamiltonian of our model is:
\begin{equation}
\label{eq:hamiltonian}
 H=-J\sum_{\langle i,j\rangle}\vec{S}_i\cdot\vec{S}_j-G\sum_i(S_i^z)^2+\frac{W}{2}\sum_ih_iS_i^z
\end{equation} 
where $\vec{S}_i=\{S_i^x,S_i^y,S_i^z\}$ is a classical Heisenberg spin with $|\vec{S}|=1$, $J$ is a positive coupling constant. The first term represents the nearest neighbor interaction of the order parameter. The second term breaks the symmetry in spin space with $G>0$ favoring a CDW; $h_i$ are statistical independent quenched random variables with a flat probability distribution between $-1$ and $+1$; also $W>0$.

As we pointed above, the system is defined on a two-dimensional square lattice, and the results described here are referred to the zero temperature state; so first of all we have to clarify the meaning of the \emph{``zero temperature state''}. In principle it is the ground state of our Hamiltonian, but how can we find it for this random field Hamiltonian? While for the Random Field Ising Model are known exact algorithms to find the ground state, to our knownledge similar algorithms don't exist for the Heisenberg situation. So every kind of algorithm we have in mind to search the ground state of our Hamiltonian necessary doesn't assure the achievement of this target; everytime, we find a local minimum and we can only ``hope'' that it is the global one. Then a proper question is why we are speaking about zero temperature properties if we are not able to guarantee the reaching of the ground state? We can try to answer this question as follow: first we have to underline that even from the experimental point of view we can't say firmly to have gained the ground state of a material whatever is the experimental protocol used; second we are not truly interested in the real ground state of the system because with our effective model we want only to reproduce stable glassy-like CDW situations observed in the pseudogap state of underdoped HTS. So our \emph{``zero temperature state''} refers simply to these stable configurations, local minima of our Hamiltonian, that we found numerically through conjugate gardient methods. Physically the last ones correspond to quench experiments from high temperature down to low temperature.

\section{Interface thickness}
\label{section2}
In this section we want to show more explicitly that the thickness of the interface between two different CDW is of the order of $\xi_g$, the length scale introduced into the previous chapter. 

\begin{figure}[htb!]
\centering
\includegraphics[clip=True,scale=0.5]{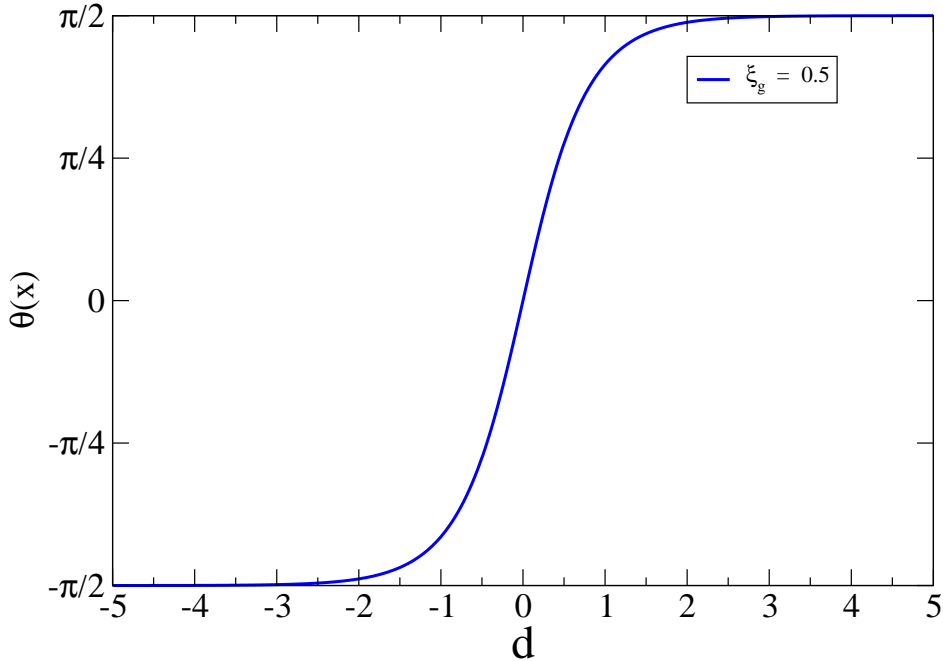}
\caption{Interface between two different CDW}
\label{fig:soliton}
\end{figure}

To solve this problem we can think to cut our system along a straight line perpendicular to the interface, 
reducing the problem itself into a one-dimensional problem, similar to that one solved for the Josephson junction. Thus taking the continuum limit of Eq. (\ref{eq:hamiltonian}), excluding the term with the disorder, we have to minimize the following functional:
\begin{eqnarray}
 F[\theta(x),\phi(x)] & = & 
\int dx \Big\{\frac{\rho}{2}\Big[\Big(\frac{d \theta}{dx}\Big)^2+\cos^2\theta\Big(\frac{d\phi}{dx}\Big)^2\Big]-g\sin^2\theta\Big\}
\end{eqnarray}
with boundary conditions:
\begin{eqnarray}
\left \{
\begin{array}{l}
\theta(-\infty)=-\frac{\pi}{2}\\
\theta(+\infty)=+\frac{\pi}{2}
\end{array}
\right.
\end{eqnarray}
We remember that the angle $\theta$ gives information about the CDW order, so when $\theta=-\pi/2$ we have a variant of the CDW, while for $\theta=+\pi/2$ we hve the other variant of the CDW. Following the strategy of the chapter (\ref{chapter4}) for the solution of the minimization problem, we can write down these Eulero-Lagrange equations:
\begin{eqnarray}
\left \{
\begin{array}{l}
I=\cos^2\theta\frac{d\phi}{dx}\equiv constant\\\\
\xi_g^2\frac{d^2\theta}{dx^2}+\xi_g^2I^2\frac{\sin\theta}{\cos^3\theta}+\sin\theta\cos\theta=0
\end{array}
\right.
\end{eqnarray}
where $I$ is the Noether current, hereafter current\footnote{As we have seen into the previous chapter, the physical current is equal to the Noether current multiplied by some constants ($2ea/\hslash$). }, related to the phase angle $\phi$, and $\xi_g^2=\rho/2g$. Moreover the boundary conditions impose $I\equiv0$\; $\forall x$, simplifying a lot the equation for $\theta$:
\begin{equation}
\label{eq:differential}
\xi_g^2\frac{d^2\theta}{dx^2}+\sin\theta\cos\theta=0
\end{equation}
The differential equation (\ref{eq:differential}) can be easily solved obtaining the following solution:
\begin{equation}
\label{eq:soliton}
 \theta(x)=2\arctan e^{x/\xi_g}-\frac{\pi}{2}
\end{equation}
Eq. (\ref{eq:soliton}) shows clearly that the thickness of the interface between two different CDW is of the order of $\xi_g$, as also sketched in Fig. (\ref{fig:soliton}).

\clearpage
\newpage

We studied also more generally the interface between two different CDW in two dimensions. The problem was defined on a square domain, and the boundary conditions were taken as:
\begin{equation}
\left \{
\begin{array}{l}
 \theta(-d/2,y)=-\pi/2\\
\theta(+d/2,y)=+\pi/2\\
\phi(x,-d/2)=0\\
\phi(x,+d/2)=\Delta\Phi
\end{array}
\right.
\end{equation}

In this case we can show that the system is invariant under translation along the $y$ direction, thus it is possible to find an expression for $\theta$ depending only by $x$. We found that (see Appendix (\ref{appendix5}) for more details):
\begin{equation}
 \theta(x)=2\arctan e^{x/\xi_g'}-\frac{\pi}{2}
\end{equation}
where
\begin{equation}
 \xi_g'=\frac{\xi_g}{\sqrt{1+(\frac{\Delta\Phi}{d})^2\xi_g^2}}
\end{equation} 
Thus we obtained the same result of the one-dimensional case, but with an amplitude of the interface between the two CDW that depends on the phase difference $\Delta\Phi$\footnote{This result is really interesting because it tells us that if into the system there are interfaces with different thickness, the current going through smaller interfaces has a value higher then current going through bigger interfaces. This is analogous to the situation for which a constant flow of water going through a pipe with a no-constant transverse section, has a velocity higher when the section is smaller.} applied along the $y$ direction (i.e. on the current along the $y$ direction). If $\Delta\Phi=0$, $\xi_g\equiv\xi_g'$.

\section{Helicity Modulus or Stiffness}
As we pointed out many times previously, an important target of our research is to see if in systems where there is competition between CDW and SC, we are able to see glassy-like CDW configurations and if a superconducting order can be found not only locally but also globally on the entire sample. We have to remember that our order parameter describes the SC when it stays in the $XY$ plane, while it is associated to the CDW when it points along the positive or negative $z-$axis (the two possible directions representing the two possible variants of the CDW in our model). So a natural way to obtain an information about the SC order in our system is to measure the in-plane magnetization:
\begin{equation}
m_{xy}=\overline{\sqrt{\frac{1}{N}\sum_i^N \Big[(S_x^i)^2+(S_y^i)^2\Big]}}
\end{equation}
where $N$ is the total number of lattice sites and $\overline{\cdots}$ is the average over the disorder. But we have to stress that this variable is not able to give us informations about the overall superconductivity in the sample, because it measures only local SC contributions, masking the possible global SC. For this reason we need a variable able to check for the overall SC, and it is given by the Helicity modulus $\Upsilon$, also called Stiffness. 


\subsection{Stiffness at T=0 as a Kirchhoff problem}
As we said in the chapter (\ref{chapter1}) solving the $XY$ toy model on three lattices, the zero temperature stiffness could be obtained finding the global conductance of the corresponding conductance network, where the single conductances are related one to one to the bonds $\{J\}$ of the $XY$ model itself. Now we want to show that this stategy can be applied also to our problem. 

We know that our model is given by the following Hamiltonian:
\begin{equation}
 H=-J\sum_{\langle i,j\rangle}\vec{S}_i\cdot\vec{S}_j-G\sum_i(S_i^z)^2+\frac{W}{2}\sum_ih_iS_i^z
\end{equation}
that in spherical coordinates can be written as:
\begin{equation}
\label{eq:spherical}
 H=-J\sum_{\langle i,j\rangle}\cos\theta_i\cos\theta_j\cos(\phi_i-\phi_j)-G\sum_i\sin^2\theta_i+\frac{W}{2}\sum_ih_i\sin\theta_i
\end{equation}
\begin{figure}[htb!]
\centering
\includegraphics[clip=True,scale=0.5]{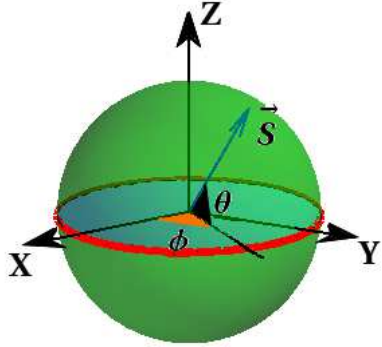}
\caption{Order parameter.}
\label{fig6-cap5}
\end{figure}

The first step in the numerical analysis of our system is finding a stable configuration, a local minumum, by the conjugate gradient method. So we want to measure the in-plane stiffness along $X$ or $Y$ direction (because of the isotropy of the problem into the $XY$ plane, considering the mean Stiffness $\Upsilon_x$ or $\Upsilon_y$ is the same, and hereafter we'll refer to $\Upsilon_x$\footnote{We have to stress that for a single configuration $\Upsilon_x$ and $\Upsilon_y$ can be clearly different, and only averaging on the disorder we find that their mean values are statistically equal.}). Only the first term of Eq.  (\ref{eq:spherical}) is important for the calculus of the Stiffness, because only here there are the phase angles essential for finding $\Upsilon$. For this reason now we can restrict ourselves to the following two-dimensional random bond $XY$ model:
\begin{equation}
 H=-\sum_{\langle i,\,j\rangle}J_{ij}\cos(\phi_i-\phi_j)
\end{equation}
where $J_{ij}=J\cos\theta_i\cos\theta_j$ are quenched random bonds obtained after the minimization.

First of all we have to remember that the current $j_{{\scriptscriptstyle ij}}$ through a bond $J_{ij}$ is defined by the following relation:
\begin{equation}
 j_{{\scriptscriptstyle ij}}=-\frac{2ea}{\hslash}\frac{\partial H}{\partial A_{ij}}
\end{equation}
where $A_{ij}$ is the line integral of the potential vector $\vec{A}$ between the point $\vec{R}_i$ and $\vec{R}_j$:
\begin{equation}
 A_{ij}=\frac{2e}{\hslash}\int_{\vec{R}_i}^{\vec{R}_j}\vec{A}\cdot d\vec{l}
\end{equation}
As we have seen in the capther (\ref{chapter3}), the potential vector enters into our model thank to the Peierls substitution in the following way:
\begin{equation}
\label{eq:rdmXY}
  H=-\sum_{\langle i,\,j\rangle} J_{ij}\cos(\phi_i-\phi_j-A_{ij})
\end{equation}
and then the current $j_{{\scriptscriptstyle ij}}$ is equal to:
\begin{equation}
\label{eq:current}
j_{{\scriptscriptstyle ij}}=\frac{2ea}{\hslash}J_{ij}\sin(\phi_i-\phi_j)
\end{equation}
where we put $A_{ij}=0$ in the absence of the external field.

\begin{figure}[htb!]
\centering
\includegraphics[clip=True,scale=1.0]{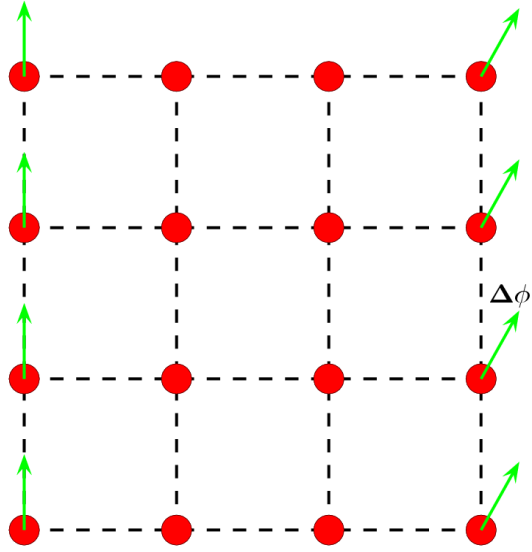}
\caption{A square lattice with an applied twist angle $\Delta\phi$ on the right side.}
\label{fig7-cap5}
\end{figure}

At this point we can start the calculation of the Stiffness as described by Jasnow (see Eq.  (\ref{eq:jasnow_stiff})), and because we are at zero temperature we have to replace the equilibrium free energy density with the equilibrium energy density. So, if we are looking for the Stiffness along the $X$ direction, we have firstly fixe all phase angles on the left side of the lattice and also apply a twist $\Delta\phi$ to all phase angles on the right as sketched in Fig. (\ref{fig7-cap5}).
\newline
Now the system has to relax this ``torsion'' finding a stable phase angle configuration $\{\phi_i\}\equiv\phi^\star$ for which it is:
\begin{equation}
\label{eq:minimum}
 \frac{\partial H}{\partial \phi_i}=0 \qquad \forall \phi_i\in\phi^\star
\end{equation}
and we indicate the energy of the system for this configuration as $H^\star$. Now we can see that the first derivative of $H^\star$ with respect to $\Delta\phi$ corresponds to the total current along the $X$ direction:
\begin{equation}
\label{eq:minimization}
 \frac{dH^\star}{d\Delta\phi} = \frac{\partial H^\star}{\partial\Delta\phi}+\sum_{i\in\mathcal L}\frac{\partial H^\star}{\partial\phi_i}\frac{\partial\phi_i}{\partial\Delta\phi}=\frac{\partial H^\star}{\partial\Delta\phi}
\end{equation}
where $\mathcal L$ is the symbol for our square lattice consisting of $N\times N$ sites. If we want to write explicitly Eq. (\ref{eq:minimization}), it is clear that the only terms of the Hamiltonian (Eq. (\ref{eq:rdmXY})) which have a non-zero derivative with respect to $\Delta\Phi$, correspond to that ones for which the bond $J_{ij}$ joins a lattice site on the last but one column of the lattice with its neighbour on the last column of the lattice itself\footnote{We are counting the columns of the lattice $\mathcal L$ from left to right, and the rows from top to bottom.}. Thus we can write down:
\begin{equation}
 \frac{dH^\star}{d\Delta\phi}=\sum_{k=1}^N j_k^{(N-1,N)}\equiv j^{TOT}_x
\end{equation}
where $j_k^{(N-1,N)}$ is the current passing through the last bond of the $k-$th row (with the couple ($N-1,N$) we indicate the column positions of the lattice sites corresponding the current we are considering; in other words the first index of this couple indicates the $(N-1)$-th column of the lattice where is located the site $i$, and the second index represents the $N$-th column of the lattice where is the site $j$. This couple whithin the row index $k$ are sufficient to identify univocally the two sites.)

We have to point out also that writing explicitly Eq. (\ref{eq:minimum}) we obtain for every node of the lattice an equation for which the sum of the currents impinguing the node itself is zero; thus the currents satisfy the Kirchhoff law:
\begin{equation}
\label{eq:kirchoff}
 \sum_{k\;{\scriptscriptstyle \in}\;n.n.} j_{{\scriptscriptstyle i,k}}=0 \qquad \forall\; i\;{\scriptstyle \in}\;{\mathcal L}
\end{equation}

In the limit that $\Delta\Phi\rightarrow0$ all phase angle differences are small, therefore Eq.  (\ref{eq:current}) becomes:
\begin{equation}
\label{eq:currentelle}
 j_{ij}=\frac{2ea}{\hslash}J_{ij}(\phi_i-\phi_j)
\end{equation}
This equation is analogous to Ohm's law if we identify $J_{ij}$ with a conductance and $(\phi_i-\phi_j)$ with a potential difference between the two sites $i$ and $j$. Also the phase difference $\Delta\Phi$ represents the total bias applied between the first and the last column of the lattice:
\begin{equation}
\label{eq:bias}
 \Delta\Phi=\phi_N-\phi_1
\end{equation}
This system of equations, within the above identification, is identical to the system of equations corresponding to a random conductance network.

So we have found a perfect analogy with a random electrical network, where the bonds $J_{i,j}$ correspond to conductances and the phase angle differences $\phi_i-\phi_j$ are similar to a bias applied at the two nodes $i$ and $j$. Moreover the Stiffness along the $X$ direction is given by\footnote{If we have not a square lattice but a rectangular one, the expression of the Stiffness is: $\Upsilon_x=\Big(\frac{L_x}{L_y}\Big)\frac{d^2H^\star}{d\Delta\phi^2}\Big|_{{\scriptscriptstyle \Delta\phi\;=\;0}}$, where $L_x$ and $L_y$ are the two dimensions of the lattice.}:
\begin{equation}
\Upsilon_x=\frac{d^2H^\star}{d\Delta\phi^2}\Big|_{{\scriptscriptstyle \Delta\phi\;=\;0}}= \frac{d\; j_{{\scriptscriptstyle x}}^{{\scriptscriptstyle TOT}}}{d\Delta\phi}\Big|_{{\scriptscriptstyle \Delta\phi\;=\;0}}
\end{equation}
and in the limit $\Delta\Phi\rightarrow0$ it reads:
\begin{equation}
\label{eq:ohm}
 j_x^{TOT}=\Upsilon_x\Delta\Phi
\end{equation}

Eq. (\ref{eq:ohm}) is exactly the Ohm's law for the entire network, where $\Delta\phi$ is the bias applied to the left and right ends of our lattice, while $\Upsilon_x$ represents the global conductance of the network itself.

So our strategy to calculate the Stiffness of the system will be to implement Eq. (\ref{eq:kirchoff}), (\ref{eq:bias})  and (\ref{eq:ohm}), i.e. first of all we have to find all phase angle $\phi_i$ solving the linear equation system defined by (\ref{eq:kirchoff}) and (\ref{eq:bias}), then using Eq.  (\ref{eq:currentelle}) it is possible to find the currents $j_{ij}$ for every bond of the network, at this point is possible to find the global current along the $X$ direction, and finally using Eq. (\ref{eq:ohm})  it is possible to find $\Upsilon_x$. The only technical difficulty to be overcome is the numerical solution of a huge sparse linear system of equations. 

\section{Some configuration snapshots}
\begin{figure}[h!]
\centering
\includegraphics[clip=True,scale=0.26]{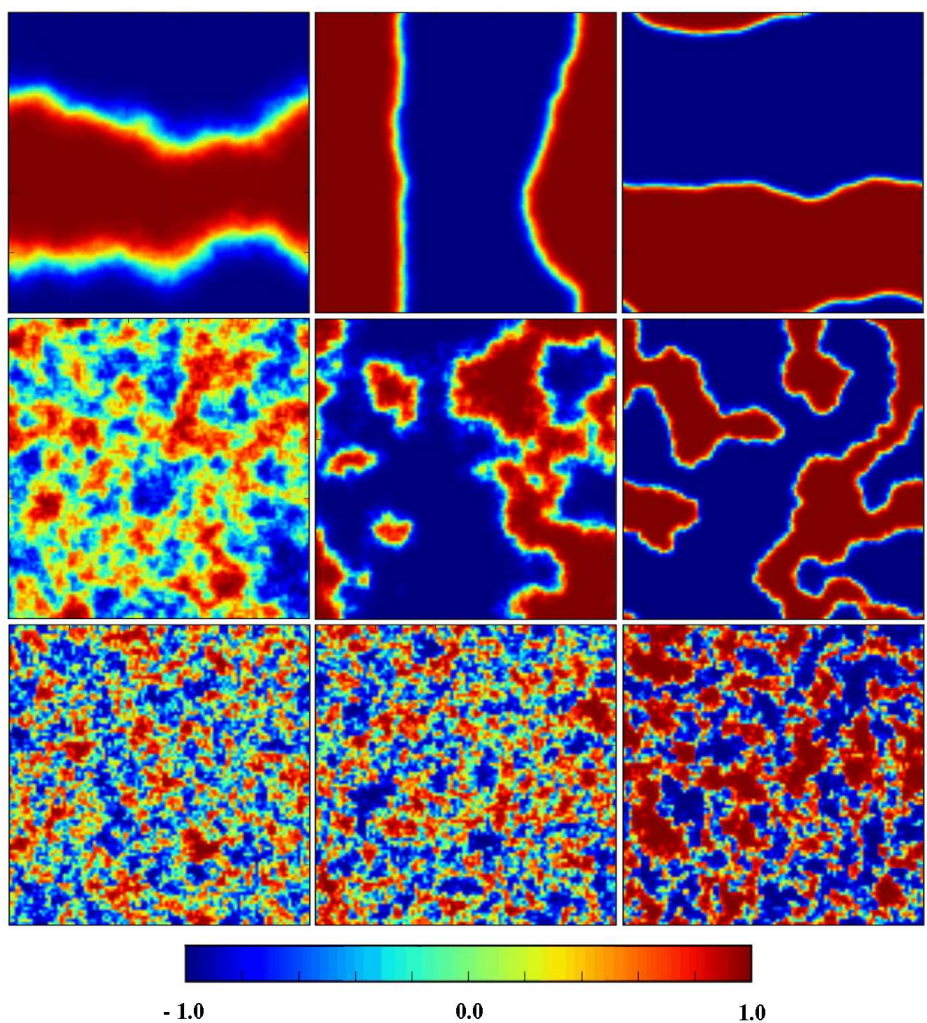}
\caption{Intensity of the $z$ component of spin variables. For every row we have from left to right the following anisotropy values: $G=0.02,\;0.1,\;0.3$. For every column we have from top to bottom the following disorder intensity values: $W=0.5,\;2.0,\;5.0$.}
\label{fig:total}
\end{figure}

\begin{figure}[h!]
\centering
\includegraphics[clip=True,scale=0.26]{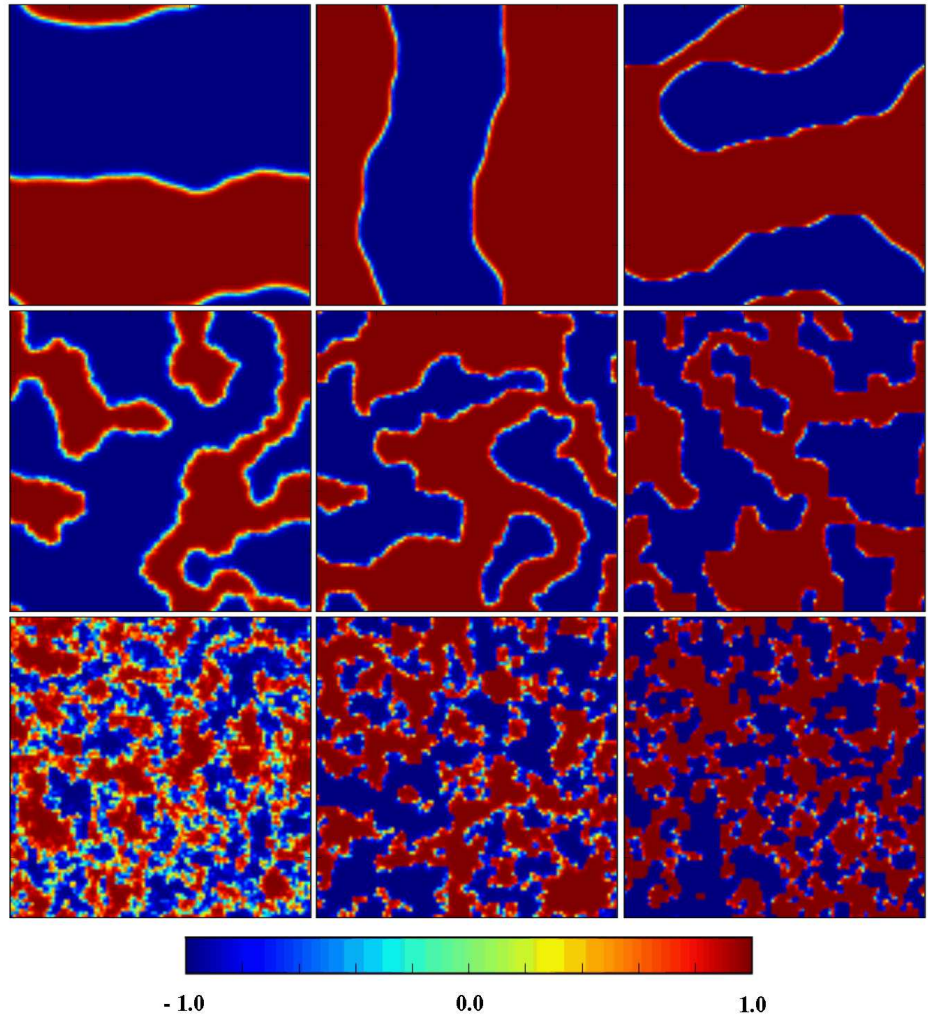}
\caption{Intensity of the $z$ component of spin variables. For every row we have from left to right the following anisotropy values: $G=0.3,\;0.5,\;0.7$. For every column we have from top to bottom the following disorder intensity values: $W=0.5,\;2.0,\;5.0$.}
\label{fig:total2}
\end{figure}

\begin{figure}[h!]
\centering
\includegraphics[clip=True,scale=0.2]{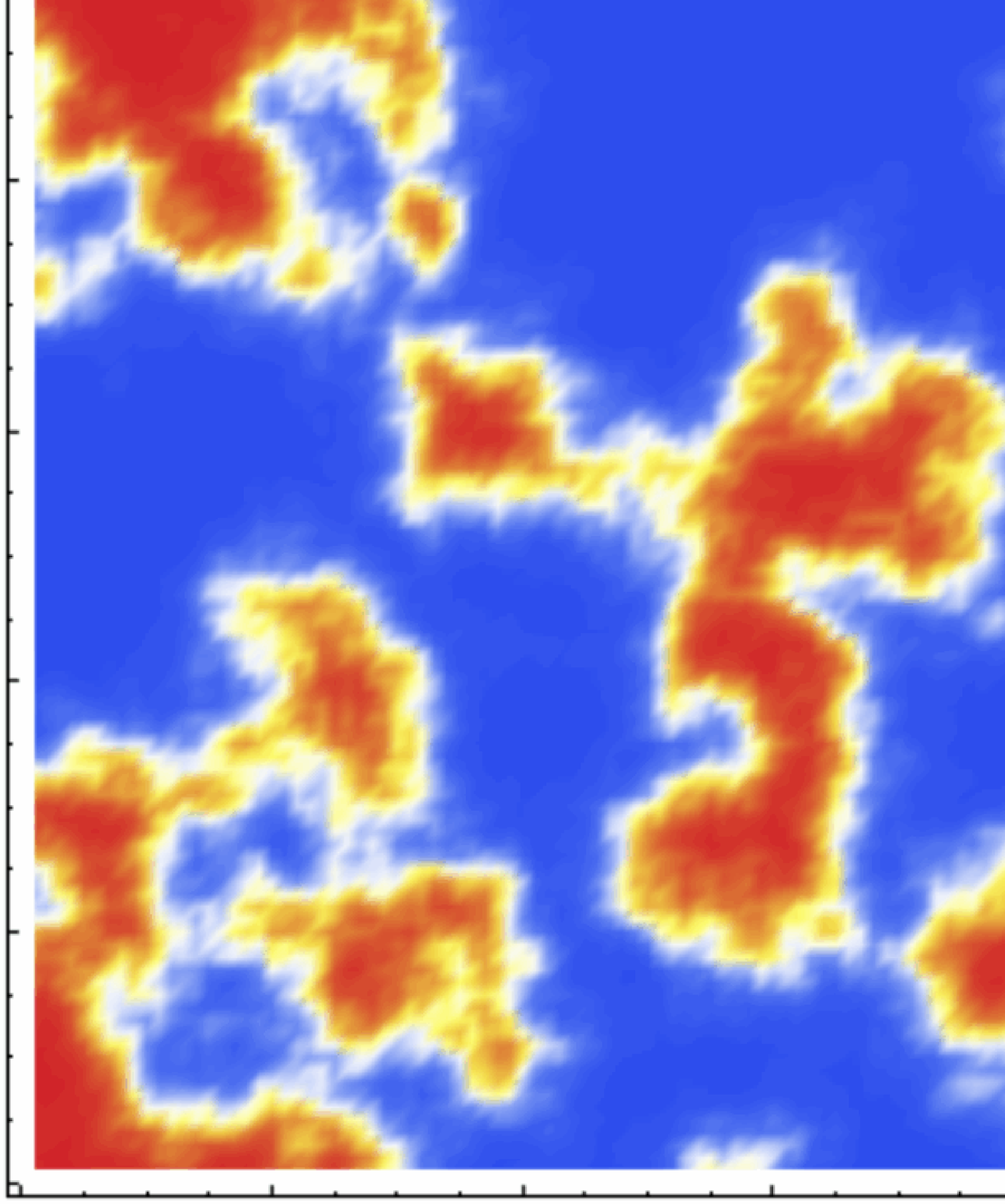}
\caption{Some configuration snapshots for a fixed random field configuration with intensity $W=2.0$, and for different anisotropy values (see Table (\ref{table}) for G values and Fig. (\ref{fig:palette}) for colorbar legend).}
\label{fig:W2.0}
\end{figure}

Here we want to show some configuration snapshots of our system. First of all we are going to show some configurations with different parameter values just only to have an idea of what kind of configurations we can obtain (see Fig. (\ref{fig:total}) and (\ref{fig:total2})). After that we shall describe a set of configurations with same realization of quenched randomness, but different anisotropy intensity (see Fig. (\ref{fig:W2.0})). 
\vspace{0.5cm} 

\clearpage
\newpage

In Fig. (\ref{fig:total}) and (\ref{fig:total2}) it is possible to see many different kinds of configurations of our system: from stripe like along $x$ or $y$ direction, to domain bubble like with different sizes. 

First of all we have to remember that we can observe domain configurations because of the presence of quenched random fields, indeed as we said in chapter (\ref{chapter2}) speaking about the Imry and Ma argument, a random field Ising model has a lower critical dimension equal to two, thus in our case because the system is defined on a two-dimensional lattice it must breaks itself into domains.  We can see that for a fixed value of the disorder intensity we are able to see that the interface (the SC state) between two different domains (two different CDW) become thicker decreasing the value of the anisotropy $G$, or in other words we can pass from a ``soft'' like to a ``sharp'' like interface behaviour tuning $G$. 
This behaviour can be explained easily taking in mind the results of the section (\ref{section2}), indeed as we have seen there the length scale over which the spin goes from the $XY$ plane to the $z-$axis (i.e. from the SC to the CDW) is related to the anisotropy parameter $G$ in an inverse proportional way (i.e. $\xi_g\sim\sqrt{J/G}$), so increasing $G$ we obtain more small interfaces and arriving to a thickness of the order of the lattice spacing we pass from a ``soft'' like to a ``sharp'' like interface. Situations for which the interfaces between different CDW domains is ``sharp'' like don't allow neither magnetization in the $XY$ plane, nor global stiffness; but clearly if we have ``soft'' like interfaces, this doesn't guarantee to observe an overall stiffness in the sample, indeed could be possible to have disconnected Heisenberg interfaces that gives a finite contribute for the $XY$ magnetization but a zero (or exponentially small) contribute for the stiffness. This situation corresponds for example to the configuration on the right panel in the central row of Fig. (\ref{fig:total}).

 Also increasing the disorder intensity $W$ it is possible to obtain much more pulverized domains; this can be understood thinking to the Binder length scale introduced in the chapter (\ref{chapter2}) speaking about of the RFIM. We remember that the Binder scale gives the order of the size of the domains, and it depends on the disorder intensity in the following way: $L_{Binder}\sim\exp[(J/W)^2]$, so increasing $W$ we decrease $L_{Binder}$. 

\clearpage
\newpage

\clearpage
\newpage

\begin{figure}[h!]
\centering
\includegraphics[clip=True,scale=0.2]{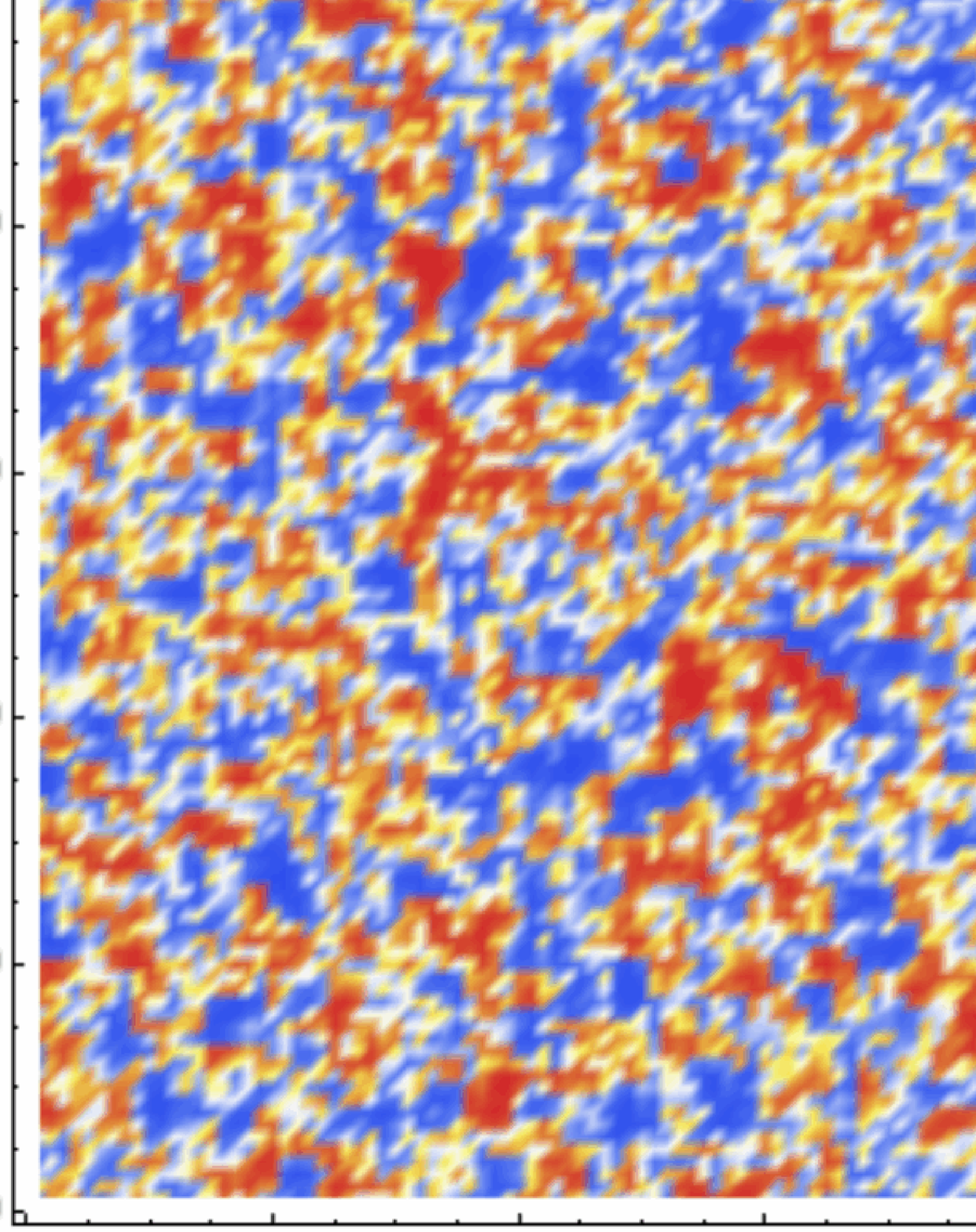}
\caption{Some configuration snapshots for a fixed random field configuration with intensity $W=8.0$, and for different anisotropy values (see Table (\ref{table}) for G values and Fig. (\ref{fig:palette}) for colorbar legend).}
\label{fig:W8.0}
\end{figure}

\clearpage
\newpage

\begin{table}[htb!]
\begin{center}
\begin{tabular}[c]{||l|l||}
\hline \hline
-0.1 & 0.0 \\ 
0.03 & 0.07 \\ 
0.1 & 0.4 \\
\hline \hline
\end{tabular}
\end{center} 
\caption{Values and positions of the anisotropies $G$ corresponding to Fig. (\ref{fig:W2.0}) and (\ref{fig:W8.0}).}
\label{table}
\end{table}

\begin{figure}[h!]
\centering
\includegraphics[clip=True,scale=0.5]{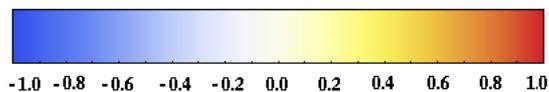}
\caption{Colorbar indicating the intensity of the $z$ component of the spin variables for Fig. (\ref{fig:W2.0}) and (\ref{fig:W8.0}).}
\label{fig:palette}
\end{figure} 

Now we can pass to describe Fig. (\ref{fig:W2.0}) and (\ref{fig:W8.0}). These configurations are obtained fixing the disorder configuration and changing the value of the anisotropy $G$ minimizing every time the energy of the system using as starting configuration that one obtained from the previous minimization. In both cases increasing $G$ we have a more ``sharp'' like scenario, but how is evident from the pictures the two sets of configurations are really different. In the weak disorder case ($W=2.0$) we have clear CDW domains, which sizes are of the order of $20-30$ lattice sites, while for strong disorder ($W=8.0$) it is difficult to distinguish differents CDW domains, which sizes are of the order of few lattice sites. Also the evolution of the configuration as $G$ increase is more evident in the weak disorder case ($W=2.0$) than in the strong disorder case ($W=8.0$). This is due to the fact that for strong disorder the length scale $\xi_g$ becomes irrelevant; indeed into the system there are a lot of small CDW domains really close each other, and for this reason it is impossible to establish a SC interface as in the weak disorder case, because there is not enough space between two different CDW. 


\section{XY Magnetization and Stiffness}
In this section we shall show the results of the numerical study of our Random Field Anistropic Heisenberg Model, in particular we shall present the behaviour of the $XY$ magnetization and Stiffness, as functions both of the anisotropy parameter $G$ and of the disorder intensity $W$.

We have to stress also that the coupling constant $J$ is fixed to one, thus every energy scale is referred to this one; also every physical observable is averaged over $200$ different realizations of the disorder\footnote{To be more precise, the Stiffness curves for high values of the disorder intensity are obtained averaging over a small set of  configuratios; indeed the error bars in those plots are not all equal. While for the magnetization curves the error bars are not equal (even if the set of configurations is fixed to $200$) because they are obtained by error propagation applied to square magnetization ($\Delta m_{xy}\propto\Delta m_{xy}^2/(2m_{xy}))$.}, and the system size is fixed to $100\times 100$ (Into the appendix \ref{appendix5} it is possible to find a finite size analysis study, which shows that finite size effects are small.).

\subsection{XY magnetization}
\begin{figure}[htb!]
\centering
\includegraphics[clip=True,scale=0.5]{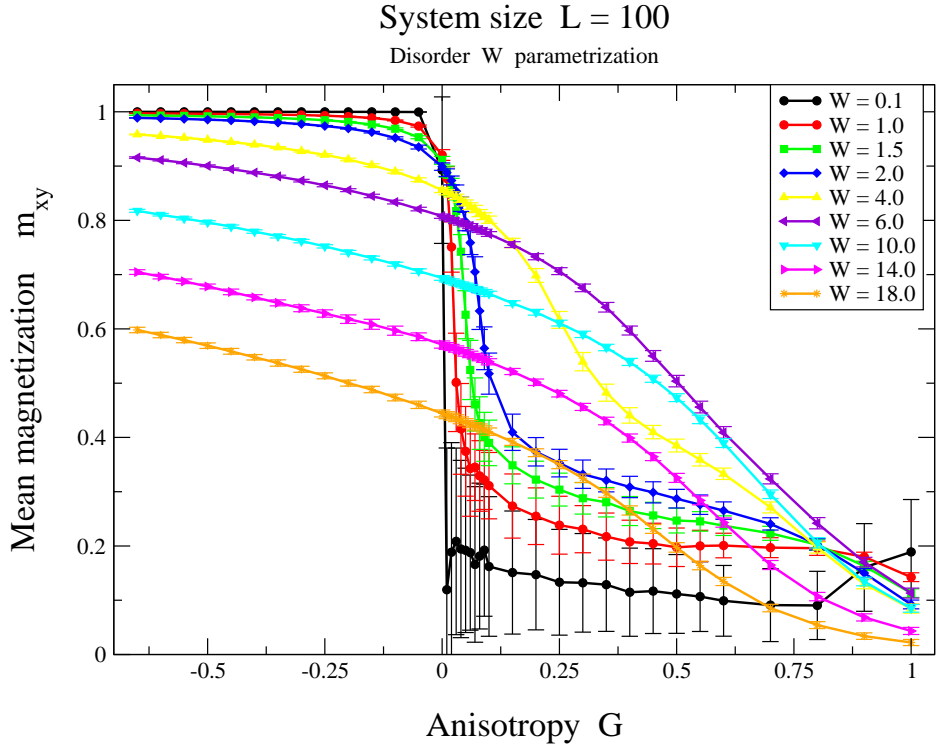}
\caption{Mean $XY$ magnetization as a function of the anisotropy $G$ for different values of the disorder intensity $W$.}
\label{fig10-cap5}
\end{figure}

If into the system there is not any kind of disorder $(W=0)$ the ground state is a simple ferromagnet along the $z-$axis for every $G>0$ (with a double degeneracy, corresponding to all spins pointing either along the positive or the negative $z-$axis), and a ferromagnet into the $XY$ plane for every $G<0$ (with an infinite degeneracy, corresponding to all spins pointing along one of the infinite direction into the $XY$ plane), so we should observe both for $|m_{xy}|$ a step function with $G$ as indipendent variable (i.e. $|m_{xy}|\equiv\Theta(-G)$). But when the disorder intensity is not zero, the above step function behaviour change immediately, allowing to the $XY$ magnetization an enhancement for positive $G$ values, and a reduction for negative $G$ values (see Fig. (\ref{fig10-cap5})). While the reduction of $|m_{xy}|$ for $G<0$ was expected introducing the disorder, it is really interesting the situation for $G>0$, for which the disorder favours the establishment of the SC. 

\begin{figure}[htb!]
\centering
\includegraphics[clip=True,scale=0.5]{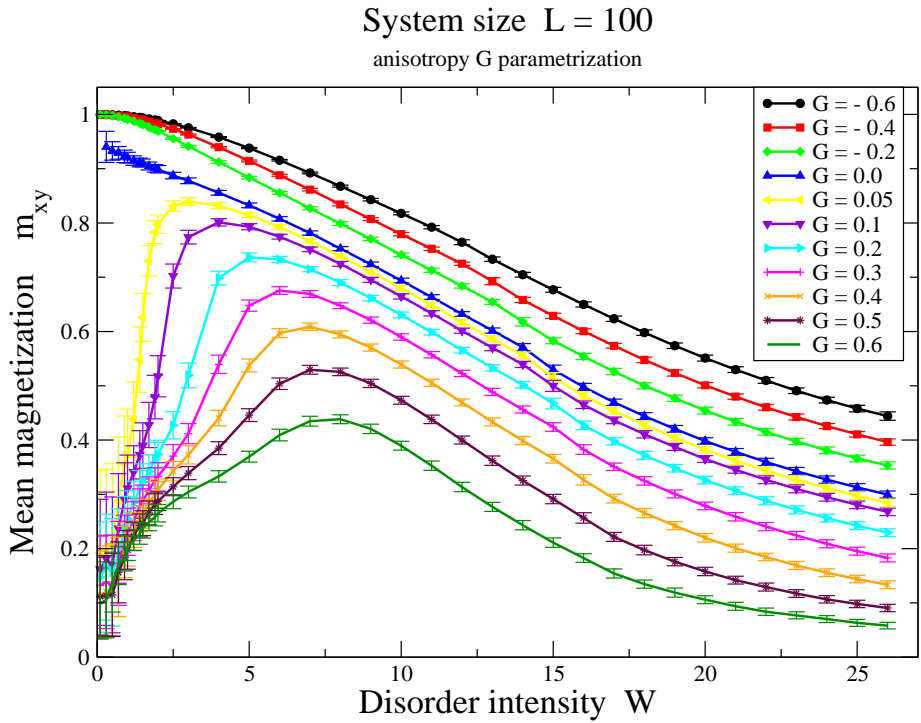}
\caption{Mean $XY$ magnetization as a function of the disorder intensity $W$ for different values of the anisotropy $G$.}
\label{fig11-cap5}
\end{figure}

Now we can also observe the behaviour of $|m_{xy}|$ as function of the disorder for a fixed value of the anisotropy. While for $G<0$ values we observe monotonic decresing functions, for wich the disorder simply weakens the $XY$ ferromagnetic order (this is the trivial effect for which the disorder degrade the superconductivity),  for $G>0$ values we have a non monotonic behaviour and there is an optimum disorder intensity for which the in-plane magnetization is maximum (see Fig. (\ref{fig11-cap5})). 

Now we can say that the quenched disorder into our system allows the formation of ferromagnetic islands (along the $z-$axis), and among them there are  ``sharp'' like or  ``soft'' like domain walls. The last ones let to the system to have non zero in-plane magnetization and local Stiffness. 

A good question could be if the superconductivity measured by $|m_{xy}|$ is only local or global into the system. Following the idea of the ``soft'' like domain walls introduced above, this question is equal to ask the existence of a percolating ``soft'' like domain wall into the system. This will be clarified in the next section, where we show the behaviour of the Stiffness $\Upsilon_x$ as a function of the anisotropy $G$ and of the disorder intensity $W$.

\subsection{Stiffness}
\begin{figure}[htb!]
\centering
\includegraphics[clip=True, scale=0.5]{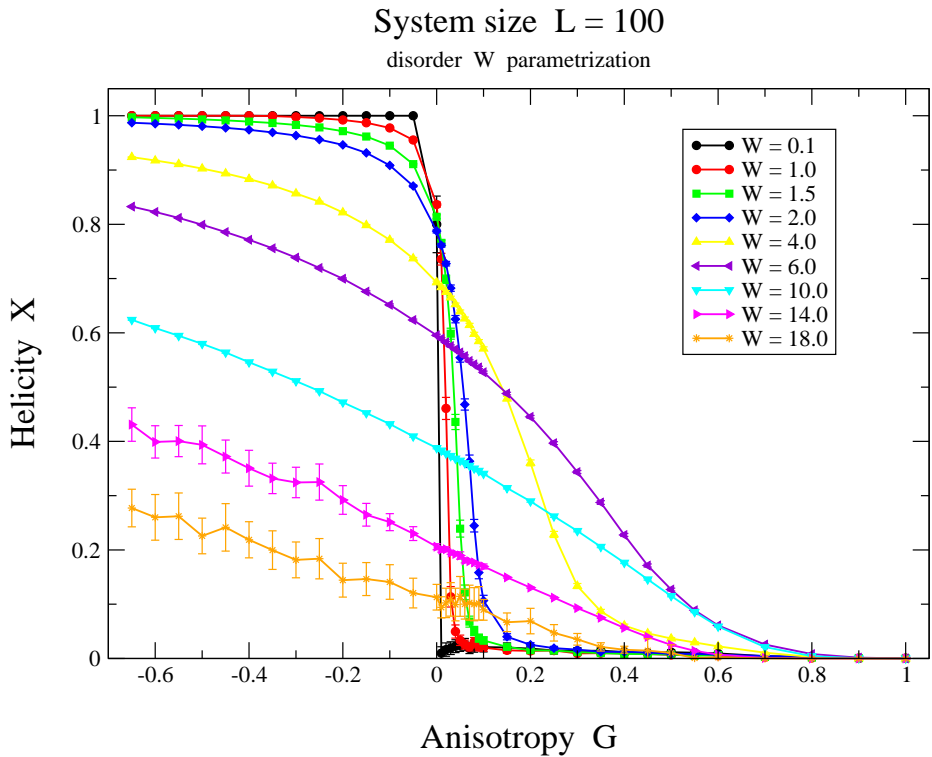}
\caption{Stiffness $\Upsilon_x$ as a function of the anisotropy $G$ for different values of the disorder intensity $W$.}
\label{fig12-cap5}
\end{figure}

\begin{figure}[htb!]
\centering
\includegraphics[clip=True,scale=0.5]{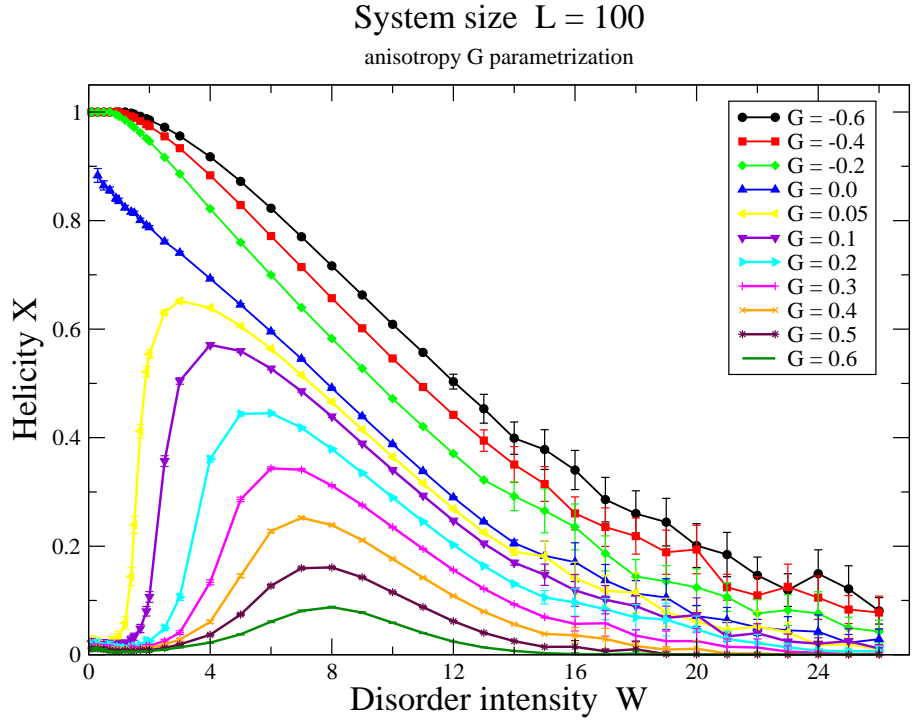}
\caption{Stiffness $\Upsilon_x$ as a function of the disorder intensity $W$ for different values of the anisotropy $G$.}
\label{fig13-cap5}
\end{figure}

The Stiffness has the same qualitative behaviour of the in-plane magnetization $|m_{xy}|$; i.e. $\Upsilon_x$, as function of the anisotropy $G$, is lowered by the presence of the disorder if $G<0$, but more surprisingly it gains a non-zero value for $G>0$ (see Fig. (\ref{fig12-cap5})). Also the behaviour of $\Upsilon_x$ as function of the disorder intensity $W$ is qualitatively similar to that one of the in-plane magnetization, i.e. or $G<0$ we have the trivial monotonic decresing behaviour for the Stiffness, while for $G>0$ we observe a non monotonic behaviour of $\Upsilon_x$ (see Fig. (\ref{fig13-cap5})), with an optimum disorder intensity where the Stiffness is maximized.  

\section{A ``Phase Diagram''}
Now we want to show better the above results, trying to go deeply into the understanding of our system. First of all we shall give a more unified vision of the $XY$ magnetization and of the Stiffness as functions of the anisotropy $G$ and of the disorder intensity $W$, and then we try to build a ``phase diagram''. We have to remember that one important question of our research is the possibility to find in our system  glassy-like CDW configurations allowing the establishment of local and global SC. Because we are able to obtain informations about local SC in our system through the magnetization $|m_{xy}|$, and about global SC through the Stiffness $\Upsilon_x$, we can ask when we have a global superconducting order, or when we have only a local SC, or when we have neither a global nor a local SC. Then could be useful to build a some kind of ``phase diagram'' in order to distinguish the previous described situations. We can achieve this aim by fixing a threshold for both $|m_{xy}|$ and $\Upsilon_x$, and considering roughly equal to zero the variable if it is below the threshold, while different from zero if it is above the threshold. It is clear that the value of the threshold is completely arbitrary, so we need to fix more than one threshold value and studying how this ``phase diagram'' changes.

\clearpage
\newpage

\begin{figure}[htb!]
\centering
\includegraphics[clip=True,scale=0.45]{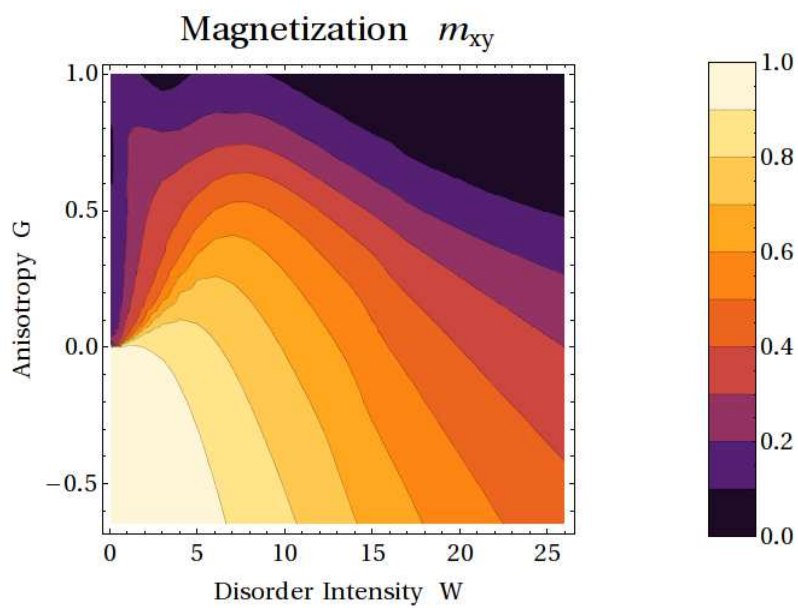}
\caption{Contour plot of the magnetization $m_{xy}$ as a function of the disorder intensity $W$ and of the anisotropy $G$.}
\label{fig14-cap5}
\end{figure}
\begin{figure}[h!]
\centering
\includegraphics[clip=True,scale=0.45]{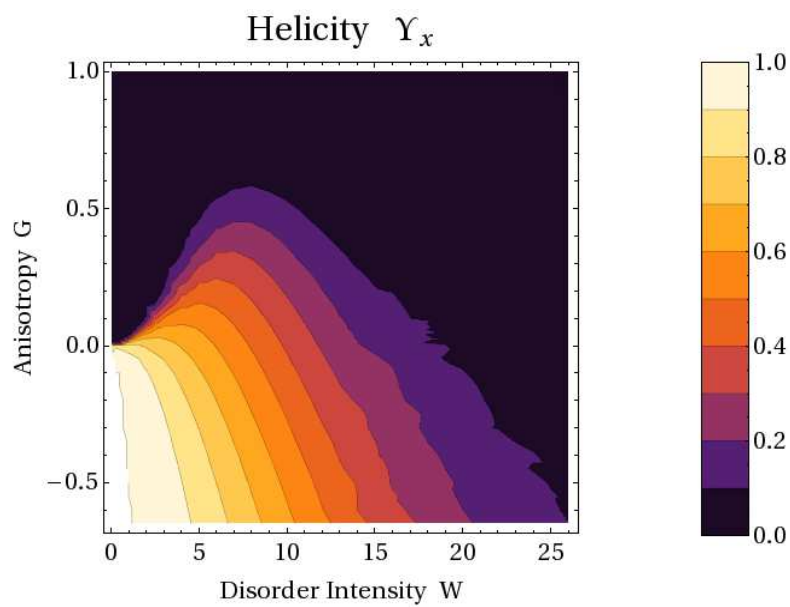}
\caption{Contour plot of the Stiffness $\Upsilon_x$ as a function of the disorder intensity $W$ and of the anisotropy $G$.}
\label{fig15-cap5}
\end{figure}

\clearpage
\newpage

Fig. (\ref{fig14-cap5}) and (\ref{fig15-cap5}) are contour plots of the magnetization $|m_{xy}|$ and of the Stiffness $\Upsilon_x$ as functions of the disorder intensity $W$ and of the anisotropy $G$. In each plot we can see that there exists a CDW region for high $G$ values (the dark region), while the SC region has qualitatively the same shape even if into the magnetization plot it represents only local SC and it is bigger then the same region into the stiffnes plot, where this region represents global SC. It is really interesting to observe also that for $G>0$ we can see superconductivity, due only to the presence of the disorder. Finally we can say that for a fixed value of the anisotropy there is a maximum value for $|m_{xy}|$ and $\Upsilon_x$ because of opposite tendencies of the system: (i) to have a high disorder intensity to gain superconductivity by increasing the density number of domains, (ii) and to have a low disorder intensity to gain superconductivity by obtaining more ``soft'' interfaces between two different CDW.  

\clearpage
\newpage

\begin{figure}[h!]
\centering
\includegraphics[clip=True,scale=0.7]{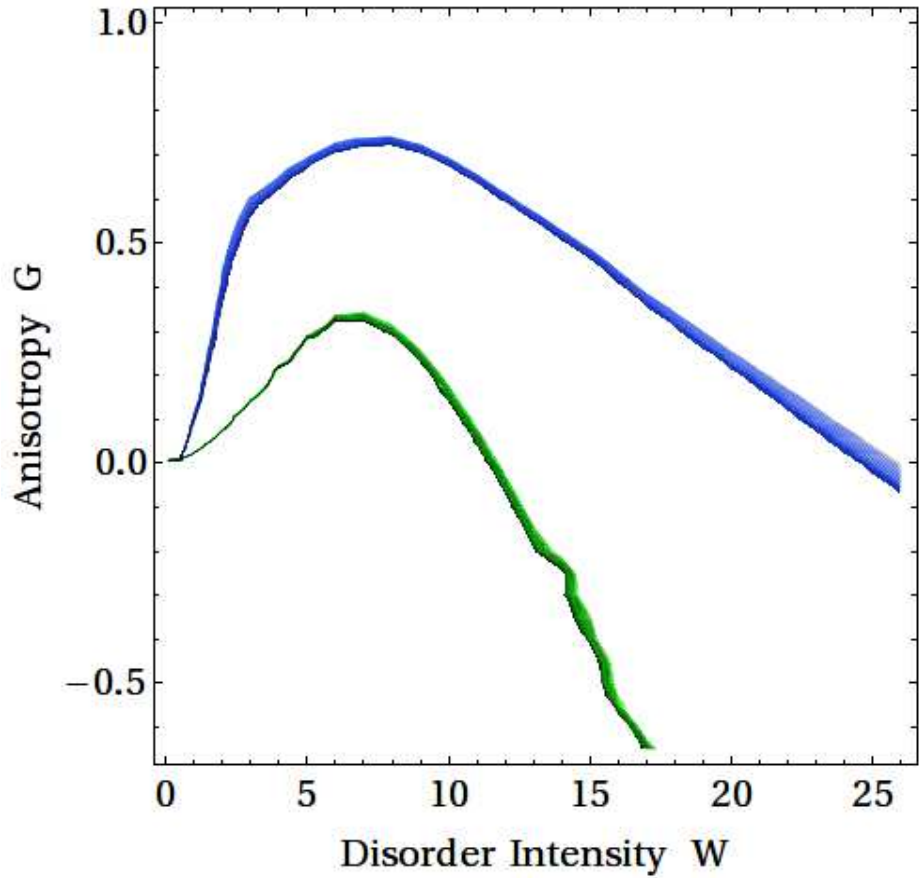}
\caption{``Phase diagram'' for a threshold value equal to $0.3$.}
\label{fig16-cap5}
\end{figure}

In the Fig. (\ref{fig16-cap5}) we show the ``phase diagram'' as described above for a fixed threshold value; the threshold is taken the same for both $|m_{xy}|$ and $\Upsilon_x$, because they are normalized in the same way. 
\vspace{1cm}

Into the ``phase diagram'' the region below the green line has a global superconducting order, the region between the green line and the blue line has only a local superconducting order, while the region above the blue line is a glassy CDW. This is exactly the behaviour that we were exptecting. We have to point out that changing the threshold value, the overall feature of the phase diagram remains the same. 

\clearpage
\newpage

  \clearpage{\pagestyle{empty}\cleardoublepage}

  \chapter*{Conclusions}
  \addcontentsline{toc}{chapter}{Conclusions}
  \markboth{Conclusions}{Conclusions}
  \label{Conclusion}
\vspace{2.5cm}

In this research we focused our attention to some open problems concerning HTS, particularly we concentrate ourselves to the pseudogap region of short coherence length HTS cuprates. Many questions about these materials are waiting for a response and a lot of experiments do not have a clear explanation yet, thus the research in this field of condensed matter physics is really exciting. 

The aim of our research was to find a possible explanation to some ``anomalous'' experiments performed in the pseudogap region of HTS, based on the study of a simplified coarse-grained model. However, before introducing the model we have revisited in the chapters (\ref{chapter1}) and (\ref{chapter2}) the basic concepts about the superconductivity and the charge-density-waves in order to have clear the context in which we have worked on. It is really important to underline that our research is based on two features observed directly by the experiments, i.e. the presence of CDW and precursor SC effects in the pseudogap region for these materials. Even if the observation of CDW in HTS comes back to years ago using different techniques, as neutron or X-ray scattering, more recently the STM has given us an important real-space probe for these materials, allowing to see more clearly their polycrystal/glassy like behaviour in the pseudogap region. Also the experiments concerning the Nerst and diamagnetic effects above the superconducting temperature $T_c$, are important and give us additional informations about the nature of the pseudogap region. 

Keeping in mind the above experimental evidences, we have built (see chapter (\ref{chapter3})) a coarse-grained model that  captures in a simple way the idea that the ``pseudo-gap'' phase is formed of bound fermion pairs which are close to a CDW instability but do not have long range order due to quenched  disorder. Following this line we modeled the charge degrees of freedom by an Ising order parameter because for simplicity we think to have only two differents CDW ``variants'' (even if in real systems they are much more than two), while the superconducting phase was modeled by an $XY$ spin variable. These degrees of freedom are been embodied together forming a single Heisenberg variable, where the order along the $z$-axis corresponds to CDW order and the order in the $xy$-plane corresponds to superconducting order. Moreover the $SO(3)$ symmetry is broken by introducing an anisotropic term favouring the CDW state; and the impurities always present in the real samples are reproduced by quenched random fields coupled to the $S_z$ component. So we wrote our model as: 
\begin{equation}
 H=-J\sum_{\langle i,j\rangle}\vec{S}_i\cdot\vec{S}_j-G\sum_i(S_i^z)^2+\frac{W}{2}\sum_ih_iS_i^z
\end{equation} 
where $\quad\vec{S}_i=\{S_i^x,S_i^y,S_i^z\}\quad$ is a classical Heisenberg spin with $|\vec{S}|=1$, $J>0$. The anisotropy $G>0$ favours a CDW, and $h_i$ are statistical indipendent quenched random variables with a flat probability distribution between $-1$ and $+1$ (also $W>0$). The policristal/glassy CDW configurations are surely obtained by this model in two-dimensions because, as descrided in chapter (\ref{chapter2}) using the Imry-Ma argument, the lower critical dimension for a random field Ising model is $d_l=2$.

As we saw in chapter (\ref{chapter3}) our model can be obtained starting from an attractive Hubbard model in the strong coupling limit at half-filling. Indeed in this limit it is possible to map the Hubbard model into a quantum antiferromagnetic Heisenberg model, for which the staggered magnetization both in the $XY$ plane and along the $z$ axis represents the order parameter corresponding to the SC and CDW order respectively. Because of our interest on long wave length physics where quantum effects can be taken into account as renormalization of the parameters, we can study the Heisenberg model considering classical spin variables and performing the transformation from antiferromagnetic to ferromagnetic Heisenberg model.

The competition between SC and CDW is a fundamental ingredient of our model, thus in chapter (\ref{chapter4}) we investigated their interplay into a simple one-dimensional geometry excluding the effect of the disorder, focusing our attention to intrisic features deriving from this competition. Taking a Josephson junction $SS'S$, where $S$ and $S'$ are HTS with $T_c'<T_c$, and putting ourselves into a temperature region $T_c'<T<T_c$, we studied if and how the superconducting order outside the CDW barrier $S'$ could propagate through the barrier itself. Another assumption was that we had strong superconductors outside the barrier, this means that the CDW order could not penetrate into the SC. Studying our model in the continous limit, we found that a characteristic length $\xi_g\sim\sqrt{J/G}$ exists; it represents the length over which our order parameter passes from the SC state to the CDW state. By the analytical and numerical solution of the problem we were able to build a phase diagram as function of the ratio $d/\xi_g$ between the length $d$ of the barrier and the characteristic length $\xi_g$, and of the superconducting phase difference $\Delta\Phi$ applied at the ends of the barrier. This phase difference $\Delta\Phi$ is directly related to the current that passes across the barrier. We found that in our phase diagram there was a transition line separating a perfect SC region from a MIXED region where our order parameter had an out of plane component. At zero $\Delta\Phi$ the pure SC region could be observed for barrier small with respect $\xi_g$, in other words the order parameter had not enough space to acquire a CDW component. Increasing $\Delta\Phi$ the energy cost of the SC solution over the entire barrier was higher than the MIXED energy solution, allowing the stabilization of the MIXED solution itself. Another important result of this study in one dimension was the behaviour of the current $I$ through the barrier as a function of the phase difference $\Delta\Phi$, that showed a non-analytical behaviour that could be tested experimentally (the standard Josephson like behaviour ($I\sim\sin\Delta\Phi$) was recovered only for high values of the ratio $d/\xi_g$). Finally we have to underline that keeping in mind the results obtained with this study, it is possible to give a possible explanation of the GPE. Indeed, even if in a Josephson junction experiment the barrier is a HTS parent compound with a short coherence length $\xi$, we can observe an anomalous proximity effect driven not by the short superconducting coherence length $\xi$, but by the length scale $\xi_g$, whose value could be much bigger than $\xi$. 

After that in chapter (\ref{chapter5}) we analyzed numerically our anisotropic random field Heisenberg model on a two dimensional square lattice. First of all we have to stress that the results are size indipendent as showed by the finite size analysis into the appendix (\ref{appendix5}), thus we concentrated ourselves on a fixed size ($100\times100$). Because we were interested in stable configurations (local minima of our Hamiltonian) we found those numerically through conjugate gardient methods, that physically  correspond to sudden experimental quenches from high temperature to low temperature. An important target of our research was to see if in systems where there was  competition between CDW and SC, we were able to see glassy-like CDW configurations and if a superconducting order could be found not only locally but also globally on the entire sample. We have to remember that our order parameter describes the SC when it stays in the $XY$ plane, while it is associated to the CDW when it points along the positive or negative $z-$axes (the two possible directions representing the two possible variants of the CDW in our model). So a natural way to obtain an information about the SC order in our system was to measure the in-plane magnetization $m_{xy}$. But we have to stress that this variable was not able to give us informations about the overall superconductivity in the sample, because it measured only local SC contributions, masking the possible global SC. For this reason we checked for the overall SC by using the stiffness $\Upsilon$, also called Helicity modulus. 

If into the system there was not disorder $(W=0)$ the ground state was a simple ferromagnet along the $z-$axis for every $G>0$ (with a double degeneracy, corresponding to all spins pointing either along the positive or the negative $z-$axis), and a ferromagnet into the $XY$ plane for every $G<0$ (with an infinite degeneracy, corresponding to all spins pointing along one of the infinite direction into the $XY$ plane). But due to the presence of quenched random fields, the system broke itself into ferromagnetic domains (along the $z-$axis), and among them there were Ising-like or Heisenberg-like domain walls. 
For a fixed value of the disorder intensity we were able to see that the interface (the SC state) between two different domains (two different CDW) became thicker decreasing the value of the anisotropy $G$. 
This behaviour is connected to the results of the chapter (\ref{chapter4}), indeed as we have seen there the length scale over which the spin goes from the $XY$ plane to the $z-$axis (i.e. from the SC to the CDW) is related to the anisotropy parameter $G$ in an inverse proportional way (i.e. $\xi_g\sim\sqrt{J/G}$), so increasing $G$ we obtain more small interfaces and arriving to a thickness of the order of the lattice spacing we pass from an Heisenberg like to an Ising like interface. While increasing the disorder intensity $W$ it was possible to obtain much more smaller domains, because the Binder scale $L_{Binder}$ (related to the size of the domains by $L_{Binder}\sim\exp(J/W)^2$) decreased. Situations for which the interfaces between different CDW domains were Ising like don't allow neither magnetization in the $XY$ plane, nor global stiffness; but clearly if we had Heisenberg like interfaces, this didn't guarantee to observe an overall stiffness in the sample, indeed could be possible to have disconnected Heisenberg interfaces that gives a finite contribute for the $XY$ magnetization but a zero contribute for the stiffness. 

We studied in a detailed way the behaviour of the magnetization $m_{xy}$ and of the stiffness $\Upsilon$
as a function both of the anisotropy $G$ and of the disorder intensity $W$. Their observed behaviour was qualitatively similar. For $G<0$ we observed a reduction of $|m_{xy}|$ and $\Upsilon$, corresponding simply to destroying the superconductivity by disorder, but for $G>0$ the disorder favours the establishment of the SC and for a fixed value of the anisotropy there was an optimum disorder intensity for which the in-plane magnetization and the stiffness were maximum. In order to clarify if the superconductivity was local or global we built a ``phase diagram'' into the $G-W$ plane, where we divided the plane into three regions: one had a global superconducting order, another one  had only a local superconducting order, while the last region had not any kind of superconducting order, and had a glassy CDW. 

Thus we can conclude saying that in strong coupling a CDW can become globally superconducting by introducing disorder, and there is an optimum value of the disorder intensity to enhance superconductivity. Also there is a region with short range superconducting correlations but not global stiffness resembling the pseudogap phase of underdoped cuprates. 

\begin{figure}[htb!]
\centering
\includegraphics[clip=True,scale=0.4]{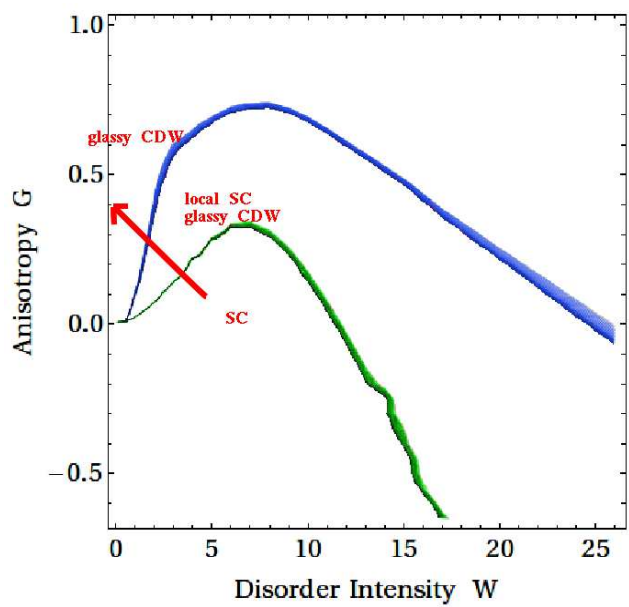}
\caption{``Phase diagram'' for our model.}
\label{fig:concl2}
\end{figure}
\begin{figure}[htb!]
\centering
\includegraphics[clip=True,scale=1.0]{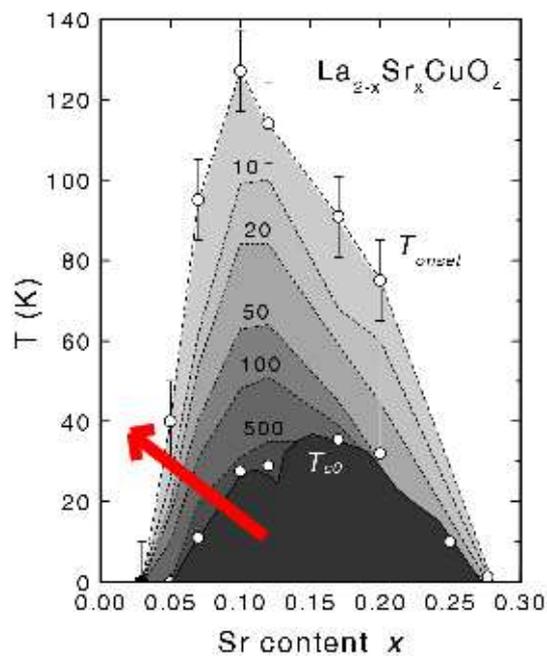}
\caption{Phase diagram of $La_{2−x}Sr_xCuO_4$ showing contour lines of the vortex Nernst signal observed above $T_{c0}$.}
\label{fig:concl1}
\end{figure}

\clearpage
\newpage

If we compare the phase diagram of our model with that one obtained for example by Ong in his Nerst experiment, we can find a strict correspondence between them. Indeed in both cases by putting disorder into the system, we can observe superconductivity; moreover, if we follow the red line drawn on Fig. (\ref{fig:concl1}) and (\ref{fig:concl2}), we pass from a completely superconducting region, to a region with only local superconductivity, and finally to an insulator region. 

We can also point out that the diamagnetic activity observed experimentally in the pseudogap phase of many HTS materials (see for example Fig: (\ref{fig:iguchi})) could be explained in our context if we think that along the ``soft'' interface of a CDW domain a superconducting current flows, expelling the magnetic field and generating a diamagnetic domain. However, this conjecture must be verified more deeply for our model studying it in the presence of a magnetic field.

  \clearpage{\pagestyle{empty}\cleardoublepage}

  \appendix
  \chapter{Some Results about Hubbard Model}
  \label{appendix3}
\vspace{3cm}

Here we shall give a little review of the basic results regarding the Hubbard model because of its importance in condensed matter physics. Indeed, this model and its many variants are widely used nowadays to investigate theoretically a lot of different features that can be observed in strongly correlated systems, particularly those concerning HTS. We shall give only a limited view of the results regarding the Hubbard model, just to give an idea of the main features that characterize the field we are working on. 

\subsection*{Some physical quantities}
We shall define some basic conserved quantities. The total-number operator $\hat{N}_e$
\begin{equation}
 \hat{N}_e=\sum_{i\in\Lambda}(n_{i,\uparrow}+n_{i,\downarrow})
\end{equation}
where $\Lambda$ indicate the lattice (while with $|\Lambda|$ we indicate the number of its sites), commutes with the Hamiltonian $H$. Since each lattice site can have at most two electrons, we have $0\leqslant N_e\leqslant2|\Lambda|$, where $N_e$ is the eigenvalue of the number operator $\hat{N}_e$, that tell us the total number of electrons that are in our lattice. 

The spin operator $\hat{S}^\alpha_i\equiv(\hat{S}^x_i,\hat{S}^y_i,\hat{S}^z_i)$ at site $i$ is defined by
\begin{equation}
 \hat{S}^\alpha_i=\frac{1}{2}\sum_{\sigma,\tau=\uparrow,\downarrow}c^\dagger_{i,\sigma}(p^\alpha)_{\sigma\tau}c_{i,\sigma}
\end{equation}
for $\alpha={x,y,z}$ where $p^\alpha$ are the Pauli matrices. The operators for the total spin of
the system are defined as: 
\begin{equation}
 \hat{S}^\alpha=\sum_i\hat{S}^\alpha_i
\end{equation}
The operators $\hat{S}^\alpha$ commute with the Hamiltonian, hence the latter is invariant under any global rotation in the spin space. On the other hand, as the operators $\hat{S}^\alpha$ do not commute with each other, we follow the standard convention in the theory of angular momenta, and simultaneously diagonalize, together with the Hamiltonian, the total-spin operator $\hat{S}^2$,
\begin{equation}
 \hat{S}^2=\sum_{\alpha=x,y,z}(\hat{S}^\alpha)^2
\end{equation}
and the $z-$component $S_z$. We denote by $S^z$ and $S(S+1)$ the eigenvalues of $\hat{S}^z$ and $\hat{S}^2$, respectively. For a given electron number $N_e$, we let:
\begin{eqnarray}
 S_{max} = \left \{
\begin{array}{l}
N_e/2\qquad {\mathrm when} \qquad 0\leqslant N_e\leqslant|\Lambda|\\\\
|\Lambda|-N_e/2\qquad {\mathrm when} \qquad |\Lambda|\leqslant N_e\leqslant2|\Lambda|
\end{array}
\right.
\end{eqnarray}
Then $S$ can be $S=0,1,\cdots,S_{max}$ (or $S=1/2,3/2,\cdots,S_{max}$). 

\subsubsection*{The Lieb-Mattis theorem}
The Lieb-Mattis theorem \cite{Lieb_1962} states that if we have an Hubbard model on a one dimensional lattice with open boundary conditions, and we assume that the hopping matrix elements satisfy $|t_{ii}|<\infty$, $|t_{ij}|<\infty$ for $\langle i,j\rangle$, and $t_{ij}=0$ otherwise, and also $U<\infty$ for any site, then the eigenvalues of the energy $E(S)$ for a fixed value of the spin satisfy the following inequality:
\begin{equation}
 E(S)<E(S+1)
\end{equation}
for any $S=0,1,\cdots,S_{max}-1$ or ($S=1/2,3/2,\cdots,S_{max}-1$). This means that the one-dimensional Hubbard model as described above has the total spin $S=0$ in its ground state. It is clear that from this fact one cannot conclude the system to be paramagnetic, but only say that there is no ferromagnetism. 

However, there are other rigorous results which show the absence of order in low-dimensional Hubbard models, among which we mention the extensions, by Ghosh \cite{Ghosh_1971} and Uhrig \cite{Uhrig_1992}, of the well known theorem of Mermin and Wagner. Ghosh proved that the Hubbard model in one or two dimensions does not exhibit symmetry breaking related to magnetic long-range order at any finite temperature. Uhrig similarly ruled out the possibility of a general planar magnetic ordering.

\subsubsection*{The Lieb's theorem}
Before describing the Lieb's theorem, let us first introduce the notion of bipartiteness. Consider a Hubbard model on a lattice $\Lambda$ with hopping matrix elements $t_{ij}$. The system is said to
be bipartite if the lattice $\Lambda$ can be decomposed into a disjoint union of two sublattices, 
$\Lambda=A\cup B$ (with $A\cap B=\oslash$), where $t_{ij}=0$ holds whenever $i,j\in A$ or $i,j\in B$. In other words, only hoppings between different sublattices are allowed. 

Then Lieb’s theorem \cite{Lieb_1989} for the repulsive Hubbard model is as follows. Consider a bipartite half-filled Hubbard model. We assume that $|\Lambda|$ is even, and that the whole of $\Lambda$ is connected through non-vanishing $t_{ij}$. We also assume that $U>0$ for any site. Then the ground states of the model are non-degenerate (apart from the trivial spin degeneracy), and have total spin $S=||A|-|B||/2$.

The total spin $S$ of the ground state determined in the theorem is exactly the same as that of the ground state of a corresponding Heisenberg antiferromagnet on the same lattice. It should be noted, however, that the knowledge of the total spin of the ground states in a finite volume does not necessarily allow one to determine the properties of the ground states in the corresponding infinite system. Despite the strong similarity between the half-filled Hubbard model and the Heisenberg antiferromagnet (expecially in the strong coupling limit as we'll see in the next section), there isn't any rigorous demonstration of the equivalence of the two models yet.

A very important corollary of Lieb's theorem is that half-filled Hubbard models on asymmetric bipartite lattices universally exhibit a kind of ferromagnetism, or more precisely, ferrimagnetism. Not all the spins in the system completely align with each other; spins on neighbouring sites have a tendency to point in opposite directions, but the big difference between the number of sites in the two sublattices cause the system to possess a bulk magnetic moment. Such a magnetic ordering is usually called ferrimagnetism.

\subsubsection*{Nagaoka's ferromagnetism}
At this point a good question could be if it does exist the possibility to observe the ferromagnetism in an Hubbard model. Up to now only few situations are known to have ferromagnetism, and one of them was  demonstrated by Nagaoka \cite{nagaoka_1966}, and it is also called Nagaoka's ferromagnetism. 

The Nagaoka's theorem states that if we take an arbitrary finite lattice $\Lambda$, and assume that $U\rightarrow\infty$ for every site, and we fix the electron number as $N_e=|\Lambda|-1$, then among the ground states of the model, there exist states with total spin $S=S_{max}=N_e/2$. The requirements that $U$ should be infinitely large and that there should be exactly one hole are admittedly rather pathological. Nevertheless, the theorem is very important since it showed for the first time in a rigorous manner that quantum mechanical motion of electrons and strong Coulomb repulsion can generate ferromagnetism. The conclusion that the system which has one less electron than the half-filled model exhibits ferromagnetism is indeed surprising. This is a very nice example which demonstrates that strongly interacting electron systems produce very rich physics.

\subsection*{Hubbard to Heisenberg mapping}
As we said in the chapter (\ref{chapter3}) the repulsive Hubbard model ($U>0$) can be transformed in the strong coupling limit ($U\gg t_{ij}$) into an effective spin Heisenberg model with antiferromagnetic coupling. Here we will describe in more detail the calculations performed in order to obtain this result as also showed in \cite{Chao_1978}. 

We remember that the starting idea is to divide the Hamiltonian into two pieces, one ($H_0$) associated to the intra-band dynamics, and the other one ($H_1$) associated to the inter-band dynamics; then performing a canonical transformation it is possible to obtain as a perturbation in $H_1$ the effective Heisenberg Hamiltonian. The decomposition of the starting Hamiltonian will be:
\begin{equation}
 H=H_0+\epsilon H_1
\end{equation}
Now we have to find an operator ${\mathcal S}$ for which the following canonical transformation:
\begin{equation}
\tilde H = e^{-i\epsilon\mathcal S}He^{i\epsilon\mathcal S}
\end{equation}
does not have linear $\epsilon$-terms, so expanding the exponential factors and using the following notation:
\begin{equation}
 [[A,B]]_n=[[\cdots[[A,B]]\cdots],B],B]
\end{equation}
with $n$ commutators in the r.h.s., the transformed Hamiltonian can be written as:
\begin{equation}
 \tilde H = H_0+\epsilon(H_1+i[H_0,\mathcal S])+\sum_{n=2}^\infty\frac{(-i\epsilon)^n}{n\!}(in[[H_1,\mathcal S]]_{n-1}-[[H_0,\mathcal S]]_n)
\end{equation}
In order to eliminate linear $\epsilon$ dependence we need to put:
\begin{equation}
 H_1+i[H_0,\mathcal S]=0
\end{equation}
so the transformed Hamiltonian up to $2^{nd}$ order in $\epsilon$, becomes:
\begin{equation}
 \tilde H=H_0+\frac{i}{2}[H_1,\mathcal S]
\end{equation}
At this point it is really useful to introduce the projector operators $P_1$ and $P_2$ respectively for the subspace with no double occupied states and for the subspace with empty or single occupied states. But first of all we introduce the $m$ electron projector $p(i,m)$ on the site $i$ (obviously $m$ can take only the values $\{0,1,2\}$), and for two single sites\footnote{Hereafter we will use a two sites Hamiltonian because we can observe that the full Hamiltonian can be written as a sum of two sites Hamiltonians, where the two sites of these Hamiltonians are obviously nearest neighbour. Only at the end we will write the result for the full lattice.} the complete Fock space is given by $\mathcal P_1\cup \mathcal P_2$:
\begin{eqnarray}
 \mathcal P_1 =\left \{
\begin{array}{l}
|0\,,0\rangle\\
|0\,,\uparrow\rangle\quad |0\,,\downarrow\rangle\quad |\uparrow\,,0\rangle\quad |\downarrow\,,0\rangle\quad\\
|\uparrow\,,\uparrow\rangle\quad |\uparrow\,,\downarrow\rangle\quad |\downarrow\,,\uparrow\rangle\quad |\downarrow\,,\downarrow\rangle\quad
\end{array}
\right.
\end{eqnarray}
\begin{eqnarray}
 \mathcal P_2 =\left \{
\begin{array}{l}
|\uparrow\downarrow\,,0\rangle\quad |0\,,\uparrow\downarrow\rangle\quad |\uparrow\downarrow\,,\uparrow\downarrow\rangle\quad\\
|\uparrow\downarrow\,,\uparrow\rangle\quad |\uparrow\downarrow\,,\downarrow\rangle\quad |\uparrow\,,\uparrow\downarrow\rangle\quad |\downarrow\,,\uparrow\downarrow\rangle\quad 
\end{array}
\right.
\end{eqnarray}
So we can now define the projectors $P_1$ and $P_2$ for the Fock subspaces $\mathcal P_1$ and $\mathcal P_2$ using the single site projectors $p(i,m)$:
\begin{equation}
 P_1=p(1,0)p(2,0)+\sum_{i\neq j}p(i,1)p(j,0)+p(1,1)p(2,1)
\end{equation}
\begin{equation}
 P_2=p(1,2)p(2,2)+\sum_{i\neq j}p(i,2)[p(j,0)+p(j,1)]
\end{equation}
Now we can calculate the expressions $P_\mu H P_\nu$. Using the relation \newline $p(i,m)p(i,n)=p(i,m)\delta_{m,n}$ and a lot of algebra we arrive to:
\begin{eqnarray}
P_1HP_1&=&t\sum_{i\neq j,\sigma}(1-n_{i\overline{\sigma}})c_{i\sigma}^\dagger c_{j\sigma}(1-n_{j\overline{\sigma}})\\
P_1HP_2&=&t\sum_{i\neq j,\sigma}(1-n_{i\overline{\sigma}})c_{i\sigma}^\dagger c_{j\sigma}n_{j\overline{\sigma}}\\
P_2HP_1&=&t\sum_{i\neq j,\sigma}n_{i\overline{\sigma}}c_{i\sigma}^\dagger c_{j\sigma}(1-n_{j\overline{\sigma}})\\
P_2HP_2&=&t\sum_{i\neq j,\sigma}n_{i\overline{\sigma}}c_{i\sigma}^\dagger c_{j\sigma}n_{j\overline{\sigma}}+U\sum_in_{i\uparrow}n_{i\downarrow}
\end{eqnarray}
where $P_1HP_1$ and $P_2HP_2$ represent the intra-band dynamics, while $P_1HP_2$ and $P_2HP_1$ bescribe the inter-band dynamics. So we can identify:
\begin{eqnarray}
 H_0&=&P_1HP_1+P_2HP_2\\
H_1&=&P_1HP_2+P_2HP_1
\end{eqnarray}
but in order to write down the expression of the effective Hamiltonian we need to know the operator $\mathcal S$ acts, so we apply the projectors $P_1$ and $P_2$ to the equality $ H_1+i[H_0,\mathcal S]=0$, obtaining:
\begin{equation}
 P_\mu HP_\nu(1-\delta_{\mu\nu})+iP_\mu HP_\mu P_\mu \mathcal SP_\nu-iP_\mu \mathcal SP_\nu P_\nu HP_\nu=0
\end{equation}
There are two possibilities:
\begin{enumerate}
 \item $\mu=\nu$ 
\begin{eqnarray}
&& P_\mu HP_\mu (P_\mu \mathcal SP_\mu)=(P_\mu \mathcal SP_\mu) P_\mu HP_\mu\nonumber\\\nonumber\\
&& P_\mu \mathcal SP_\mu=\gamma P_\mu\qquad con\quad\gamma\in \mathbb R
\end{eqnarray}
\item $\mu\neq\nu$
\begin{equation*}
 iP_\mu HP_\nu=P_\mu HP_\mu(P_\mu \mathcal SP_\nu)-(P_\mu \mathcal SP_\nu) P_\nu HP_\nu
\end{equation*}
approximating $P_\mu HP_\mu$ and $P_\nu HP_\nu$ with their expectation values $H_{\mu\mu}$ and $H_{\nu\nu}$ we obtain:
\begin{equation*}
 iP_\mu HP_\nu=(H_{\mu\mu}-H_{\nu\nu})P_\mu \mathcal SP_\nu
\end{equation*}
but $H_{\mu\mu}-H_{\nu\nu}=\pm U$ (thisi is the energy that divide the two subspaces), so:
\begin{eqnarray}
 P_1\mathcal SP_2 &=& -\frac{i}{U}P_1HP_2\\
P_2\mathcal SP_1 &=& \frac{i}{U}P_2HP_1
\end{eqnarray}
\end{enumerate}

Then the effective Hamiltonian is:
\begin{eqnarray}
 \tilde H &=& H_0+\frac{i}{2}[H_1,\mathcal S]=\nonumber\\
&=& H_0+\frac{i}{2}H_1\mathcal S-\frac{i}{2}\mathcal SH_1=\nonumber\\
&=& H_0+\frac{i}{2}(P_1HP_2+P_2HP_1)\mathcal S-\frac{i}{2}\mathcal S(P_1HP_2+P_2HP_1)\nonumber
\end{eqnarray}
or:
\begin{equation}
 \tilde H=P_1HP_1+P_2HP_2-\frac{1}{U}[P_1HP_2HP_1-P_2HP_1HP_2]
\end{equation}
At this point we have:
\begin{eqnarray}
 P_1HP_2HP_1&=&P_1HP_2P_2HP_1=\nonumber\\\nonumber\\
&=&t^2\sum_{i\neq j,\sigma}\sum_{l\neq m,\sigma'}(1-n_{i\overline{\sigma}})c^\dagger_{i\sigma}c_{j\sigma}n_{j\overline{\sigma}}n_{l\overline{\sigma'}}c^\dagger_{l\sigma'}c_{m\sigma'}(1-n_{m\overline{\sigma'}})\nonumber\\
\end{eqnarray}
\begin{eqnarray}
 P_2HP_1HP_2&=&P_2HP_1P_1HP_2=\nonumber\\\nonumber\\
&=&t^2\sum_{i\neq j,\sigma}\sum_{l\neq m,\sigma'}n_{i\overline{\sigma}}c^\dagger_{i\sigma}c_{j\sigma}(1-n_{j\overline{\sigma}})(1-n_{l\overline{\sigma'}})c^\dagger_{l\sigma'}c_{m\sigma'}n_{m\overline{\sigma'}}\nonumber\\
\end{eqnarray}
Now we have to consider every possibility, so for the two-sites interaction: either  $\delta_{i,l}\delta_{j,m}\delta_{\sigma,\sigma'}$ or $\delta_{i,m}\delta_{j,l}\delta_{\sigma,\sigma'}$    or $\delta_{i,l}\delta_{j,m}\delta_{\sigma,\overline{\sigma'}}$ or  $\delta_{i,m}\delta_{j.l}\delta_{\sigma\overline{\sigma'}}$, while for the three-sites interaction: either  $\delta_{j.l}\delta_{\sigma,\sigma'}$ or $\delta_{j,l}\delta_{\sigma,\overline{\sigma'}}$.

\begin{description}
 \item[Case $\delta_{i,l}\delta_{j,m}\delta_{\sigma,\sigma'}$]
\begin{equation}
 P_1HP_2HP_1=t^2\sum_{i\neq j,\sigma}(1-n_{i\overline{\sigma}})c^\dagger_{i\sigma}c_{j\sigma}n_{j\overline{\sigma}}n_{i\overline{\sigma}}c^\dagger_{i\sigma}c_{j\sigma}(1-n_{j\overline{\sigma}})=0\end{equation}
\begin{equation}
P_2HP_1HP_2=t^2\sum_{i\neq j,\sigma}n_{i\overline{\sigma}}c^\dagger_{i\sigma}c_{j\sigma}(1-n_{j\overline{\sigma}})(1-n_{i\overline{\sigma}})c^\dagger_{i\sigma}c_{j\sigma}n_{j\overline{\sigma}}=0 
\end{equation}
\end{description}

\begin{description}
 \item[Case $\delta_{i,l}\delta_{j,m}\delta_{\sigma,\overline{\sigma'}}$]
\begin{equation}
 P_1HP_2HP_1=t^2\sum_{i\neq j,\sigma}(1-n_{i\overline{\sigma}})c^\dagger_{i\sigma}c_{j\sigma}n_{j\overline{\sigma}}n_{i\sigma}c^\dagger_{i\overline{\sigma}}c_{j\overline{\sigma}}(1-n_{j\sigma})=0
\end{equation}
\begin{eqnarray}
P_2HP_1HP_2 &=& t^2\sum_{i\neq j,\sigma}n_{i\overline{\sigma}}c^\dagger_{i\sigma}c_{j\sigma}(1-n_{j\overline{\sigma}})(1-n_{i\sigma})c^\dagger_{i\overline{\sigma}}c_{j\overline{\sigma}}n_{j\sigma}=\nonumber\\
&=& -t^2\sum_{i\neq j,\sigma}n_{i\overline{\sigma}}c^\dagger_{i\overline{\sigma}}c^\dagger_{i\sigma}(1-n_{j\overline{\sigma}})(1-n_{i\sigma})c_{j\overline{\sigma}}c_{j\sigma}n_{j\sigma}=\nonumber\\
&=& -t^2\sum_{i\neq j,\sigma}c^\dagger_{i\overline{\sigma}}c^\dagger_{i\sigma}(1-n_{j\overline{\sigma}})(1-n_{i\sigma})c_{j\overline{\sigma}}c_{j\sigma}=\nonumber\\
&=& -t^2\sum_{i\neq j,\sigma}c^\dagger_{i\overline{\sigma}}c^\dagger_{i\sigma}c_{j\overline{\sigma}}c_{j\sigma}=-2t^2\sum_{i\neq j}c^\dagger_{i\uparrow}c^\dagger_{i\downarrow}c_{j\uparrow}c_{j\downarrow}\nonumber\\
\end{eqnarray}
\end{description}

\begin{description}
 \item[Case $\delta_{i,m}\delta_{j,l}\delta_{\sigma,\sigma'}$]
\begin{eqnarray}
 P_1HP_2HP_1 &=& t^2\sum_{i\neq j,\sigma}(1-n_{i\overline{\sigma}})c^\dagger_{i\sigma}c_{j\sigma}n_{j\overline{\sigma}}n_{j\overline{\sigma}}c^\dagger_{j\sigma}c_{i\sigma}(1-n_{i\overline{\sigma}})=\nonumber\\
&=& t^2\sum_{i\neq j,\sigma}n_{i\sigma}(1-n_{i\overline{\sigma}})n_{j\overline{\sigma}}(1-n_{j\sigma})\nonumber\\
\end{eqnarray}
\begin{eqnarray}
P_2HP_1HP_2 &=& t^2\sum_{i\neq j,\sigma}n_{i\overline{\sigma}}c^\dagger_{i\sigma}c_{j\sigma}(1-n_{j\overline{\sigma}})(1-n_{j\overline{\sigma}})c^\dagger_{j\sigma}c_{i\sigma}n_{i\overline{\sigma}}=\nonumber\\
&=& t^2\sum_{i\neq j,\sigma}n_{i\sigma}n_{i\overline{\sigma}}(1-n_{j\sigma})(1-n_{j\overline{\sigma}})\nonumber\\
\end{eqnarray}
\end{description}

\begin{description}
 \item[Case $\delta_{i,m}\delta_{j,l}\delta_{\sigma,\overline{\sigma'}}$]
\begin{eqnarray}
 P_1HP_2HP_1 &=& t^2\sum_{i\neq j,\sigma}(1-n_{i\overline{\sigma}})c^\dagger_{i\sigma}c_{j\sigma}n_{j\overline{\sigma}}n_{j\overline{\sigma}}c^\dagger_{j\overline{\sigma}}c_{i\overline{\sigma}}(1-n_{i\sigma})=\nonumber\\
&=& -t^2\sum_{i\neq j,\sigma}(1-n_{i\overline{\sigma}})c^\dagger_{i\sigma}c_{j\sigma}c^\dagger_{j\overline{\sigma}}c_{i\overline{\sigma}}(1-n_{i\sigma})=\nonumber\\
&=& -t^2\sum_{i\neq j,\sigma}c^\dagger_{i\sigma}c_{j\sigma}c^\dagger_{j\overline{\sigma}}c_{i\overline{\sigma}}=\nonumber\\
&=& -t^2\sum_{i\neq j}\big(c^\dagger_{i\uparrow}c_{i\downarrow}c^\dagger_{j\downarrow}c_{j\uparrow}+c^\dagger_{i\downarrow}c_{i\uparrow}c^\dagger_{j\uparrow}c_{j\downarrow}\big)=\nonumber\\
&=& -t^2\sum_{i\neq j}\big(S_i^+S_j^-+S_i^-S_j^+\big)
\end{eqnarray}
where $S^+$ and $S^-$ are the rising and lowering spin operators.
\begin{equation}
P_2HP_1HP_2 = t^2\sum_{i\neq j,\sigma}n_{i\overline{\sigma}}c^\dagger_{i\sigma}c_{j\sigma}(1-n_{j\overline{\sigma}})(1-n_{j\sigma})c^\dagger_{j\overline{\sigma}}c_{i\overline{\sigma}}n_{i\sigma}=0
\end{equation}
\end{description}

Now we can consider the three-sites terms (We underline also that the situations $\delta_{jm}\delta_{\sigma\sigma'}$, $\delta_{jm}\delta_{\sigma\overline{\sigma'}}$, $\delta_{il}\delta_{\sigma\sigma'}$, $\delta_{il}\delta_{\sigma\overline{\sigma'}}$ are all equal to zero as can be checked easily, while the situations $\delta_{im}\delta_{\sigma\sigma'}$ and $\delta_{im}\delta_{\sigma\overline{\sigma'}}$ are equal to the following if we rename the indeces).

\begin{description}
 \item[Case $\delta_{jl}\delta_{\sigma\sigma'}$]
 \end{description}
\begin{eqnarray}
 P_1HP_2HP_1&=&t^2\sum_{\langle ijm\rangle\sigma}(1-n_{i\overline{\sigma}})c^\dagger_{i\sigma}c_{j\sigma}n_{j\overline{\sigma}}c^\dagger_{j\sigma}c_{m\sigma}(1-n_{m\overline{\sigma}})\\
P_2HP_1HP_2&=&t^2\sum_{\langle ijm\rangle\sigma}n_{i\overline{\sigma}}c^\dagger_{i\sigma}c_{j\sigma}(1-n_{j\overline{\sigma}})c^\dagger_{j\sigma}c_{m\sigma}n_{m\overline{\sigma}}
\end{eqnarray}

\begin{description}
 \item[Case $\delta_{jl}\delta_{\sigma\overline{\sigma'}}$]
 \end{description}
\begin{eqnarray}
 P_1HP_2HP_1&=&t^2\sum_{\langle ijm\rangle\sigma}(1-n_{i\overline{\sigma}})c^\dagger_{i\sigma}c_{j\sigma}c^\dagger_{j\overline{\sigma}}c_{m\overline{\sigma}}(1-n_{m\sigma})\\
P_2HP_1HP_2&=&0
\end{eqnarray}

Before writing down the final effective Hamiltonian, we can point out that if our system is defined onto a bipartite lattice, and we can assume this, there exist the particle-hole symmetry and then it is possible to describe the system only for electron density up to half-filling. This condition enables to trash these kind of terms $P_2\cdots P_2$. So:
\begin{eqnarray}
 \tilde H&=&t\sum_{\langle ij\rangle,\sigma}(1-n_{i\overline{\sigma}})c_{i\sigma}^\dagger c_{j\sigma}(1-n_{j\overline{\sigma}})+\nonumber\\
&&-\frac{t^2}{U}\sum_{\langle ij\rangle,\sigma}n_{i\sigma}(1-n_{i\overline{\sigma}})n_{j\overline{\sigma}}(1-n_{j\sigma})+\nonumber\\
&&+\frac{t^2}{U}\sum_{\langle ij\rangle}\big(S_i^+S_j^-+S_i^-S_j^+\big)+\nonumber\\
&&-\frac{t^2}{U}\sum_{\langle ijm\rangle\sigma}(1-n_{i\overline{\sigma}})c^\dagger_{i\sigma}c_{j\sigma}n_{j\overline{\sigma}}c^\dagger_{j\sigma}c_{m\sigma}(1-n_{m\overline{\sigma}})+\nonumber\\
&&-\frac{t^2}{U}\sum_{\langle ijm\rangle\sigma}(1-n_{i\overline{\sigma}})c^\dagger_{i\sigma}c_{j\sigma}c^\dagger_{j\overline{\sigma}}c_{m\overline{\sigma}}(1-n_{m\sigma})
\end{eqnarray}
Introducing the following operators:
\begin{eqnarray}
 \left \{
\begin{array}{l}
\widehat{c}^\dagger_{i\sigma}=c^\dagger_{i\sigma}(1-n_{i\overline{\sigma}})\\
\widehat{c}_{i\sigma}=c_{i\sigma}(1-n_{i\overline{\sigma}})\\
\widehat{n}_{i\sigma}=\widehat{c}^\dagger_{i\sigma}\widehat{c}_{i\sigma}
\end{array}
\right.
\end{eqnarray}
it is possible to write the following effective Hamiltonian:
\begin{eqnarray}
 \tilde H &=& t\sum_{\langle ij\rangle,\sigma}\big(\widehat{c}_{i\sigma}^\dagger \widehat{c}_{j\sigma}+h.c\big)+\frac{4t^2}{U}\sum_{\langle ij\rangle}\Big(\vec{S}_i\cdot\vec{S}_j-\frac{\widehat{n}_i\widehat{n}_j}{4}\Big)+\nonumber\\
&&-\frac{t^2}{U}\sum_{\langle ijm\rangle\sigma}\big(\widehat{c}^\dagger_{i\sigma}\widehat{n}_{j\overline{\sigma}}\widehat{c}_{m\sigma}-\widehat{c}^\dagger_{i\sigma}\widehat{c}^\dagger_{j\overline{\sigma}}\widehat{c}_{j\sigma}\widehat{c}_{i\overline{\sigma}}+h.c.\big)
\end{eqnarray}

If this effective Hamiltonian is studied for the special half-filling $(n_{i\uparrow}+n_{i\downarrow}=1)$ case, it reduces to a quantum antiferromagnet Heisenberg model:
\begin{equation}
 \tilde H = J\sum_{\langle ij\rangle}\Big(\vec{S}_i\cdot\vec{S}_j-\frac{1}{4}\Big)
\end{equation}
where $J=4t^2/U$.

  \clearpage{\pagestyle{empty}\cleardoublepage}
  \chapter{GPE in Competing CDW-SC Systems}
  \label{appendix4}
\vspace{3cm}

In this appendix we show more explicitly the calculations performed in the chapter (\ref{chapter4}) regarding the MIXED solutions, because the others solutions (the MIXED/CDW and the SC) are trivial to solve.
\vspace{1cm}
\newline
Willing to solve the differential equation 
 \begin{eqnarray}
& &
\xi_g^2\frac{d^2\theta}{dx^2}+\xi_g^2I^2\frac{\sin\theta}{\cos^3\theta}+\frac{1}{2}\sin(2\theta)=0
\end{eqnarray}
with rigid superconducting boundary conditions, we can simply integrate one time obtaining:
\begin{eqnarray}
\frac{\xi_g^2}{2}\Big(\frac{d\theta}{dx}\Big)^2+\frac{\xi_g^2I^2}{2}\frac{1}{\cos^2\theta}+\frac{1}{2}\sin^2\theta+C=0
\end{eqnarray}

 The integration constant $C$ could be expressed as a function of the maximum (or minimum) $\theta_0$, which will be taken, thank to the symmetry of the problem, for $x=0$:
 \begin{equation}
  C=-\frac{\xi_g^2I^2}{2}\frac{1}{\cos^2\theta_0}-\frac{1}{2}\sin^2\theta_0
 \end{equation}

 Now we can write our differential equation as:
 \begin{eqnarray}
 \xi_g^2\Big(\frac{d\theta}{dx}\Big)^2+\xi_g^2I^2\frac{1}{\cos^2\theta}+\sin^2\theta-\xi_g^2I^2\frac{1}{\cos^2\theta_0}-\sin^2\theta_0=0
 \end{eqnarray}
and integrating by variables' separation the above equation, we have:
 \begin{equation}
 \int_{-d/2}^{x} \frac{dx}{\xi_g}=\int_{0}^{|\theta(x)|} \frac{d\theta}{\sqrt{\sin^2\theta_0-\sin^2\theta+\xi_g^2I^2\frac{1}{\cos^2\theta_0}-\xi_g^2I^2\frac{1}{\cos^2\theta}}}
 \end{equation}
 We use $|\theta(x)|$ to ensure r.h.s. to be positive because of the l.h.s. is obviously not negative. So we have:
 \begin{equation}
 \frac{2x+d}{2\xi_g}=\int^{|\theta(x)|}_{0} \frac{d\theta}{\sqrt{\sin^2\theta_0-\sin^2\theta+\xi_g^2I^2\frac{1}{\cos^2\theta_0}-\xi_g^2I^2\frac{1}{\cos^2\theta}}}
 \end{equation}
 We can simplify the above integral in the following way:
 \begin{eqnarray}
 & &
 \int^{|\theta(x)|}_{0} \frac{d\theta}{\sqrt{\sin^2\theta_0-\sin^2\theta+\xi_g^2I^2\frac{1}{\cos^2\theta_0}-\xi_g^2I^2\frac{1}{\cos^2\theta}}}=\nonumber\\
 = & &
 \int^{|\theta(x)|}_{0} \frac{d\theta}{\sqrt{\sin^2\theta_0-\sin^2\theta+\omega^2\frac{1}{\cos^2\theta_0}-\omega^2\frac{1}{\cos^2\theta}}}=\nonumber\\
 = & &
 \int^{|\theta(x)|}_{0} \frac{\cos\theta\cos\theta_0 d\theta}{\sqrt{(\sin^2\theta_0-\sin^2\theta)(\cos^2\theta\cos^2\theta_0+\omega^2)}}=\nonumber\\
 = & &
 \int_{0}^{|\sin\theta(x)|} \frac{\cos\theta_0dt}{\sqrt{(\sin^2\theta_0-t^2)(\omega^2+\cos^2\theta_0-\cos^2\theta_0t^2)}}=\nonumber\\
 = & &
 \int_{0}^{|\sin\theta(x)|} \frac{dt}{\sqrt{(\sin^2\theta_0-t^2)(\frac{\omega^2+\cos^2\theta_0}{\cos^2\theta_0}-t^2)}}
 \end{eqnarray}
 we can observe that we put $\sin|\theta(x)|=|\sin\theta(x)|$, and it is true for the range of variability of $\theta$; we also introduce the adimensional parameter $\omega$, that represents the current $I$ rescaled by the length scale $\xi_g$, so from here in we will refer to it as current too:
 \begin{equation}
  \omega=\xi_g I\,.
 \end{equation}
 If now we put:
 \begin{eqnarray}
 &&
 p^2=\frac{\cos^2\theta_0+\omega^2}{\cos^2\theta_0}\\
 &&
 q^2=\sin^2\theta_0
 \end{eqnarray}
 our integral can be written as:
 \begin{equation}
  \int_{0}^{|\sin\theta(x)|} \frac{dt}{\sqrt{(p^2-t^2)(q^2-t^2)}}
 \end{equation}
 After a change of variable and keeping in mind the expression of the elliptic integral of first kind:
 \begin{equation}
  F(\varphi;k^2)=\int_0^y \frac{dt}{\sqrt{(1-t^2)(1-k^2t^2)}}=\int_0^\varphi\frac{d\theta}{\sqrt{1-k^2\sin^2\theta}}
 \end{equation}
 we can write down:
 \begin{equation}
  \int_{0}^{|\sin\theta(x)|} \frac{dt}{\sqrt{(p^2-t^2)(q^2-t^2)}}=\frac{1}{p}F(\varphi;k^2)
 \end{equation}
 where we have:
 \begin{eqnarray}
 & &
 \sin\varphi=\Big|\frac{\sin\theta(x)}{\sin\theta_0}\Big|\\
 & &
 k^2=\frac{q^2}{p^2}=\frac{\cos^2\theta_0\sin^2\theta_0}{\cos^2\theta_0+\omega^2}
 \end{eqnarray}
 So we can write now an implicit expression for the solution $\theta(x)$ of our problem:
 \begin{equation}
 \frac{2x+d}{2\xi_g}=\frac{\cos\theta_0}{\sqrt{\cos^2\theta_0+\omega^2}}F(\varphi;k^2)
 \end{equation}
 
 Now we have to find a similar relation for the other component $\phi(x)$ of our order parameter. Starting from the conservation law:
 \begin{equation}
  I=\cos^2\theta\frac{d\phi}{dx}
 \end{equation}
 we can integrate by variables' separation, obtaining:
 \begin{equation}
  \int_0^{\phi(x)}d\phi=I\int_{-d/2}^{x}\frac{dx}{\cos^2\theta(x)}
 \end{equation}
 and then:
 \begin{equation}
  \phi(x)=I\int_{0}^{\theta(x)}\Big(\frac{d\theta}{dx}\Big)^{-1}\frac{d\theta}{\cos^2\theta}
 \end{equation}
 Using the differential equation that define $\theta(x)$ we obtain:
 \begin{equation}
  \phi(x)=\omega\int_{0}^{|\theta(x)|}\frac{d\theta}{\cos^2\theta\sqrt{\sin^2\theta_0-\sin^2\theta+\omega^2\frac{1}{\cos^2\theta_0}-\omega^2\frac{1}{\cos^2\theta}}}
 \end{equation}
 We can simplify this integral using again an elliptic integral; specifically:
 \begin{eqnarray}
 & &
 \int_{0}^{|\theta(x)|}\frac{d\theta}{\cos^2\theta\sqrt{\sin^2\theta_0-\sin^2\theta+\omega^2\frac{1}{\cos^2\theta_0}-\omega^2\frac{1}{\cos^2\theta}}}=\nonumber\\
 = & &
 \int_{0}^{|\theta(x)|}\frac{\cos\theta\cos\theta_0d\theta}{\cos^2\theta\sqrt{(\sin^2\theta_0-\sin^2\theta)(\cos^2\theta\cos^2\theta_0+\omega^2)}}=\nonumber\\
 = & &
 \int^{|\sin\theta(x)|}_{0} \frac{\cos\theta_0dt}{(1-t^2)\sqrt{(p^2-t^2)(q^2-t^2)}}
 \end{eqnarray}
 and now with the help of the elliptic integral of third kind,
 \begin{eqnarray}
 \Pi(\varphi;\alpha,k^2) & = &\int_0^y \frac{dt}{(1-\alpha^2t^2)\sqrt{(1-t^2)(1-k^2t^2)}}=\nonumber\\
 & = &
 \int_0^\varphi\frac{d\theta}{(1-\alpha\sin^2\theta)\sqrt{1-k^2\sin^2\theta}}
 \end{eqnarray}
 we can write:
 \begin{equation}
  \phi(x)=\omega\frac{\cos\theta_0}{\sqrt{\omega^2+\cos^2\theta_0}}\Pi(\varphi;\sin^2\theta_0,k^2)
 \end{equation}
 
 We summarize the above results:
 \begin{eqnarray}
 &&
 \frac{2x+d}{2\xi_g}=\frac{\cos\theta_0}{\sqrt{\cos^2\theta_0+\omega^2}}F(\varphi;k^2)\\
 &&
 \phi(x)=\omega\frac{\cos\theta_0}{\sqrt{\omega^2+\cos^2\theta_0}}\Pi(\varphi;\sin^2\theta_0,k^2)\\
 &&
 \sin\varphi=\Big|\frac{\sin\theta(x)}{\sin\theta_0}\Big|\\
 &&
 k^2=\frac{\cos^2\theta_0\sin^2\theta_0}{\cos^2\theta_0+\omega^2}\\
 &&
 \omega=I\xi_g\\
 &&
 \xi_g^2=\frac{\rho}{2g}
 \end{eqnarray}
 
 At this point we have two equations that implicitally define our order parameter, but we have to observe that these solutions are be found with the underlying assumption for which a $\theta_0\neq 0$ exists for $x=0$. So it is necessary to understand when $\theta_0$ really exists. In order to achieve this we can solve a self-consinstent equation for $\theta_0$, indeed taking the equation that define $\theta(x)$ and imposing $x=0$ we have a new equation that implicitally defines $\theta_0$:
 \begin{equation}
 \frac{d}{2\xi_g}=\frac{\cos\theta_0}{\sqrt{\cos^2\theta_0+\omega^2}}K(k^2)\,.
 \end{equation}
 $K(k^2)=F(\pi/2,k^2)$ represents the complete elliptic integral of first kind. 
  \clearpage{\pagestyle{empty}\cleardoublepage}
  \chapter{CDW vs SC: size analysis}
  \label{appendix5}
\vspace{3cm}

\section*{Interface thickness in two dimensions}
Here we will show the analytical solution of the problem of the interface thickness between two different CDW in two dimensions. 

First of all we have to write the rigth expression of the functional to be minimized:
\begin{eqnarray}
 F[\theta,\phi] & = & 
\int dxdy \Big\{\frac{\rho}{2}\Big[\Big(\frac{d \theta}{dx}\Big)^2+\Big(\frac{d \theta}{dy}\Big)^2+\cos^2\theta\Big(\frac{d\phi}{dx}\Big)^2+\nonumber\\
& & +\cos^2\theta\Big(\frac{d\phi}{dy}\Big)^2\Big]-g\sin^2\theta\Big\} \,.
\end{eqnarray}
thus the Eulero-Lagrange equations are:
\begin{equation}
\left \{
\begin{array}{l}
 \frac{\partial}{\partial x}\Big(\frac{\partial F}{\partial\theta_x}\Big)+\frac{\partial}{\partial y}\Big(\frac{\partial F}{\partial\theta_y}\Big)=\frac{\partial F}{\partial \theta}\\\\
 \frac{\partial}{\partial x}\Big(\frac{\partial F}{\partial\phi_x}\Big)+\frac{\partial}{\partial y}\Big(\frac{\partial F}{\partial\phi_y}\Big)=\frac{\partial F}{\partial \phi}
\end{array}
\right.
\end{equation}
that give us:
\begin{equation}
\left \{
\begin{array}{l}
 \rho\frac{\partial^2\theta}{\partial x^2}+\rho\frac{\partial^2\theta}{\partial y^2}+\rho\sin\theta\cos\theta\Big[\Big(\frac{\partial\phi}{\partial x}\Big)^2+\Big(\frac{\partial\phi}{\partial y}\Big)^2\Big]+2g\sin\theta\cos\theta=0\\\\
\frac{\partial}{\partial x}\Big[\rho\cos^2\theta\frac{\partial\phi}{\partial x}\Big]+\frac{\partial}{\partial y}\Big[\rho\cos^2\theta\frac{\partial\phi}{\partial y}\Big]=0
\end{array}
\right.
\end{equation}
If we define the Noether current vector $\vec{I}\equiv(I_x,I_y)$:
\begin{equation}
\left \{
\begin{array}{l}
 I_x=\cos^2\theta\frac{\partial\phi}{\partial x}\\\\
I_y=\cos^2\theta\frac{\partial\phi}{\partial y}\\\\
\end{array}
\right.
\end{equation}
we can rewrite the Eulero-Lagrange equations as:
\begin{equation}
\left \{
\begin{array}{l}
 \xi_g^2\bigtriangleup\theta+\xi_g^2I^2_x\frac{\sin\theta}{\cos^3\theta}+\xi_g^2I^2_y\frac{\sin\theta}{\cos^3\theta}+\sin\theta\cos\theta=0\\\\
\vec{\bigtriangledown}\cdot\vec{I}=0
\end{array}
\right.
\end{equation}

Now we have to specify the geometry and the boundary conditions for our problem. We take a square domain $d\times d$ with axes origin coicident with the centre of the square. The boundary conditions are taken as following:
\begin{equation}
\left \{
\begin{array}{l}
 \theta(-d/2,y)=-\pi/2\\
\theta(+d/2,y)=+\pi/2\\
\phi(x,-d/2)=0\\
\phi(x,+d/2)=\Delta\Phi
\end{array}
\right.
\end{equation}
and they imply that $I_x(-d/2,y)=I_x(+d/2,y)=I_y(-d/2,y)=I_y(+d/2,y)=0$. Moreover, using the equation $\vec{\bigtriangledown}\cdot\vec{I}=0$ we can write:
\begin{equation}
 \int_{-d/2}^{d/2}\vec{\bigtriangledown}\cdot\vec{I}dx=0
\end{equation}
\begin{equation}
 \frac{\partial}{\partial y}\int_{-d/2}^{d/2}I_ydx=-\int_{-d/2}^{d/2}\frac{\partial}{\partial x}I_xdx=\Big[I_x\Big]_{-d/2}^{d/2}\equiv0
\end{equation}

Thus the ``charge'' related to the current $I_y$ is conserved:
\begin{equation}
 Q_y=\int_{-d/2}^{d/2}I_ydx\equiv constant
\end{equation}
and this means that the current $I_y\equiv I_y(x)$ is a function only of the $x$ variable. At the same way we have that $I_x\equiv I_x(y)$ is a function only of the $y$ variable, but we know by the boundary conditions that $I_x(-d/2,y)=0$, so $I_x\equiv 0$ $\forall x,y$. 

Using the expression $I_y=\cos^2\theta(\partial\phi/\partial y)$, and keeping in mind that $I_y\equiv I_y(x)$, we will have necessarily:
\begin{equation}
\left \{
\begin{array}{l}
\theta(x,y)\equiv\theta(x)\\\\
\phi(x,y)\equiv f(x)+g(x)y
\end{array}
\right.
\end{equation}
This shows that there is translational invariance along the $y$ direction for the angle $\theta$. But we know also that the current $I_x=\cos^2\theta(\partial\phi/\partial x)\equiv0$, and also that $I_x\equiv I_x(y)$, thus $\theta\equiv\theta(x)$ and $(\partial\phi/\partial x)\equiv0$, i.e.:
\begin{equation}
 \phi(x,y)\equiv\phi(y)=A+By
\end{equation}
where $A$ and $B$ are two real numbers. We have obtained translational invariance along the $x$ direction for the angle $\phi$, while its variation along the $y$ direction is linear. The constants $A$ and $B$ can be easily obtained by the boundary conditions, so at the end we have:
\begin{equation}
 \phi(x,y)\equiv\phi(y)=\frac{\Delta\Phi}{2}+\frac{\Delta\Phi}{d}y
\end{equation}

Using these results the differential equation that defines the angle $\theta$ is simplyfied into:
\begin{equation}
\xi_g^2\frac{\partial^2\theta}{\partial x^2}+(\xi_g^2B^2+1)\sin\theta\cos\theta=0
\end{equation}

We have obtained the same equation solved for the one-dimensional case with the difference that the caracteristic length $\xi_g$ has to take into account of the boundary phase difference $\Delta\phi$:
\begin{equation}
 \theta(x)=2\arctan e^{x/\xi_g'}-\frac{\pi}{2}
\end{equation}
\begin{equation}
 \xi_g'=\frac{\xi_g}{\sqrt{1+B^2\xi_g^2}}=\frac{\xi_g}{\sqrt{1+(\frac{\Delta\Phi}{d})^2\xi_g^2}}
\end{equation}

\section*{Finite size effects}
Another important point to be discussed is the size dependence of the physical observables; we would not want that our results are strongly influenced by the size of our system. Now we shall display the behaviour of the square magnetization in the XY plane and stiffness as a function of the lattice size. We find that for large enough lattice the size effects are negligible, in fact all our results refer to a $100\times100$ square lattice for which the size is not important. 

In all plots it is possible to see that for every measured variable the lattice size is important only for small samples, and even if this behaviour is showed only for one value of the disorder intensity, it is valid in general. 

%
%
\clearpage
\newpage

\vspace{1cm}
\begin{figure}[htb!]
\centering
\includegraphics[clip=True,totalheight=0.9\textwidth,angle=90]{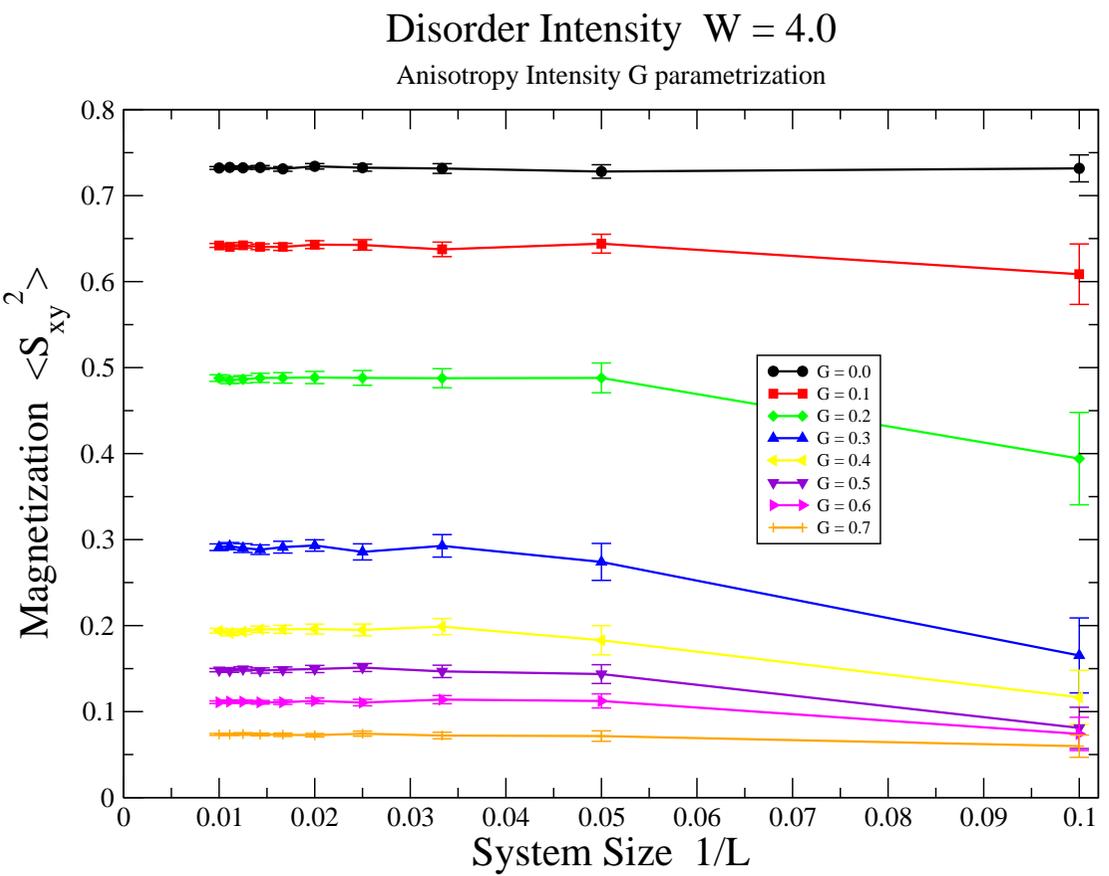}
\caption{Mean square magnetization in the $XY$ plane for a fixed disorder intensity and different anisotropy values.}
\label{fig2-cap5}
\end{figure}

\clearpage
\newpage

\vspace{1cm}
\begin{figure}[htb!]
\centering
\includegraphics[clip=True,totalheight=0.9\textwidth,angle=90]{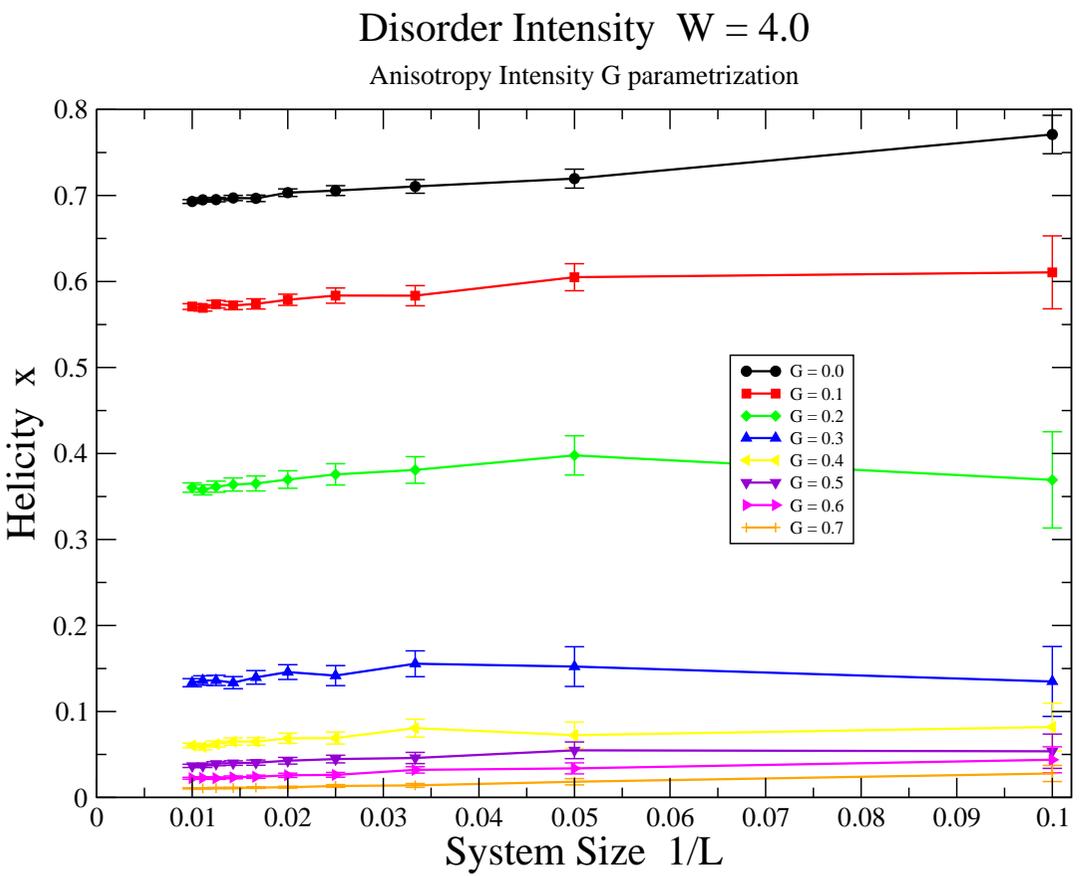}
\caption{Stiffness in the $XY$ plane along the X direction for a fixed disorder intensity and different anisotropy values.}
\label{fig3-cap5}
\end{figure}

  \clearpage{\pagestyle{empty}\cleardoublepage}
  

  \addcontentsline{toc}{chapter}{Bibliography}

\end{document}